\documentclass[amssymb,nobibnotes,nofootinbib,aps,prd]{revtex4}

 \usepackage{here}

\usepackage[dvipdfmx]{graphicx}
\usepackage{amssymb,amsfonts,amsmath,bm,color,multirow,cases,empheq,hyperref}
\usepackage{mathrsfs}
\definecolor{darkergreen}{rgb}{0,0.5,0}

\usepackage{enumerate}
\usepackage{ulem}
\usepackage{afterpage}

\hypersetup{%
 setpagesize=false,
 bookmarksnumbered=true,%
 bookmarksopen=true,%
 colorlinks=true,%
 linkcolor=blue,%
 citecolor=red}

\usepackage{tikz}
\usetikzlibrary{positioning,fit}
\usetikzlibrary{calc,intersections,through,backgrounds}
\usetikzlibrary{calc,arrows,decorations.markings,shapes.arrows,patterns, decorations.pathmorphing}
\usetikzlibrary{positioning,matrix,arrows,decorations.pathmorphing,decorations.pathreplacing}
\usetikzlibrary{arrows,matrix,positioning,shapes,arrows}
\usetikzlibrary{shapes.geometric}
\usetikzlibrary{decorations.text}
\usetikzlibrary{arrows.meta}

\tikzset{
  ->-/.style={decoration={markings, mark=at position 0.5 with {\arrow{to}}},
              postaction={decorate}},}
\tikzset{
  -<-/.style={decoration={markings, mark=at position 0.5 with {\arrow{to reversed}}},
              postaction={decorate}},}
              
\tikzset{
  pics/torus/.style n args={3}{
    code = {
      \providecolor{pgffillcolor}{rgb}{1,1,1}
      \begin{scope}[
          yscale=cos(#3),
          outer torus/.style = {draw,line width/.expanded={\the\dimexpr2\pgflinewidth+#2*2},line join=round},
          inner torus/.style = {draw=pgffillcolor,line width={#2*2}}
        ]
        \draw[outer torus] circle(#1);\draw[inner torus] circle(#1);
        \draw[outer torus] (180:#1) arc (180:360:#1);\draw[inner torus,line cap=round] (180:#1) arc (180:360:#1);
      \end{scope}
    }
  }
}

\newcommand{\tikznode}[2]{\relax
    \ifmmode%
    \tikz[remember picture,baseline=(#1.base),inner sep=0pt] \node (#1) {$#2$};
    \else
    \tikz[remember picture,baseline=(#1.base),inner sep=0pt] \node (#1) {#2};%
    \fi
}

\newcommand{\no}{\nonumber}

\newcommand{\cA}{\mathcal A}

\newcommand{\cF}{\mathcal F}

\newcommand{\cK}{\mathcal K}
\newcommand{\cM}{\mathcal M}\newcommand{\cN}{\mathcal N}

\newcommand{\cQ}{\mathcal Q}

\newcommand{\sfA}{\mathsf A}

\newcommand{\la}{\lambda}
\newcommand{\Tr}{{\rm Tr}}

\allowdisplaybreaks

\begin{document}

\begin{flushright}
TTI-MATHPHYS-40\\
OU-HET-1308
\end{flushright}
\vspace*{0.5cm}

\title{Monodromy-Matrix Description of Extremal Multi-centered Black Holes
}

\author{Jun-ichi Sakamoto$^{1}$ }
\email{jsakamoto@het.phys.sci.osaka-u.ac.jp}
\author{Shinya Tomizawa$^{2}$ }
\email{tomizawa@toyota-ti.ac.jp}
\affiliation{\vspace{3mm}$^{1}$Department of Physics, The University of Osaka\vspace{2mm}\\
Machikaneyama-Cho 1-1, Toyonaka, Japan 560-0043\vspace{2mm}\\
$^{2}$Mathematical Physics Laboratory, Toyota Technological Institute\vspace{2mm}\\Hisakata 2-12-1, Tempaku-ku, Nagoya, Japan 468-8511
\vspace{3mm}}

\begin{abstract}
We study solution-generating techniques based on the Breitenlohner--Maison linear system for extremal, stationary biaxisymmetric black hole solutions in five-dimensional $U(1)^3$ supergravity.
Focusing on multi-center configurations over a Gibbons--Hawking base, we analyze both BPS and almost-BPS solutions, including rotating single-center black holes and two-center black rings.
After dimensional reduction to three dimensions, the system is described by a coset sigma model with target space $SO(4,4)/[SO(2,2)\times SO(2,2)]$, where solutions are encoded in coset and monodromy matrices.
For Bena--Warner BPS solutions, we construct the coset and monodromy matrices and show that they admit an exponential representation governed by nilpotent elements.
Although the monodromy matrices generically exhibit double poles, they can be factorized explicitly using the nilpotent algebra of $\mathfrak{so}(4,4)$, reconstructing the solutions.
We extend this to almost-BPS solutions and derive the corresponding matrices.
While the single-center case exhibits commuting residues, the two-center black ring leads to a more intricate structure with a third-order pole, which disappears when regularity is imposed.
Finally, we analyze the extremal limits of the Rasheed--Larsen solution, where the fast-rotating branch is governed by idempotent elements.
We also construct an explicit $SO(4,4)$ duality transformation relating the slowly-rotating branch to a single-center almost-BPS solution.
These results will provide the BM formalism as a unified framework for extremal multi-center black holes.
\end{abstract}

\date{\today}

\maketitle

\section{Introduction}

The study of black hole solutions in Einstein gravity provides a fundamental testing ground for both classical and quantum aspects of gravity. 
In particular, higher-dimensional black holes have attracted considerable attention over the past two decades, motivated by developments such as the microscopic derivation of Bekenstein--Hawking entropy~\cite{Strominger:1996sh} and the possibility of black hole production in scenarios with large extra dimensions~\cite{Argyres:1998qn}.
Despite this progress, our understanding of higher-dimensional gravity remains limited due to its  complexity and richer structure. 
For instance, in five dimensions, the topology theorem for stationary black holes~\cite{Galloway:2005mf,Cai:2001su,Hollands:2007aj} admits the horizon cross-section to $S^3$, $S^1 \times S^2$, or lens spaces $L(p;q)$ under suitable symmetry and asymptotic conditions. 
Exact vacuum solutions are known for the first two cases~\cite{Tangherlini:1963bw,Myers:1986un,Emparan:2001wn,Pomeransky:2006bd}, whereas solutions with lens space topology remain elusive.
Although various black hole solutions have been constructed, largely through advances in solution-generating techniques, a complete classification is still lacking, even in 5D vacuum Einstein gravity.
Emparan and Reall~\cite{Emparan:2001wn} first constructed the exact solution describing an $S^1$-rotating black ring, revealing that 5D vacuum Einstein gravity admits multiple distinct solutions---including a spherical black hole and two black rings---with identical conserved charges, thereby providing a clear example of non-uniqueness in higher dimensions. 
Since 5D black holes can in general rotate along two independent planes, it is natural to seek more general black ring solutions carrying both $S^1$ and $S^2$ angular momenta.
An $S^2$-rotating black ring was subsequently obtained independently by Mishima and Iguchi~\cite{Mishima:2005id} and Figueras~\cite{Figueras:2005zp}, although these solutions suffer from conical singularities. 
A major breakthrough was achieved using the inverse scattering method (ISM)~\cite{Belinsky:1979mh}, which provides a systematic framework for constructing higher-dimensional solutions. 
While the ISM successfully generates $S^2$-rotating rings~\cite{Tomizawa:2005wv}, constructing $S^1$-rotation remains subtle due to the appearance of singularities from regular seeds. 
This issue was resolved by identifying an appropriate singular seed~\cite{Iguchi:2006rd,Tomizawa:2006vp}, leading to the construction of the balanced doubly rotating black ring by Pomeransky and Sen'kov~\cite{Pomeransky:2006bd}. 
More general unbalanced solutions~~\cite{Morisawa:2007di} and their compact forms~~\cite{Chen:2011jb} were later obtained.

\medskip
The presence of supersymmetry imposes strong constraints on charged solutions.
In particular, in 5D minimal supergravity, BPS (supersymmetric) solutions admit a systematic construction based on the bilinear formalism developed by Gauntlett {\it et al.}~\cite{Gauntlett:2002nw}. 
In this framework, solutions with a timelike Killing vector are described as a $\mathbb{R}$-bundle over a 4D hyper-K\"ahler base space, and the governing equations reduce to a set of linear equations, allowing the explicit construction of a wide class of solutions. 
In particular, supersymmetric black holes are subject to topological constraints: Reall showed that the horizon cross-section can be $S^3$, $S^1\times S^2$, $T^3$, or their quotients~\cite{Reall:2002bh}. 
Explicit realizations include the BMPV black hole~\cite{Breckenridge:1996is} and supersymmetric black rings~\cite{Elvang:2004rt}. 
Moreover, a supersymmetric black lens solution with horizon topology $L(2;1)=S^3/{\mathbb Z}_2$ was constructed~\cite{Kunduri:2014kja}, and more general black lens solution with $L(n;1)=S^3/{\mathbb Z}_n$ was obtained~\cite{Tomizawa:2016kjh,Breunholder:2017ubu}. 
In $U(1)^3$ supergravity, such supersymmetric configurations can be expressed in terms of harmonic functions on a 3D base space, enabling a systematic construction of multi-center solutions~\cite{Bena:2004de,Bena:2005ni,Gutowski:2004yv,Elvang:2004ds,Gauntlett:2004qy}.
As is to be expected, 
this construction is limited to BPS black hole solutions and cannot be extended to non-BPS cases.

\medskip
Meanwhile,  for non-BPS exact solutions~\cite{exact},  the sigma-model approach has been established as a powerful framework.
In 4D vacuum gravity with a single Killing vector, dimensional reduction leads to a non-linear sigma model with target space $SL(2,\mathbb{R})/SO(1,1)$ if the Killing vector is timelike, and $SL(2,\mathbb{R})/SO(2)$ if it is spacelike~\cite{Ernst:1967wx}. 
In the Einstein--Maxwell case, the corresponding coset becomes $SU(2,1)/S(U(1,1)\times U(1))$ for timelike reduction and $SU(2,1)/S[U(2)\times U(1)]$ for spacelike reduction~\cite{Ernst:1967by}.
These symmetries generate transformations such as the Ehlers and Harrison transformations, which add NUT charge and electromagnetic charge, respectively. 
Although rotation cannot be generated directly within this symmetry while preserving asymptotic flatness, it can be achieved by combining transformations associated with different choices of Killing vectors~\cite{Clement:1997tx,Clement:1999bv}.
More generally, higher-dimensional vacuum gravity with multiple commuting Killing vectors admits a sigma-model description with target space given by the coset $SL(D-2,\mathbb{R})/SO(2,D-4)$ or $SL(D-2,\mathbb{R})/SO(D-2)$, depending on the signature of the reduced directions~\cite{Maison:1979kx}.
In five dimensions, this structure allows the generation of rotating solutions such as the Myers--Perry black hole from static seeds by the Ehlers transformation~\cite{Giusto:2007fx}.
In 5D minimal supergravity, the coset structure is given by $G_{2(2)}/[SL(2,\mathbb{R})\times SL(2,\mathbb{R})]$ or $G_{2(2)}/SO(4)$~\cite{Mizoguchi:1998wv,Mizoguchi:1999fu}. 
Using this symmetry, charged rotating solutions can be generated via Harrison transformations~\cite{Bouchareb:2007ax}, although applying them to doubly rotating black rings generally leads to Dirac--Misner string singularities. 
Recently, Refs~\cite{Suzuki:2024coe,Suzuki:2024vzq} succeeded in resolving this issue by applying the transformation to a singular seed, thereby constructing a charged rotating black ring and obtaining a non-BPS asymptotically flat black hole with a nontrivial domain of communication (DOC) topology~\cite{Suzuki:2023nqf,Suzuki:2024phv,Suzuki:2024abu}. 
Moreover, the sigma-model approach has also been generalized to 5D $U(1)^3$ supergravity~\cite{Galtsov:2008bmt}, which can be regarded as a truncation of 11D supergravity compactified on $T^6$, and reduces to 5D minimal supergravity under an appropriate ansatz. 
Upon further dimensional reduction, the theory is described by 3D gravity coupled to a sigma model with target space given by the coset $SO(4,4)/[SO(2,2)\times SO(2,2)]$ or $SO(4,4)/[SO(4)\times SO(4)]$, depending on whether one of the reduced directions is timelike or all reduced directions are spacelike. 

\medskip
In these Einstein gravity and supergravity, assuming the existence of additional Killing vectors reduces the field equations to those of a 2D integrable coset sigma model~\cite{Maison:1979kx} defined on a conformally flat 2D space. 
The corresponding action takes the form 
\[
S=\int d\rho\, dz\, \rho\, \mathrm{Tr}\left(M^{-1}\partial_m M\, M^{-1}\partial^m M\right),
\]
where $M=M(x)$ ( $x=(\rho,z)$: Weyl--Papapetrou coordinates) is a coset matrix depending on each gravity theory. 
One of the most powerful solution-generating techniques in this framework is based on the Breitenlohner--Maison (BM) linear system~\cite{Breitenlohner:1986um}, further developed in~\cite{Chakrabarty:2014ora,Katsimpouri:2012ky,Katsimpouri:2013wka,Katsimpouri:2014ara}. 
This approach provides a unified framework that encompasses various solution-generating methods, including the inverse scattering method and sigma-model transformations such as the Ehlers and Harrison transformations.
A key advantage of this approach is that it does not rely on a specific choice of seed solutions. 
Instead, the central object is the monodromy matrix ${\cal M}(w)$ associated with the BM linear system. 
 The matrix is a meromorphic function of an auxiliary complex variable $w$, called the spectral parameter, and takes values on the Geroch group, which is an infinite-dimensional symmetry group underlying the 2D integrable coset sigma model. 
 For 5D vacuum Einstein theory, asymptotically flat, stationary and bi-axisymmetric black holes are uniquely determined by the asymptotic charges, the mass and two angular momenta and the rod data~\cite{Hollands:2007aj},  which includes the information on the topologies of the event horizon and the DOC.
Therefore,  clarifying how such rod data is encoded in the monodromy matrix would be useful, when  attempting to establish a systematic procedure to construct new black hole solutions. 
The exact gravitational solutions can be systematically constructed by solving a Riemann-Hilbert problem that involves factorizing the monodromy matrix, namely 
${\cal M}(w) = {\cal V}^{\sharp}(\lambda,x){\cal V}(\lambda,x)=X_-(\lambda,x)M(x)X_+(\lambda,x)$ ($\sharp$: anti-involution) with $X_+(\lambda,x):=V^{-1}(x){\cal V}(\lambda,x)$ and $X_-(\lambda,x):=X^\sharp_+(-1/\lambda,x)$, 
where ${\cal V}(\lambda,x)$ is the coset element of the BM linear system and $M(x)=V^\sharp(x)V(x)$ is the coset matrix on a symmetric space  (for example, $SO(4,4)/[SO(2,2)\times SO(2,2)]$ in $U(1)^3$ supergravity), obeying the field equations, with another spectrum parameter $\lambda$ and 2D coordinates $x$.  
Thus, it is expected that once one obtains the monodromy matrix ${\cal M}(w)$ from the rod data, the coset matrix $M(x)$ can be reconstructed by means of this factorization procedure. 
However, (1) the procedure for constructing the monodromy matrix from the rod data has not yet been clarified. 
Moreover, in general, (2) solving this factorization procedure is highly nontrivial. 
For 5D non-extremal black holes with spherical horizon topology, it is known that the monodromy matrix is a matrix-valued meromorphic function with only simple poles in $w$, and the associated residues are constant matrices independent of the coordinates $x=(z,\rho)$~\cite{Chakrabarty:2014ora}, where the factorization problem reduces to solving certain algebraic equations. 

\medskip

Motivated by these backgrounds, 
in Ref.~\cite{Sakamoto:2025xbq}, we constructed the monodromy matrix associated with the BM linear system, which provides a unified framework for describing three distinct asymptotically flat, vacuum black hole solutions with a single angular momentum in five dimensions, each with a different horizon topology: (i) the singly rotating Myers-Perry black hole, (ii) the Emparan-Reall black ring, and (iii) the Chen-Teo rotating black lens.
Conversely, by solving the corresponding Riemann-Hilbert problem using the procedure developed by Katsimpouri et al., we demonstrate that factorization of the monodromy matrix exactly reproduces these vacuum solutions, thereby reconstructing the three geometries.
Moreover, in Ref.~\cite{Sakamoto:2025sjq}, we extended this work to the BM linear system for the doubly rotating Myers-Perry black holes and the Pomeransky-Sen'kov black rings~\cite{Sakamoto:2025sjq}.
In this work, extending our previous studies~\cite{Sakamoto:2025xbq,Sakamoto:2025sjq} on non-extremal black holes to extremal ones, we investigate solution-generating techniques based on the BM linear system for extremal black hole solutions in 5D $U(1)^3$ supergravity. 
In particular, we focus on stationary, biaxisymmetric extremal solutions formulated as a $\mathbb{R}$-bundle over a 4D Gibbons--Hawking base space. 
We consider both BPS (supersymmetric) solutions, including horizonless solitons, and almost-BPS (non-supersymmetric) black hole solutions~\cite{Bena:2009ev} in this theory.
As is well known, BPS solutions are characterized by eight harmonic functions defined on a 3D flat space. 
At first sight, this may suggest that there is limited motivation to construct such solutions using integrable techniques, such as the inverse scattering method. 
However, there are two motivations for pursuing this approach.
\begin{itemize}
\item First, our goal is to develop a framework for the systematic construction of new almost-BPS black hole solutions with extremal horizons using the Breitenlohner--Maison (BM) formalism. 
A direct application of this approach to extremal solutions is technically challenging, since the corresponding monodromy matrices typically exhibit higher-order poles in the spectral parameter, in contrast to the non-extremal case where only simple poles appear. 
In the latter case, one can apply the standard factorization procedure for monodromy matrices developed in~\cite{Breitenlohner:1986um,Katsimpouri:2012ky,Chakrabarty:2014ora,Katsimpouri:2013wka}. 
For BPS solutions, previous analyses of Bena--Warner multi-center configurations within the BM framework~\cite{Roy:2018ptt} have been restricted to bubbling geometries, for which the monodromy matrix can be expressed as a sum of simple poles. 
Therefore, as a first step, we consider BPS black hole solutions with extremal horizons.

\item Second, it remains unclear how the rod structure and the asymptotic charges that characterize a black hole are encoded in the monodromy matrix. 
Since extremal BPS black holes possess a simpler structure than non-BPS solutions, it is expected that the relation between these quantities becomes more transparent in this case, thereby providing useful insights into the non-extremal black hole and almost BPS black hole solutions.

\end{itemize}
We summarize three main results of this paper as follows: 
\begin{itemize}
\item We first show that  the corresponding monodromy matrices can be factorized for a wide class of biaxisymmetric extremal BPS black hole solutions, even in the presence of higher-order poles, in a relatively simple manner.
While the monodromy matrices associated with BPS black hole solutions generally exhibit double poles, they are governed by nilpotent subalgebras of $\mathfrak{so}(4,4)$ that satisfy simple algebraic relations. 
Exploiting these relations, we show that the monodromy matrices for the most general biaxisymmetric Bena--Warner multi-center solutions can be factorized explicitly through elementary algebraic manipulations. 
This, in turn, enables the reconstruction of the original coset matrix and hence the corresponding gravitational fields.

\item Next, we extend  the associated monodromy-matrix description to almost-BPS solutions. 
After dimensional reduction to three dimensions, the system can be formulated as a symmetric coset space $SO(4,4)/(SO(2,2)\times SO(2,2))$.
We derive explicit coset matrices and monodromy matrices for two classes of almost-BPS (non-BPS) extremal solutions: 
a single-center rotating extremal black hole, and a two-center black ring. 
The resulting monodromy matrices again exhibit higher-order pole structures, as in the BPS case. 
For the single-center black hole, the residue matrices commute, allowing a direct factorization analogous to that of the BPS solutions. 
In contrast, for the black ring, the monodromy matrix naively contains a third-order pole at the horizon location. 
We show that this higher-order pole disappears precisely when the parameters are tuned to ensure the regularity of the horizon. 
This provides a concrete example of how the regularity conditions of a gravitational solution are encoded in the analytic structure of the monodromy matrix.
Thus, we can see that  in contrast to the BPS and almost-BPS black holes, the coset and monodromy matrices for the black ring are governed by a more intricate nilpotent algebraic structure.

\item Finally, we present an instructive example in which the algebraic structure of the monodromy matrix for an extremal black hole differs from the nilpotent algebra characteristic of BPS or almost-BPS solutions, and is instead governed by idempotent algebra. 
Such an example arises in the fast-rotating extremal limit of the Rasheed--Larsen solution, which describes a dyonic rotating black hole in 5D Kaluza--Klein theory~\cite{Rasheed:1995zv,Larsen:1999pp}. 
This solution admits two distinct extremal branches: a fast-rotating limit with an ergoregion and a slowly rotating limit without an ergoregion. 
We analyze the monodromy matrices associated with these two extremal limits. 
The slowly rotating branch is known to lie in the same duality orbit as the single-center almost-BPS solution~\cite{Bena:2009ev}, and its monodromy matrix is correspondingly characterized by nilpotent algebra. 
Furthermore, at the level of the coset matrix, we present an explicit $SO(4,4)$ duality transformation that maps the slowly rotating extremal solution to that of the single-center almost-BPS solution.
\end{itemize}

The rest of this paper is organized as follows. 
In Section~\ref{sec:MMKP}, we review the BPS and almost-BPS families of solutions in the M2--M5--KK6--P system and fix our conventions. 
In Section~\ref{sec:sigma}, we perform the dimensional reduction to three dimensions, construct the coset matrix. 
In Section~\ref{sec:mono}, we then derive the corresponding monodromy matrix and explain our method for explicitly factorizing monodromy matrices with double poles.
In Section~\ref{sec:example}, we focus on 5D minimal supergravity and analyze how the rod structure and regularity conditions are encoded in the residue (or charge) matrices for several asymptotically flat supersymmetric black hole solutions. 
In Section~\ref{sec:RL}, we study the monodromy matrix of the Rasheed--Larsen solution and its two extremal limits, and examine the associated algebraic structures. 
We also construct an explicit $SO(4,4)$ duality map relating the slowly rotating extremal limit to an almost-BPS configuration at the level of the coset matrix. 
Finally, in Section~\ref{sec:conclusion}, we conclude with a discussion of open problems and possible extensions to more general extremal and non-extremal multi-center solutions.

\section{Supergravity solutions of M2-M5-KK6-P brane system}\label{sec:MMKP}

In this section, we summarize a particular class of gravitational solutions described by the M2-M5-KK6-P brane system, which we consider in this work. The solutions include the Bena--Warner multi-center solutions \cite{Bena:2004de} and the almost-BPS solutions \cite{Goldstein:2008fq, Bena:2009ev}.

\subsection{M2-M5-KK6-P brane system}

In this paper, we consider a class of supersymmetric and non-supersymmetric gravitational solutions in eleven-dimensional supergravity that carry various M2, M5, KK6 monopole, and momentum charges.
The solutions have an internal structure of the form $(T^2)^3\sim T^6/(\mathbb{Z}_2\times \mathbb{Z}_2)$. 
We fix the volume of the internal space to unity and freeze all complex structure deformations. Then, the corresponding eleven dimensional metric and the 3-form gauge potential take the expressions
\begin{align}
\begin{split}
          ds^2_{11}&=ds_5^2+ds_{T^6}^2\,,\qquad
        \cA_3=\sum_{I=1}^{3}A^I\wedge dy^{2I-1}\wedge dy^{2I}\,.
\end{split}
\end{align}
The $T^6$ part with the coordinates $y^i\,(i=1,2,\dots,6)$ is 
\begin{align}
    ds_{T^6}^2&=\sum_{I=1}^{3}h^{I}\left(\left(dy^{2I-1}\right)^2+\left(dy^{2I}\right)^2\right)\,.
\end{align}
where the scalar functions $h^{I}$ satisfy the constraint $h^1h^2h^3=1$ and are parameterized by
\begin{align}
    h^{I}=\frac{Z^{\frac{1}{3}}}{Z_I}\,,\qquad Z=Z_1Z_2Z_3\,.
\end{align}
Performing a dimensional reduction along the $T^6$ direction, the 11D supergravity reduces to the 5D $U(1)^3$ supergravity with the action
\begin{align}
    S_{\rm 5D}=\int R_5\star_51-\frac{1}{2}G_{IJ}\star_5 dh^I\wedge dh^J
    -\frac{1}{2}G_{IJ}\star_5 F^I\wedge F^J-\frac{1}{6}C_{IJK}F^I\wedge F^J\wedge A^K\,,
\end{align}
where $F^I=dA^I$, and $C_{IJK}$ is the magnitude of the totally antisymmetric tensor i.e. $ C_{IJK}=\lvert\epsilon_{IJK}\lvert$, and $G_{IJ}$ is a diagonal matrix with $G_{II}=(h^I)^{-2}$. 
The 5D metric $ds_{5}^2$ and the gauge fields $A^I$ are
\begin{align}\label{5d-metric}
    ds^2_5=-\frac{1}{Z^{\frac{2}{3}}}\left(dt+\omega\right)^2
    +Z^{\frac{1}{3}}ds_{4}^2\,,\qquad
    A^I=-\varepsilon\frac{1}{Z_I}(dt+\omega)+B^I\,,
\end{align}
where the functions $Z^I$ depend on the coordinates $\{x^m\}$ on $ds_{4}^2$, and the 1-form fields $\omega$ and $B^I$ are defined on the 4D base space $ds_{4}^2=h_{mn}dx^mdx^n$, which describes a 4D hyper-K\"ahler metric. 
The real constant $\varepsilon$ takes the values $\varepsilon=\pm 1$, depending on the self-duality or anti-self-duality properties of the fields, as explained below.

\medskip

Depending on whether the solution is supersymmetric or non-supersymmetric, the curvature tensor of the 4D hyper-K\"ahler metric may be chosen to be either self-duality or anti-self-duality conditions. 
To write down the conditions, we introduce the field strength for the 1-form potential $B^I$,
\begin{align}
    \Theta^I=dB^I\,,\qquad I=1,2,3\,.
\end{align}
The self-duality condition is formulated as
\begin{align}\label{bps-eq}
    \Theta^I=\star_4\Theta^I\,,\qquad d\star_4dZ_I=\frac{C_{IJK}}{2}\Theta^J\wedge \Theta^K\,,\qquad d\omega+\star_4 d\omega=Z_I \Theta^I\,.
\end{align}
The condition corresponds to supersymmetric solutions describing an ansatz for a 1/8 BPS solution with three charges in five dimensions \cite{Bena:2004de}. The BPS conditions express all fields in terms of the eight harmonic functions which we will denote by $(V,M,K^I,L_I)$ with $I=1,2,3$ \cite{Gauntlett:2002nw,Gauntlett:2004qy,Bena:2005ni}.

\medskip

On the other hand, the anti-self-duality relation is given by \cite{Goldstein:2008fq}\footnote{The sign conventions used for the almost-BPS equations differ among previous studies. In this work, we follow the sign convention in \cite{Bena:2009qv}. At least within this convention, we can construct the associated coset matrix for almost-BPS solutions, and we show that these solutions are associated with the null orbit, in agreement with the observation of \cite{Bossard:2011kz}.}
\begin{align}\label{abps-eq}
    \Theta^I=-\star_4\Theta^I\,,\qquad d\star_4dZ_I=-\frac{C_{IJK}}{2}\Theta^J\wedge \Theta^K\,,\qquad d\omega-\star_4 d\omega=-Z_I \Theta^I\,.
\end{align}
While the geometry locally coincides with that of a supersymmetric solution, the absence of a globally well-defined Killing spinor renders the non-supersymmetric configuration.
In this sense, the non-supersymmetric solutions are called the almost-BPS solutions \cite{Goldstein:2008fq, Bena:2009ev}.
The almost-BPS equations (\ref{abps-eq}) cannot be solved in general only in terms of harmonic forms, but for certain special configurations, they can be solved using only harmonic functions \cite{Bena:2009ev} as we will see later.

\medskip

In the following, we restrict on a special case that the 4D base space $ds_4^2$ in (\ref{5d-metric}) describes a Gibbons--Hawking space with the metric
\begin{align}
    ds^2_4&=\frac{1}{V}(d\psi+\varpi)^2+V ds_{\mathbb{R}^3}^2\,.\label{4dGH}
\end{align}
The periodicity of the coordinate $\psi$ is taken to be $\psi \simeq \psi+4\pi$.
The 1-form field $\varpi$ on $\mathbb{R}^3$ satisfies
\begin{align}\label{V-def}
    \star_3d\varpi=+dV\,,
\end{align}
where the scalar function $V$ is a harmonic function defined on $\mathbb{R}^3$, and the Hodge star operator $\star_3$ is taken as the 3D base metric
\begin{align}\label{3dmetric}
    ds_{\mathbb{R}^3}^2=\sum_{i=1}^{3}(dx^i)^2=dr^2+r^2d\theta^2+r^2\sin^2\theta d\phi^2,
\end{align}
where $0\leq \theta\leq \pi$ and $0\leq \phi <2\pi$.

\subsection{Bena--Warner's supersymmetric multi-center solution}

Here, for later analysis, we summarize the explicit expressions of the Bena--Warner's supersymmetric black hole solutions in terms of harmonic functions \cite{Bena:2004de}.

\subsubsection{5D supersymmetric solutions}

The supersymmetric solutions correspond to the self-dual case of the 4D Gibbons--Hawking base space. Under this choice, the 5D gravitational solution (\ref{5d-metric}) takes the following form:
\begin{align}\label{5d-BW}
    ds^2_5&=-\frac{1}{Z^{\frac{2}{3}}}\left(dt+\omega\right)^2
    +Z^{\frac{1}{3}}\biggl[\frac{1}{V}(d\psi+\varpi)^2+V ds_{\mathbb{R}^3}^2\biggr]\,,\\
    A^I&=-\frac{1}{Z_I}(dt+\omega)+\frac{K^I}{V}(d\psi+\varpi)+\xi^I\,,\label{5dA-BW}\\
     \star_3d\varpi&= dV\,.\label{5d-BW-V}
\end{align}
The scalar functions $Z_I$ and the 1-forms $\xi^I$ are expressed in terms of the harmonic functions as
\begin{align}\label{5d-BW-xi}
      Z_I=\frac{1}{2}C_{IJK}V^{-1}K^JK^K+L_I\,,\qquad  \star_3d\xi^{I}=-dK^I\,.
\end{align}
The 1-form $\omega$ on the 4D Gibbons--Hawking space takes the form
\begin{align}\label{omega-ex}
    \omega&=\mu(d\psi+\varpi)+\omega_{\rm BW}\,,
\end{align}
where $\mu$ and $\omega_{\rm BW}$ are given by
\begin{align}
    \mu&=\frac{1}{6}C_{IJK}\frac{K^IK^JK^K}{V^2}+\frac{1}{2V}K^IL_I+M\,,\\
    \star_3d\omega_{\rm BW}&=VdM-MdV+\frac{1}{2}(K^IdL_I-L_IdK^I)\,.\label{omegabw}
\end{align}
Thus, the Bena--Warner's multi-center solutions are described by the eight harmonic functions $(V,M,K^I,L_I)$ on the 3D flat base space.

\subsubsection{Choice of harmonic functions}

\begin{figure}
\centering
\begin{tabular}{lcccccccccccccc}
\hline
   11D & $t$ & $\rho$ & $z$ & $\phi$ & $\psi$ & $y^1$ & $y^2$ & $y^3$ & $y^4$ & $y^5$ & $y^6$ & IIA & charge\\ \hline
   M2${}_{1}$ & - &  &    & &  & - & - & &  &  & &D2${}_{1}$ & $l^1_j$ \\
   M2${}_{2}$ & - &  &    & &  &  &  & - & - &  & &D2${}_{2}$& $l^2_j$ \\
   M2${}_{3}$ & - &  &    & &  &  &  & &  & - & -&D2${}_{3}$& $l^3_j$ \\
   M5${}_{1}$ & - &  &    & & - &  &  & - & - & - & - &D4${}_{1}$& $k_j^1$ \\
   M5${}_{2}$ & - &  &    & & - & - & - &  &  & - & - &D4${}_{2}$& $k_j^2$ \\
   M5${}_{3}$ & - &  &    & & - & - & - & - & - &  &  &D4${}_{3}$& $k_j^3$ \\
   KK6 & - &  &    & & - & - & - & - & - & - & - &D6& $q_j$ \\
   KK0(P) & - &  &    & & - &  &  &  &  &  &  &D0& $m_j$\\
 \hline
\end{tabular}
 \caption{Relation between brane charges and charges in harmonic functions. The angle variable $\psi$ expresses M-theory circle. }\label{fig.charge-brane}
 \end{figure}

We consider supersymmetric solutions described by the set of eight harmonic functions $(V,M,K^I,L_I)$ which have $N$ point sources (co-dimension 3) in the 3D Euclidean base $\mathbb{R}^3$, and are expressed as
\begin{align}
\begin{split}\label{hf-source}
    V&=q_0+\sum_{i=1}^{N}\frac{q_i}{r_i}\,,\qquad
    K^I=k_0^I+\sum_{i=1}^{N}\frac{k_i^I}{r_i}\,,\qquad
    L_I=l_0^I+\sum_{i=1}^{N}\frac{l_i^I}{r_i}\,,\qquad
    M=m_0+\sum_{i=1}^{N}\frac{m_i}{r_i}\,.
\end{split}
\end{align}
Here, $r_i$ denotes the distance between a point $\vec{x}$ and the position $\vec{x}_i$ of $i$-th center in the 3D base space $\mathbb{R}^3$,
\begin{align}
    r_i=|\vec{x}-\vec{x}_{i}|\,.
\end{align}
Since the Gibbons--Hawking metric (\ref{4dGH}) at $r_i=0$ is locally $\mathbb{R}^4/\mathbb{Z}_{q_i}$, we impose $q_i\in \mathbb{Z}$ in order for the quotient to be well-defined\footnote{This point describes a $\mathbb{Z}_{|q_j|}$ orbifold singularity. If we consider these solutions in string theory, it has been discussed that such singularities can be resolved and hence may be treated as acceptable singularities of the space.
Therefore, we do not necessarily regard such solutions as unphysical.}.
The residues at each center in the harmonic functions correspond to the charges of the associated branes, and this relationship is summarized in Fig.\ref{fig.charge-brane}.
In this coordinate system, we assume that all
centers of the harmonic functions are aligned along the z-axis, $\vec{x}_i=(0,0,w_i)$. 
This enhances the Gibbons--Hawking geometry to admit a $U(1)\times U(1)$ isometry. Solving the Hodge duality relations (\ref{5d-BW-V}) and (\ref{5d-BW-xi}) give the 1-form fields
\begin{align}
    \varpi=\sum_{i=1}^{N}q_i\cos \theta_i\,d\phi\,,\qquad \xi=-\sum_{i=1}^{N}k_i \cos\theta_i\,d\phi\,,
\end{align}
where
\begin{align}
    r_i=\sqrt{r^2+w_i^2-2w_ir\cos\theta}\,,\qquad \cos\theta_i=\frac{r\cos\theta-w_i}{\sqrt{r^2+w_i^2-2w_ir\cos\theta}}\,.
\end{align}
The 1-form $\omega_{\rm BW}$ is relatively complicated. For the choice (\ref{hf-source}) of harmonic functions, the solution to the Hodge duality relation (\ref{omegabw}) was constructed in \cite{Bena:2005va}\footnote{In 5D minimal supergravity, which is a special case of $U(1)^3$ supergravity, the explicit form of the 1-form $\omega_{\rm BW}$ for more general configurations of centers is given, for example, in appendix A of \cite{Dunajski:2006vs}.}. Here we present an explicit expression following \cite{Cassani:2025iix}, and refer the reader to appendix A of \cite{Cassani:2025iix} for the details of the computation:
\begin{align}\label{eq:omega-bw}
    \omega_{\rm BW}=\sum_{i=1}^{N}s_i\cos\theta_i d\phi+\sum_{i=1}^{N}\sum_{j>i}\frac{C_{ij}}{w_i-w_j}(1+\cos\theta_i)\left(1-\frac{r_i+w_i-w_j}{r_j}\right)d\phi\,,
\end{align}
where
\begin{align}
\begin{split}
    s_i&=\beta_i-\sum_{j\neq i}\frac{C_{ij}}{|w_i-w_j|}\,,\\
    \beta_i&=q_0 m_i-m_0 q_i+\frac{1}{2}\sum_{I=1}^{3}(k_0^Il_i^I-l_0^Ik_i^I)\,,\\
    C_{ij}&=q_im_j-m_i q_j+\frac{1}{2}\sum_{I=1}^{3}(k_i^Il_j^I-l_i^I k_j^I)\,.\label{omegabw-rep}
\end{split}
\end{align}
The corresponding gravitational solutions exhibit significantly different behaviors depending on the values taken by the $9N+8$ parameters $(q_0,q_i,k_0^I,k_i^I,l_0^I,l_0^I,m_0,m_I,w_i)$ appearing in the eight harmonic functions\footnote{The number of independent parameters here is counted without assuming any symmetries, such as translational invariance of the centers or the gauge symmetry of the harmonic functions \cite{Bena:2005va}.}.
These differences manifest in their asymptotic structures, regularity, and the presence or absence of horizons.
In the following, we briefly summarize the relations between the constraints on these parameters and the corresponding gravitational solutions.

\medskip

For arbitrary real parameters of the harmonic functions, the Bena--Warner multi-center solutions are not asymptotically 5D Minkowski spacetime.
The condition for the solution (\ref{5d-BW}) to describe an asymptotically 5D Minkowski spacetime is given by \cite{Bena:2005va}
\begin{align}\label{flat-con}
    q_0=0\,,\quad k_0^I=0\,,\quad l_0^I=1\,,\quad m_0=-\frac{1}{2}\sum_{j=1}^{N}\sum_{I=1}^{3}k_j^I\,,\quad  q_{\text{total}}=\sum_{i=1}^{N}q_i=\pm 1\,.
\end{align}
For the parameter regions of the harmonic functions corresponding to other major asymptotic geometries, see e.g. \cite{Bena:2007kg}.

\subsubsection{Regularity conditions of solutions}

For generic choices of the harmonic functions, the Bena--Warner multi-center solutions typically develop closed timelike curves (CTCs), which are physically unacceptable. In the 5D ansatz (\ref{5d-BW}), the absence of CTCs is ensured by requiring that the spatial components of the metric remain non-negative everywhere.
In this case, the harmonic functions must obey the inequalities \cite{Bena:2005va}
\begin{align}\label{noCTC1}
    VZ_I&=\frac{1}{2}C_{IJK}K^JK^K+L_IV \geq 0\,,\qquad I=1,2,3\,,\\
    &Z_1 Z_2 Z_3\, V - \mu^2\, V^2 \ge 0\,.\label{noCTC2}
\end{align}
The first condition becomes particularly nontrivial when $V$ can change sign. In such regions, one must check that the combinations $VZ_I$ remain non-negative everywhere, even though $V$ itself may be negative.

\subsubsection*{Absence of Dirac--Misner string and singularities}

Even after imposing these conditions, CTCs can still arise near each Gibbons--Hawking center if the 1-form field $\omega_{\rm BW}$ takes a nonzero value at the polar axes $\theta_i=0$ or $\theta_i=\pi$. This type of pathology arises from the presence of Dirac--Misner strings in the metric and must be removed in order to obtain a regular solution.
From the expression (\ref{eq:omega-bw}) of $\omega_{\rm BW}$, we need to impose \cite{Tomizawa:2016kjh}
\begin{align}\label{local_noCTC-0}
    s_i=q_0m_i-m_0\, q_i + \frac{1}{2}\sum_{I=1}^3 (l_i^Ik_0^I-k^I_il_0^I)-\sum_{j\neq i}\frac{C_{ij}}{r_{ij}}=0\,.
\end{align}
Furthermore, we need to impose restrictions on the parameters of the harmonic functions at centers corresponding to no horizon, since nonzero sources of brane charges generally give rise to singularities or black hole horizons.
The condition for eliminating the brane sources was given in \cite{Bena:2005va}, which imposes the following constraints on the parameters of the harmonic functions 
\begin{align}\label{bubbing}
    l_j^I=-\frac{1}{2}C_{IJK}\frac{k_j^Jk_j^K}{q_j}\,,\qquad m_j=\frac{k_j^1k_j^2k_j^3}{2q_j^2}\,.
\end{align}
This ensures that no curvature singularities appear in the domain of outer communications.
one requires that all centers other than horizons to be regular, together with the conditions (\ref{bubbing}), the constraint (\ref{local_noCTC-0}) reduces to \cite{Bena:2005va}
\begin{align}\label{local_noCTC}
\frac{1}{2}\sum_{j(\neq i)} C_{IJK}
\Pi^{(I)}_{ij}\,\Pi^{(J)}_{ij}\,\Pi^{(K)}_{ij}
\frac{q_i\,q_j}{r_{ij}}
=
q_0m_i-m_0\, q_i + \frac{1}{2}\sum_{I=1}^3 (l_i^Ik_0^I-k^I_il_0^I)\,,
\end{align}
where $r_{ij} = |\vec x_i-\vec x_j|$ and we imposed the regular condition (\ref{bubbing}).
The quantities $\Pi^{(I)}_{ij}$ are the flux differences across the non-contractible two-cycles, which are defined by an arbitrary curve connecting centers $i$ and $j$ together with the $U(1)$ fiber of the Gibbons--Hawking base space (\ref{4dGH}) :
\begin{align}
\Pi^{(I)}_{ij} = \frac{k^I_j}{q_j} -\frac{k^I_i}{q_i}\,, \qquad I=1,2,3\,.
\end{align}
These quantities are invariant under the shift $K^I\to K^I+c^I V$ with constants $c^I$ \cite{Bena:2005va}.
When the asymptotically flatness condition (\ref{flat-con}) is imposed, the condition (\ref{local_noCTC}) is reduced to (4.22) in \cite{Bena:2005va}.

\subsection{Almost-BPS solutions}

In general, an almost-BPS solution cannot be expressed in terms of harmonic functions, making the construction of analytic solutions a challenging problem.
Here, we present a 5D rotating five-charge extremal non-BPS black hole \cite{Bena:2009ev} and a non-BPS black ring solution as examples of almost-BPS solutions.

\subsubsection{5D almost-BPS solutions}

We begin by summarizing the 5D non-supersymmetric solutions on the 4D Gibbons--Hawking base space obtaining from the $T^6$ reduction of the almost-BPS solutions, together with the conditions that follow from the anti-self-duality condition.
The 5D metric and the abelian gauge fields are given by
\begin{align}\label{5d-abps}
\begin{split}
    ds^2_5&=-\frac{1}{Z^{\frac{2}{3}}}\left(dt+\omega\right)^2
    +Z^{\frac{1}{3}}\biggl[\frac{1}{V}(d\psi+\varpi)^2+V ds_{\mathbb{R}^3}^2\biggr]\,,\\
    A^I&=+\frac{1}{Z_I}(dt+\omega)+K^I(d\psi+\varpi)+\xi^I\,,\\
     \star_3d\varpi&= +dV\,,
\end{split}
\end{align}
where $K^I$ are harmonic function of the 3D flat base space and are compatible with the anti-self-duality condition.
The scalar functions $(Z_I, K^I)$ and 1-forms $(\omega,\varpi,\xi^I)$ are chosen to satisfy the anti-self-dual constraints. For completeness, we briefly summarize the corresponding conditions below. For the details, see for example \cite{Bena:2009ev}.
The scalar functions $Z_I$ and the 1-forms $\xi^I$ are expressed as
\begin{align}\label{abps-eq1}
      d\star_3dZ_I=\frac{1}{2}C_{IJK}V\,d\star_3d(K^JK^K)\,,\qquad  \star_3d\xi^{I}=V dK^I-K^I dV\,,
\end{align}
The 1-form $\omega$ on the 4D Gibbons--Hawking space takes the form
\begin{align}
    \omega&=\mu(d\psi+\varpi)+\omega_{\rm aBPS}\,,
\end{align}
One of the anti-self duality condition of the Gibbons--Hawking space gives a condition
\begin{align}\label{abps-omega-eq}
    d(V\,\mu)+\star_3 d\omega_{\rm aBPS}=-V Z_I dK^I\,.
\end{align}
Acting $d\star_3$ on the equation (\ref{abps-omega-eq}) leads to
\begin{align}\label{abps-eq3}
   d \star_3d(V\mu)=-d(V\,Z_{I})\wedge \star_3dK^I\,.
\end{align}
It is difficult to obtain the general solution to the above conditions, but they can be solved in certain special cases.
In the following, we consider two special cases, namely an extremal non-BPS rotating black hole with a single-center and a non-BPS black ring solution constructed in \cite{Bena:2009ev}.

\subsubsection{General extremal non-BPS rotating black holes}

In \cite{Bena:2009ev}, it has been shown that a rotating extremal non-BPS black hole with a single center, which is an asymptotically Taub-NUT solution, can be also characterized by the following eight harmonic functions:
\begin{align}
\begin{split}\label{hf-nonbps}
    V&=q_0+\frac{Q_6}{r}\,,\qquad K^I=0\,,\qquad L_I=l_0^I+\frac{Q_I}{r}\,,\qquad 
    M=m_0+\frac{m}{r}+\alpha\frac{\cos\theta}{r^2}\,,
\end{split}
\end{align}
where the real parameters $q_0$ and $m_0$ satisfy
\begin{align}\label{I-con}
    q_0-m_0^2=1\,.
\end{align}
Here, we introduced the new scalar function $M$ defined by $M=\mu V$, which is a harmonic function on $\mathbb{R}^3$ and satisfies
\begin{align}
    \star_3d\omega_{\rm aBPS}&=-dM\,.\label{omegaaBPS}
\end{align}
This follows from the equation (\ref{abps-eq3}) with $K^I=0$.
The real parameter $Q_6$ describes the D6 charge, $Q_I (I=1,2,3)$ describe the set of three distinct D2 brane charges, and $\alpha$ describes a dipole term which generates angular momentum of the solution.
Absence of Dirac--Misner strings of $\omega_{\rm aBPS}$ at $\theta=0,\pi$ requires
\begin{align}
    m=0\,.
\end{align}
We can confirm that under the choice (\ref{hf-nonbps}) of harmonic functions, the almost-BPS equations (\ref{abps-eq1}), (\ref{omegaaBPS}) and (\ref{abps-eq3}) can be solved by taking
\begin{align}
\begin{split}\label{abps-single}
    Z_I&=L_I\,,\qquad 
    \mu=\frac{M}{V}=\frac{m_0}{V}+\alpha \frac{\cos\theta}{Vr^2}\,,\qquad \xi^I=0\,,\\
    \varpi&=Q_6\,\cos\theta d\phi\,,\qquad \omega_{\rm aBPS}=\alpha\frac{\sin^2\theta}{r}d\phi\,,
\end{split}
\end{align}
where we used the Hodge-duality relation
\begin{align}
    \star_3d\left(\frac{\sin^2\theta}{r}d\phi\right)=-d\left(\frac{\cos\theta}{r^2}\right)\,.
\end{align}
The solution has a regular horizon identical to that of the BMPV black hole carrying the same charges \cite{Bena:2009ev}.
Furthermore, performing the dimensional reduction along the $\psi$-direction gives a 4D rotating extremal non-BPS black hole, which can serve as the seed solution for the most generic under-rotating non-BPS extremal black hole in the STU model and in $\cN=8$ supergravity in four dimensions. In particular, for special values of the charges it reduces to the under-rotating D0-D6 extremal black hole \cite{Rasheed:1995zv,Larsen:1999pp} constructed by Rasheed and Larsen.
This allows us to construct the corresponding monodromy matrix with a simpler structure.
Indeed, Teo and Wan \cite{Teo:2023wfd} and authors \cite{Tomizawa:2025tvb} have constructed a generalization of under-rotating Rasheed-Larsen's black hole solutions.

\subsubsection*{Almost-BPS black ring}

The second example is a non-BPS black ring solution with two centers \cite{Bena:2009ev}.
The non-BPS black ring solution is described by the eight scalar functions $(V,K^I,Z_I,\mu)$ given by
\begin{align}
    \begin{split}
V&=q_0+\frac{Q_6}{r}\,,\qquad K^I=\frac{d_I}{r_2}\,,\qquad 
Z_I^{-1}=l_0^I+\frac{Q_I}{r_2}+\frac{1}{6}\frac{C_{IJK}d_Jd_K}{r_2^2}\left(q_0+\frac{Q_6}{R^2} r\right)\,,
\label{scalar-ringabps}
\end{split}
\end{align}
and
\begin{align}
  \begin{split}\label{scalar-ringabps-mu}
    \mu&=\mu_L+\mu_P\,,\\
\mu_L&=\frac{1}{V}\left(m_0+\frac{m_1}{r}+\frac{m_2}{r_2}\right)+\alpha\frac{d_1 d_2d_3}{R V r_2^2}\left(q_0^2+\frac{Q_6^2}{R^2}\right)\left(\frac{r\cos\theta-R}{r_2}\right)\,,\\
\mu_P&=-\frac{\sum_{I=1}^{3}l_0^I d_I}{2r_2}-\frac{1}{Vr^2_2}\biggl[\frac{\sum_{I=1}^{3}l_0^I d_I}{2}\left(q_0+\frac{Q_6}{R}\cos\theta\right)+\frac{d_1d_2d_3}{r_2}\left(\left(q_0^2+\frac{Q_6^2}{R^2}\right)\frac{r\cos\theta}{R}+\frac{q_0 Q_6}{2 R^2}\frac{3r^2+R^2}{r}\right)\biggr]\,,
\end{split}
\end{align}
where $r_1$ and $r_2$ in the spherical coordinates are taken as
\begin{align}
   r_1=r\,,\qquad  r_2=\sqrt{r^2+R^2-2 r R\cos\theta}\,.
\end{align}
The scalar function $\mu$ is decomposed into two parts, $V\mu_L$ and $V\mu_P$, where $V\mu_L$ is a solution of the Laplace equation and $V\mu_P$ is a solution of the Poisson equation,
\begin{align}
    d\star_3 d(V \mu_L)=0\,,\qquad d\star_3 d(V \mu_P)+\sum_{I=1}^{3}d\left(V Z_I \star_3 d K^I\right)=0\,.
\end{align}
For a real parameter $\alpha$, the set of scalar functions satisfies the almost-BPS equations, but the gravitational solution has a regular horizon with finite area only when $\alpha$ takes the following values:
\begin{align}\label{regular-alpha}
    \alpha=\frac{q_0^2R^2}{q_0^2R^2+Q_6^2}\,.
\end{align}
The corresponding 1-form fields $\varpi$ and $\xi^I$ are given in
\begin{align}
    \varpi=Q_6\,\cos\theta d\phi\,,\qquad \xi^I=d_I\left(q_0\frac{r\cos\theta-R}{r_2}+\frac{Q_6}{R}\frac{r-R\cos\theta}{r_2}\right)d\phi\,.
\end{align}
Corresponding to the decomposition of the function $\mu$ into two parts, we similarly decompose $\omega_{\rm aBPS}$ into two parts $\omega_{{\rm aBPS},L}$ and $\omega_{{\rm aBPS},P}$ such that they satisfy
\begin{align}
    d(V\mu_{L})+\star_3 d\omega_{{\rm aBPS},L}=0\,,\qquad d(V\mu_{P})+\star_3 d\omega_{{\rm aBPS},P}=-V Z_I dK^I\,.
\end{align}
Solving these equations gives \cite{Bena:2009ev}
\begin{align}
\begin{split}
    \omega_{\rm aBPS}&=\omega_{{\rm aBPS},L}+\omega_{{\rm aBPS},P}\,,\\
    \omega_{{\rm aBPS},L}&=\biggl[\kappa-m_1\cos\theta-m_2\frac{r\cos\theta-R}{r_2}+\alpha\left(q_0^2+\frac{Q_6^2}{R^2}\right)d_1d_2d_3\frac{r^2\sin^2\theta}{R r_2^3}\biggr]d\phi\,,\\
    \omega_{{\rm aBPS},P}&=-\frac{1}{2}\sum_{I=1}^{3}l_0^I \xi^I-\biggl[Q_6\sum_{I=1}^{3}Q_Id_I\frac{r\sin^2\theta}{2 R r_2^2}+\left(q_0^2+\frac{Q_6^2}{R^2}\right)d_1d_2d_3\frac{r^2\sin^2\theta}{R r_2^3}\\
    &\quad+q_0Q_6d_1d_2d_3\left(\frac{r-R\cos\theta}{2R^3r_2}+\frac{r \sin^2\theta}{R r_2^3}\right)\biggr]d\phi\,,
\end{split}
\end{align}
where $\kappa$ is an integration constant.
Absence of the Dirac--Misner string requires to set 
\begin{align}
\begin{split}\label{nonbps-ring-DM}
    m_1&=\kappa=Q_6\left(\frac{1}{2R}\sum_{I=1}^{3}l_0^Id_I+\frac{q_0 d_1d_2d_3}{2R^3}\right)\,,\\
    m_2&=-\frac{1}{2}\left(q_0+\frac{Q_6}{R}\right)\sum_{I=1}^{3}l_0^Id_I-\frac{q_0Q_6 d_1d_2d_3}{2R^3}\,.
\end{split}
\end{align}
Unlike the BPS case, in which the balance equations include at most terms proportional to the inverse of the distance between centers, the balance condition for the almost-BPS solution involves terms proportional to $R^{-3}$, and therefore exhibits a more intricate structure.

\newpage

\section{Sigma model description of M2-M5-KK6-P brane system}\label{sec:sigma}

In this section, following \cite{Roy:2018ptt}, we perform a dimensional reduction of the supersymmetric and non-supersymmetric solutions discussed in the previous section to three dimensions, and formulate the resulting theory as a 3D coset sigma model with a symmetric coset space as the target space, coupled to 3D Einstein gravity.
For the supersymmetric solutions, this has already been performed in \cite{Roy:2018ptt}, and we give a brief review for completeness. We then perform a similar analysis for the almost-BPS case and discuss the algebraic differences between the two classes of the gravitational solutions.

\subsection{Coset space description}

After dimensional reduction of 5D $U(1)^3$ supergravity to three dimensions, the scalar moduli space has an $SO(4,4)$ isometry, and the resulting model is described by a 3D coset sigma model coupled with 3D Einstein gravity. Since the parametrization of the coset space in terms of scalar moduli depends on the order of dimensional reductions, we follow the prescription of \cite{Roy:2018ptt}, in which the reduction from five to three dimensions is performed in the following sequence:
\begin{align}\label{5d-d1d5p-e-gen}
\begin{split}
       ds_5^2&=-f^2(dt+\check{A}^0)^2+f^{-1}ds_4^2\,,\\
    A^{I}&=\chi^I(dt+\check{A}^0)+\check{A}^I\,,
\end{split}
\end{align}
and then
\begin{align}
\begin{split}\label{4d-3d-m-iib}
    ds_4^2&=e^{2U}(d\psi+\omega_3)^2+e^{-2U}ds_3^2\,,\\
     \check{A}^{\Lambda}&=\zeta^{\Lambda}(d\psi+\omega_3)+\hat{A}^{\Lambda}\,.
\end{split}
\end{align}
Here, the indices take values $I=1,2,3$, and $\Lambda=0,1,2,3$.
The field strengths $\hat{F}_2^{\Lambda}$ and $\hat{F}_2$ for the 1-form fields $\hat{A}^{\Lambda}$ and $\omega_3$ are defined by
\begin{align}\label{fs-aomega}
   \hat{F}_2^{\Lambda}=d\hat{A}^{\Lambda} \,,\qquad \hat{F}_2=d\omega_3\,.
\end{align}
We utilize the fact that in 3D space, the field strengths $\hat{F}_2^{\Lambda}$ and $\hat{F}_2$ can be dualized to scalar fields $\tilde{\zeta}_{\Lambda}$ and $\sigma$ using the Hodge duality as follows \cite{Sahay:2013xda}
\begin{align}
    d\tilde{\zeta}_{\Lambda}&=e^{2U}({\rm Im}\,N)_{\Lambda \Sigma}\star_{3}(\hat{F}_2^{\Sigma}+\zeta^{\Sigma}\hat{F}_2)-({\rm Re}\,N)_{\Lambda\Sigma}d\zeta^{\Sigma}\,,\label{dual-tzeta-eq}\\
   d\sigma&= -2e^{4U}\star_{3}\hat{F}_2-\tilde{\zeta}_{\Lambda}d\zeta^{\Lambda}+\zeta^{\Lambda}d\tilde{\zeta}_{\Lambda} \,,\label{dual-sig-eq}
\end{align}
where the matrix $N_{\Lambda\Sigma}\,(\Sigma=0,1,2,3)$ is a complex symmetric matrix and is computed from the prepotential of the 4D Euclidean STU model \cite{Roy:2018ptt}.
The explicit expressions of the real and the imaginary parts of $N_{\Lambda\Sigma}$ are given by
\begin{align}
   {\rm Re}N&=
    \begin{pmatrix}
       2\chi^1\chi^2\chi^3&\chi^2\chi^3&\chi^1\chi^3&\chi^1\chi^2\\
       \chi^2\chi^3&0&\chi^3&\chi^2\\
      \chi^1\chi^3&\chi^3&0&\chi^1\\
      \chi^1\chi^2&\chi^2&\chi^1&0
    \end{pmatrix}
    \,,\qquad
   {\rm Im}N=f
    \begin{pmatrix}
       f^2+\sum_{i=1}^{3}\frac{(\chi^i)^2}{(h^i)^2}&-\frac{\chi^1}{(h^1)^2}&-\frac{\chi^2}{(h^1)^2}&-\frac{\chi^3}{(h^1)^2}\\
       -\frac{\chi^3}{(h^1)^2}&-\frac{1}{(h^1)^2}&0&0\\
     -\frac{\chi^2}{(h^1)^2}&0&-\frac{1}{(h^2)^2}&0\\
      -\frac{\chi^3}{(h^1)^2}&0&0&-\frac{1}{(h^3)^2}
    \end{pmatrix}\,.
\end{align}
Thus, gravitational solutions are characterized in sixteen scalar fields
\begin{align}
    \{U,x^I:=-\chi^I,y^I:=fh^I, \tilde{\zeta}_\Lambda,\zeta^\Lambda,\sigma\}\,,
\end{align}
and their dynamics is described by a 3D symmetric coset sigma model coupled to 3D gravity.

\medskip

The resulting 3D action is given by~\cite{Roy:2018ptt}
\begin{align}\label{3d-sigma}
    S_3=\int d^3x\sqrt{g_3}\biggl(R_3-\frac{1}{2}\Tr(M^{-1}\partial_m M M^{-1}\partial^m M)\biggr)\,.
\end{align}
The sigma-model field $M(\vec{x})$ ($\vec{x}$ is a point the 3D base space $\mathbb{E}^3$) take values in a symmetric coset space
\begin{align}\label{sym-coset}
    \frac{G}{H}=\frac{SO(4,4)}{SO(2,2)\times SO(2,2)}\,.
\end{align}
Each Lie group has a $8\times 8$ matrix realization defined by
\begin{align}
    G&=SO(4,4)=\{~g \in GL(8,\mathbb{R}) ~\lvert~ g^{T}\eta g=\eta\,,~ {\rm det}g=1~\}\,,\\
    H&=SO(2,2)\times SO(2,2)=\{~g \in SO(4,4)~ \lvert ~g^{T}\eta' g=\eta'~\}\,,
\end{align}
where the invariant metrics $\eta$ and $\eta'$ for $G$ and $H$ are
\begin{align}
    \eta=
    \begin{pmatrix}
        0_4&1_4\\
        1_4&0_4
    \end{pmatrix}\,,\qquad
    \eta'=\text{diag}(-1,1,-1,1,-1,1,-1,1)\,.
\end{align}
By following \cite{Roy:2018ptt}, we express the coset matrix $M(\vec{x})$ as
\begin{align}\label{ginv-M}
    M(\vec{x})=V^{\natural}(\vec{x})V(\vec{x}) \in G\,,
\end{align}
where $\natural:G\to G$ is an anti-involutive automorphism 
\begin{align}
    x^{\natural}=\eta' x^{T}\eta'\qquad \text{for}\quad x\in G\,.
\end{align}
The element $V\in G$ in terms of the 16 scalar fields $\{U,x^I,y^I, \tilde{\zeta}_\Lambda,\zeta^\Lambda,\sigma\}$ is in the Iwasawa gauge
\begin{align}\label{iwasawa-rep}
    V&=e^{-U\,\mathbb{H}_0}\cdot\left(\prod_{I=1}^{3}e^{-\frac{1}{2}(\log y^I)\mathbb{H}_I}\cdot e^{-x^I\mathbb{E}_{I}}\right)\cdot e^{-\zeta^{\Lambda}\mathbb{E}_{q_{\Lambda}}-\tilde{\zeta}_{\Lambda}\mathbb{E}_{p^{\Lambda}}}\cdot e^{-\frac{1}{2}\sigma \mathbb{E}_0}\,,
\end{align}
where $\{\mathbb{H}_{\Lambda}, \mathbb{E}_{\Lambda},\mathbb{E}_{q_{\Lambda}}, \mathbb{E}_{p^{\Lambda}}, \mathbb{F}_{\Lambda},\mathbb{F}_{q_{\Lambda}},\mathbb{F}_{p^{\Lambda}}\}\,(\Lambda=0,1,2,3)$ are the 28 generators  of the semisimple Lie algebra $\mathfrak{g}=\mathfrak{so}(4,4)$, and the $8\times 8$ matrix representation is used in the appendix of \cite{Roy:2018ptt}.
The 3D base space in the Weyl-Papapetrou coordinate takes the form 
\begin{align}
    ds_{3}^2&=e^{2\nu}\left(d\rho^2+dz^2\right)+\rho^2d\phi^2\,.
\end{align}
The relation between the Weyl-Papapetrou coordinates $(\rho,z)$ and the spherical coordinates $(r,\theta)$ is given by
\begin{align}\label{WP-def}
    \rho=r\,\sin\theta\,,\qquad z=r\,\cos\theta\,.
\end{align}
Since any solutions with 4D Gibbons--Hawking base have the 3D flat base space, the conformal factor $e^{2\nu}$ is trivial 
\begin{align}\label{conf-fac}
    e^{2\nu}=1\,.
\end{align}

\subsection{Bena--Warner's supersymmetric multi-center solution}

We start with presenting the 16 scalar fields for the Bena--Warner's supersymmetric multi-center solutions Eqs.~(\ref{5d-BW})--(\ref{omegabw}), and then construct the corresponding coset matrix, denoted by $M_{\rm BW}(\vec{x})$. The derivation of these scalar fields can be seen in \cite{Roy:2018ptt}, and we only summarize the results here. The scalar fields are given by
\begin{align}
\begin{split}\label{scalar-hf}
    e^{2U}&=V^{-1}\,,\qquad
    \sigma=2V^{-1}\,,\qquad
    \tilde{\zeta}_{\Lambda}=0\,,\\
    \zeta^0&=\mu=\frac{1}{6}C_{IJK}\frac{K^IK^JK^K}{V^2}+\frac{1}{2V}K^IL_I+M\,,\quad
    \zeta^I=V^{-1}K^I\,,\\
   x^I&=-\chi^I=Z_I^{-1}=\left(\frac{1}{2}C_{IJK}V^{-1}K^JK^K+L_I\right)^{-1}\,,\\
   y^I&=f h^I=Z_I^{-1}=\left(\frac{1}{2}C_{IJK}V^{-1}K^JK^K+L_I\right)^{-1}\,.
\end{split}
\end{align}
Here, the scalar fields $\tilde{\zeta}^{\Lambda}$ and $\sigma$ are obtained by solving the Hodge-duality relations (\ref{dual-tzeta-eq}) and (\ref{dual-sig-eq}), which require the explicit expressions of the field strengths $\hat{F}_2^{\Lambda}$ and $\hat{F}_2$, as well as the complex symmetric matrix $N_{\Lambda\Sigma}$.
These quantities can be read off from the Bena--Warner ansatz (\ref{5d-BW}), and the field strengths take the form
\begin{align}
   \hat{F}_2^{0}=d\hat{A}^{0}=d\omega_{\rm BW}\,,\qquad \hat{F}_2^{I}=d\hat{A}^{I}=d\xi^I \,,\qquad \hat{F}_2=d\omega_3=d\varpi\,,
\end{align}
and the real and the imaginary parts of the period matrix $N_{\Lambda\Sigma}$ are given by \cite{Roy:2018ptt}
\begin{align}\label{bw-coset}
   {\rm Re}N&=
    \begin{pmatrix}
       -2Z^{-1}&Z^{-1}Z_1&Z^{-1}Z_2&Z^{-1}Z_3\\
       Z^{-1}Z_1&0&-Z_3^{-1}&-Z_2^{-1}\\
      Z^{-1}Z_2&-Z_3^{-1}&0&-Z_1^{-1}\\
      Z^{-1}Z_3&-Z_2^{-1}&-Z_1^{-1}&0
    \end{pmatrix}
    \,,\qquad
   {\rm Im}N=
    \begin{pmatrix}
       -2Z^{-1}&Z^{-1}Z_1&Z^{-1}Z_2&Z^{-1}Z_3\\
       Z^{-1}Z_1&-\frac{Z_1^2}{Z}&0&0\\
      Z^{-1}Z_2&0&-\frac{Z_2^2}{Z}&0\\
      Z^{-1}Z_3&0&0&-\frac{Z_3^2}{Z}
    \end{pmatrix}\,.
\end{align}
Using these scalar fields (\ref{scalar-hf}), we can compute the corresponding coset matrix (\ref{ginv-M}) constructed from the group element $V(\vec{x})$ with the parametrization (\ref{iwasawa-rep}). 
As shown in Ref.~\cite{Roy:2018ptt}, this coset matrix where is written in an exponential representation
\begin{align}\label{bw-gauge-mat2}
    M_{\rm BW}(\vec{x})=Y_{\rm BW}\exp\left(\sum_{j=1}^{N}\frac{\sfA_j}{r_j}\right)\,.
\end{align}
The constant matrix $Y_{\rm BW}$ encodes the asymptotic behavior of the corresponding solution, and its components are characterized by the constant terms of the harmonic functions as follows
\begin{align}\label{Y-BW}
    Y_{\rm BW}= \begin{pmatrix}
        Y_{11}&l_0^2&Y_{13}&k^1_0&-1&l_0^3&0&-2m_0\\
        -l_0^2&0&k_0^3&0&0&-1&0&0\\
        Y_{13}&-k_0^3&Y_{33}&q_0&0&-k_0^2&-1&l_0^1\\
        -k_0^1&0&-q_0&0&0&0&0&-1\\
        -1&0&0&0&0&0&0&0\\
        -l_0^3&-1&k_0^2&0&0&0&0&0\\
        0&0&-1&0&0&0&0&0\\
        2m_0&0&-l_0^1&-1&0&0&0&0
    \end{pmatrix}
    \,,
\end{align}
where the components $Y_{11}\,,Y_{13}\,,Y_{33}$ are given by
\begin{align}
\begin{split}
    Y_{11}&=l_0^2l_0^3-2k_0^1m_0\,,\\
    Y_{13}&=\frac{1}{2}(k_0^1l_0^1-k_0^2l_0^2-k_0^3l_0^3-2m_0q_0)\,,\\
    Y_{33}&=k_0^2k_0^3+l_0^1q_0\,.
\end{split}
\end{align}
In particular, for the asymptotically flat solutions, the asymptotic matrix (\ref{Y-BW}) becomes
\begin{align}
    Y_{\rm flat}= \begin{pmatrix}
        1&1&0&0&-1&1&0&-2m_0\\
        -1&0&0&0&0&-1&0&0\\
        0&0&0&0&0&0&-1&1\\
        0&0&0&0&0&0&0&-1\\
        -1&0&0&0&0&0&0&0\\
        -1&-1&0&0&0&0&0&0\\
        0&0&-1&0&0&0&0&0\\
        2m_0&0&-1&-1&0&0&0&0
    \end{pmatrix}
    \,,
\end{align}
where the condition (\ref{flat-con}) are used.
The matrices $\sfA_j \in \mathfrak{so}(4,4)$ associated with each center are nilpotent of degree three i.e.
\begin{align}\label{nil-con}
    \mathsf{A}_j^3=0\,,
\end{align}
and are expanded as
\begin{align}\label{sfA-mat}
    \mathsf{A}_j&=-q_j \mathbb{F}_0-\sum_{I=1}^{3}l_{j}^I\mathbb{F}_I-\beta_j \mathbb{F}_{p^0}+\sum_{I=1}^{3}k_{j}^{I}\mathbb{F}_{p^I}-2m_j\mathbb{E}_{q_0}\,,
\end{align}
where the constants $\beta_j$ are defined in (\ref{omegabw-rep}). The matrix rank of $\sfA_j$ are $\text{Rank}\,\sfA_j=4$ for arbitrary parameters of the harmonic functions.
When the regular condition \eqref{bubbing} is imposed at the $j$-th center, the corresponding matrix $\sfA_j$ becomes nilpotent of degree two  
\begin{align}
    \mathsf{A}_j^2=0
\end{align}
with $\text{Rank}\,\sfA_j=2$.
For later analysis, it is convenient to expand the exponential, and it takes the form
\begin{align}\label{bw-gauge-mat}
    M_{\rm BW}(\vec{x})&=Y_{\rm BW}\biggl(1+\sum_{j=1}^{N}\frac{1}{r_j}\mathsf{A}_j+\frac{1}{2}\sum_{k,l=1}^{N}\frac{1}{r_kr_l}\{\mathsf{A}_k,\mathsf{A}_l\}\biggr)\,,
\end{align}
where the symbol $\{x,y\}$ denotes the anti-commutator of matrices $x,y$. 
The $1/r_j^2$ terms of the coset matrix (\ref{bw-gauge-mat}) vanish when the regular condition (\ref{bubbing}) is imposed (i.e. $\mathsf{A}_j^2=0$).
The fact that extremal black hole solutions are characterized by a nilpotent algebra has been widely discussed in the literature \cite{Gaiotto:2007ag}.

\subsubsection*{Conserved current and charge matrix}

For the coset matrix $M_{\rm BW}(\vec{x})$, the equations of motion of the 3D sigma model are $d\star_3(M_{\rm BW}^{-1}dM_{\rm BW})=0$.
The conserved current $J_{\rm BW}=M_{\rm BW}^{-1}dM_{\rm BW}$ is expanded as
\begin{align}\label{bw-current}
    J_{\rm BW}=-dV\mathbb{F}_0+\sum_{I=1}^{3}dL_I\mathbb{F}_I-\star_3d\omega_{\rm BW} \mathbb{F}_{p^0}+\sum_{I=1}^{3}dK^I\mathbb{F}_{p^I}-2dM\mathbb{E}_{q_0}
\end{align}
We can define the charge matrix $\cQ_{(i)}$ by integrating the current over a two-cycle $\Sigma_i$ around the $i$-th center in 3D flat space:
\begin{align}
    \cQ_{(i)}=\frac{1}{4\pi}\int_{\Sigma_i}\star_3 J_{\rm BW}=-\sfA_i-\frac{1}{2}\sum_{\substack{j=1\\j\neq i}}^{N}\frac{[\sfA_i,\sfA_j]}{|\vec{x}_i-\vec{x}_j|}\,.
\end{align}
Using (\ref{omegabw-rep}) and the expression (\ref{sfA-mat}) of $\sfA_i$, the charge matrix is rewritten as
\begin{align}
    \cQ_{(i)}=q_j \mathbb{F}_0+\sum_{I=1}^{3}l_{j}^I\mathbb{F}_I+s_j \mathbb{F}_{p^0}-\sum_{I=1}^{3}k_{j}^{I}\mathbb{F}_{p^I}+2m_j\mathbb{E}_{q_0}\,,
\end{align}
where we used
\begin{align}
    [\sfA_i,\sfA_j]=2C_{ij}\mathbb{F}_{p^0}\,.
\end{align}
This shows that the vanishing condition $s_i=0$ of the Dirac--Misner strings at the $i$-th center is naturally encoded in the vanishing of the $\mathbb{F}_{p^0}$ component of the charge matrix $\cQ_{(i)}$.
Each component of the charge matrix except for $\mathbb{F}_{p^0}$, corresponds to a brane charge as shown in Fig.~\ref{fig.charge-brane}. On the other hand, the coefficient of $\mathbb{F}_{p^0}$ is proportional to the NUT charge associated with the Dirac--Misner string, which is dual to the Komar mass \cite{Bossard:2008sw}.

\subsection{Almost-BPS solution}

Next, we compute the coset matrix $M_{\rm aBPS}(\vec{x})$ corresponding almost-BPS black hole solutions.
By comparing (\ref{5d-abps}) with (\ref{5d-d1d5p-e-gen}) and (\ref{4d-3d-m-iib}), we easily obtain
\begin{align}
\begin{split}
    e^{2U}&=V^{-1}\,,\quad \zeta^{0}=\mu\,,\quad \zeta^I=K^I\,,\quad
    x^I=-\chi^I=-Z_I^{-1}\,,\quad y^I=f h^I=Z_I^{-1}\,.
\end{split}
\end{align}
Since both the BPS and almost-BPS solutions share the same Gibbons--Hawking space as their 4D base space, the scalar field $e^{2U}$ is same as that in the BPS case.
The relative sign between $x^I$ and $y^I$ is opposite to that in the BPS case.
Furthermore, while we use the same notation $Z_I$ and $\mu$ as in the BPS configuration, these quantities are no longer expressed in terms of the harmonic functions. Instead, they are scalar functions determined by the differential equations (\ref{abps-eq3}) and (\ref{5d-abps}).
This difference comes from the relative sign in the $dt$ component of the gauge fields $A^I$ (\ref{5d-metric}). As a result, the symmetric matrix $N_{\Lambda\Sigma}$ has the same real part as in the BPS case, whereas its imaginary part contains slightly different components,
\begin{align}\label{N-abps}
   {\rm Re}N&=
    \begin{pmatrix}
       -2Z^{-1}&Z^{-1}Z_1&Z^{-1}Z_2&Z^{-1}Z_3\\
       Z^{-1}Z_1&0&-Z_3^{-1}&-Z_2^{-1}\\
      Z^{-1}Z_2&-Z_3^{-1}&0&-Z_1^{-1}\\
      Z^{-1}Z_3&-Z_2^{-1}&-Z_1^{-1}&0
    \end{pmatrix}
    \,,\qquad
   {\rm Im}N=
    \begin{pmatrix}
       -2Z^{-1}&-Z^{-1}Z_1&-Z^{-1}Z_2&-Z^{-1}Z_3\\
       -Z^{-1}Z_1&-\frac{Z_1^2}{Z}&0&0\\
     -Z^{-1}Z_2&0&-\frac{Z_2^2}{Z}&0\\
      -Z^{-1}Z_3&0&0&-\frac{Z_3^2}{Z}
    \end{pmatrix}\,.
\end{align}
The remaining task is to compute the potentials $\tilde{\zeta}_{\Lambda}$ and $\sigma$.
As we will see, we obtain the same result with the BPS case (\ref{scalar-hf}). 

\medskip

To see this, we express the field strengths (\ref{fs-aomega}) as
\begin{align}
   \hat{F}_2^{0}=d\hat{A}^{0}=d\omega_{\rm aBPS}\,,\qquad \hat{F}_2^{I}=d\hat{A}^{I}=d\xi^I \,,\qquad \hat{F}_2=d\omega_3=d\varpi\,.
\end{align}
Using the Hodge duality relations (\ref{5d-abps}), (\ref{abps-eq1}) and (\ref{abps-omega-eq}), we compute the combinations $\star_{3}(\hat{F}_2^{\Lambda}+\zeta^{\Lambda}\hat{F}_2)$ as
\begin{align}
    \star_{3}(\hat{F}_2^{0}+\zeta^{0}\hat{F}_2)&=\star_3d\omega_{\rm aBPS}+\mu \star_3d\varpi\no\\
    &=-VZ_I dK^I-Vd\mu\,,
\end{align}
and
\begin{align}
    \star_{3}(\hat{F}_2^{I}+\zeta^{I}\hat{F}_2)&=\star_3 d\xi^I+K^I dV=VdK^I\,.
\end{align}
Substituting the combinations of the field strenghs and (\ref{N-abps}) into the Hodge duality relations (\ref{dual-tzeta-eq}) for $\tilde{\zeta}_{\Lambda}$ leads to
\begin{align}
    d\tilde{\zeta}_{\Lambda}&=0\,,
\end{align}
and then we set 
\begin{align}
    \tilde{\zeta}_{\Lambda}=0\,.
\end{align}
Since $e^{2U}$ and $\hat{F}_2$ take the same form as in the BPS case, we again find
\begin{align}
    \sigma&=+2V^{-1}\,.
\end{align}
To summarize, the 16 scalars corresponding to almost-BPS solutions are given by 
\begin{align}
\begin{split}\label{scalar-gabps}
    e^{2U}&=V^{-1}\,,\qquad
    \sigma=+2V^{-1}\,,\qquad
    \tilde{\zeta}_{\Lambda}=0\,,\\
    \zeta^0&=\mu\,,\qquad
    \zeta^I=K^I\,,\qquad
   x^I=-Z_I^{-1}\,,\qquad
   y^I=Z_I^{-1}\,.
\end{split}
\end{align}

\subsubsection{Conserved charge}

As a consistency check, we can verify that, with this parametrization of the coset matrix $M_{\rm aBPS}(\vec{x})$, the equations of motion $d\star_3(M_{\rm aBPS}^{-1}dM_{\rm aBPS})=0$ of the 3D sigma model are compatible with the almost-BPS equations.
The conserved current $J=M_{\rm aBPS}^{-1}dM_{\rm aBPS}$ is expanded as
\begin{align}\label{J-abps-current}
    J=-dV\,\mathbb{F}_0+ J_{\mathbb{F}_I}\mathbb{F}_I-\star_3d\omega_{\rm aBPS}\mathbb{F}_{p^0}+\star_3 d\xi^I\,\mathbb{F}_{p^I}+ J_{\mathbb{E}_{q_0}}\mathbb{E}_{q_0}-2\,dK^I\,\mathbb{E}_{q_I}\,,
\end{align}
where
\begin{align}
  J_{\mathbb{F}_I}&= dZ_I-\frac{1}{2}VC_{IJK}d(K^JK^K)+ \frac{1}{2}C_{IJK}K^JK^KdV\,,\label{JF-abps}\\
  J_{\mathbb{E}_{q_0}}&=V d(K^1K^2K^3)-K^1K^2K^3 dV+\sum_{I=1}^{3}(Z_IdK^I-K^IdZ_I)\,.\label{Je-abps}
\end{align}
The sigma model equations of motion $d\star_3J=0$ lead to $d\star_3dV=0\,, d\star_3dK^I=0$ and
\begin{align}
    d\star_3J_{\mathbb{F}_I}&=d\star_3 dZ_I-\frac{1}{2}C_{IJK}V\,d\star_3d(K^JK^K)=0\,,\\
    d\star_3J_{\mathbb{E}_{q_0}}&=\sum_{I=1}^{3}K^I\left(d\star_3 dZ_I-\frac{1}{2}C_{IJK}V\,d\star_3d(K^JK^K)\right)=0\,,
\end{align}
where in the last two equations, we have omitted terms proportional to $d\star_3dV=0$ and $d\star_3dK^I=0$.
The conserved currents $J_{\mathbb{F}_I}$ and $J_{\mathbb{E}_{q_0}}$ correspond to the electric charges $q_I$ and $q_0$ defined in Eq.~(A.25) of \cite{Bah:2021jno} (see also Eq.~(A.26) of \cite{Bah:2021jno}) in the 4D STU model obtained by dimensional reduction along the $\psi$-direction\footnote{On the other hand, the coefficients of $\mathbb{F}_0$ and $\mathbb{F}_{p^I}$ correspond to the magnetic charges $p^0$ and $p^I$ defined in Eq.~(A.25) of \cite{Bah:2021jno} in the 4D STU model.}.
As in the BPS case (\ref{bw-current}), the coefficients of $\mathbb{F}_0$ and $\mathbb{F}_{p^0}$ are naturally interpreted as the conserved currents associated with the Gibbons--Hawking charges $q_i$ and the NUT charge $s_i$, respectively. 
In particular, as in the BPS case, the absence of Dirac--Misner strings imposes the condition that the integral of the $\mathbb{F}_{p^0}$ component in (\ref{J-abps-current}) over an $S^2$ enclosing each center must vanish i.e. $s_i=0$.
In contrast, the D4-brane charges $k^I$ associated with the harmonic functions $K^I$ are encoded in the coefficients of $\mathbb{F}_{p_I}$ in the BPS case, whereas in the almost-BPS case they are instead encoded in the coefficients of the newly emerging $\mathbb{E}_{q_I}$.

\medskip

In the following, we present the explicit expressions of the coset matrices for the almost-BPS black hole with a single-center and the almost-BPS black ring discussed in Section \ref{sec:MMKP}.

\subsubsection{Almost-BPS black hole}

We compute the explicit expression of the coset matrix $M_{\rm aBPS}(\vec{x})$ corresponding to the almost-BPS black hole solution with a single center \cite{Bena:2009ev}.
The corresponding 16 scalar fields are given by 
\begin{align}
\begin{split}\label{scalar-abps}
    e^{2U}&=V^{-1}\,,\qquad
    \sigma=+2V^{-1}\,,\qquad
    \tilde{\zeta}_{\Lambda}=0\,,\\
    \zeta^0&=\mu=\frac{M}{V}=\frac{m_0}{V}+\alpha \frac{\cos\theta}{Vr^2}\,,\qquad
    \zeta^I=K^I=0\,,\\
   x^I&=-(L_{I})^{-1}\,,\qquad
   y^I=(L_{I})^{-1}\,,
\end{split}
\end{align}
where the harmonic functions $V$ and $L_I$ are given in (\ref{hf-nonbps}).
As in the BPS case, the coset matrix $M_{\rm aBPS}(r,\theta)$ can be expressed in an exponential representation
\begin{align}\label{abps-coset}
    M_{\rm aBPS}(r,\theta)=Y_{\rm aBPS}\exp\left(\frac{1}{r} \sfA^{(1)}-\frac{\cos\theta}{r^2}\tilde{\sfA}^{(2)}\right)\,,
\end{align}
where the asymptotic matrix $Y_{\rm aBPS}$ is
\begin{align}\label{Y-aBPS}
    Y_{\rm aBPS}= \left(
\begin{array}{cccccccc}
 l_0^2 l_0^3 & -l_0^2 & m_0 & 0 & -1 & -l_0^3 & 0 & 0 \\
 l_0^2 & 0 & 0 & 0 & 0 & -1 & 0 & 0 \\
 m_0 & 0 & q_0 l_0^1 &-q_0 & 0 & 0 & 1 & l_0^1 \\
 0 & 0 & q_0 & 0 & 0 & 0 & 0 & 1 \\
 -1 & 0 & 0 & 0 & 0 & 0 & 0 & 0 \\
 l_0^3 & -1 & 0 & 0 & 0 & 0 & 0 & 0 \\
 0 & 0 & 1 & 0 & 0 & 0 & 0 & 0 \\
 0 & 0 & -l_0^1 & 1 & 0 & 0 & 0 & 0 \\
\end{array}
\right)
    \,.
\end{align}
The matrices $\sfA^{(1)}$ and $\tilde{\sfA}^{(2)}$ are nilpotent matrices of degree three and two, respectively, given by
\begin{align}
    \sfA^{(1)}&=-Q_6 \mathbb{F}_0+\sum_{I=1}^{3}Q_{I}\mathbb{F}_I\,,\qquad \tilde{\sfA}^{(2)}=-\alpha \,\mathbb{F}_{p^{0}}\,.
\end{align}
In contrast with the BPS case, the coset matrix contains a term proportional to $r^{-2}$. The conserved current $J_{\rm aBPS}$ is expanded as
\begin{align}\label{abps-bh-current}
    J_{\rm aBPS}=-dV\,\mathbb{F}_0+dL_I\mathbb{F}_I+d\left(\alpha \frac{\cos\theta}{r^2}\right)\mathbb{F}_{p^0}\,,
\end{align}
and then the charge matrix at $r=0$ is
\begin{align}
    \cQ=Q_6\mathbb{F}_{0}-\sum_{I=1}^{3}Q_I\mathbb{F}_{I}\,.
\end{align}
As in the BPS case, the absence of the $\mathbb{F}_{p^0}$ component indicates that the non-BPS black hole is free from causal pathologies associated with Dirac--Misner strings. 
Since the $\mathbb{F}_{\Lambda}$ components of the conserved current (\ref{abps-bh-current}) coincide with the BPS case (\ref{bw-current}), this black hole can be regarded as a bound state of a D2$_1$-D2$_2$-D2$_3$-D6 brane system carrying D2$_I$ brane charges $Q_I$ and a D6-brane charge $Q_6$, from the charge-brane correspondence shown in Fig.~\ref{fig.charge-brane}.

\medskip

Before proceeding to the next example, we briefly comment on the extension to multi-center configurations.
Since the generators $\{\mathbb{F}_0, \mathbb{F}_I, \mathbb{F}_{p^0}\}$ mutually commute, a multi-center generalization of the black hole solution can be performed straightforwardly. In particular, we replace the single-center contributions involving $\sfA^{(1)}$ and $\tilde{\sfA}^{(2)}$ by a sum over centers. The corresponding coset matrix becomes
\begin{align}\label{mabps-coset}
    M_{\rm aBPS}(r,\theta)=Y_{\rm aBPS}\exp\left(\sum_{i=1}^{N}\frac{1}{r_i} \sfA_i^{(1)}-\sum_{i=1}^{N}\frac{\cos\theta_i}{r^2_i}\tilde{\sfA}_i^{(2)}\right)\,,
\end{align}
where $\sfA_i^{(1)}$ and $\tilde{\sfA}_i^{(2)}$ are 
\begin{align}
    \sfA_i^{(1)}&=-Q_{6,i} \mathbb{F}_0+\sum_{I=1}^{3}Q_{I,i}\mathbb{F}_I\,,\qquad \tilde{\sfA}_i^{(2)}=-\alpha_i \,\mathbb{F}_{p^{0}}\,.
\end{align}
This coset matrix clearly describes a solution to the Einstein equations, since all coefficient functions are harmonic functions and $[\sfA_i^{(1)}, \tilde{\sfA}_j^{(2)}]=0$. Moreover, we find that the corresponding multi-center solutions do not exhibit any pathologies associated with Dirac--Misner strings around each center arising from the $\mathbb{F}_{p^0}$ component of the conserved current $J$.
In addition, since the 3D base space is simply 3D Euclid space $\mathbb{E}^3$, there is no conical singularities on the solutions.
However, other regularity conditions (e.g., the absence of curvature singularities on the horizons due to the presence of multiple horizons \cite{Welch:1995dh,Candlish:2007fh} , as well as the absence of curvature singularities in the domain of outer communication) cannot, at this stage, be extracted from the coset matrix directly.

\medskip

Recently, using similar methods, multi-center solutions describing Rasheed-Larsen's slowly rotating extremal black holes \cite{Rasheed:1995zv,Larsen:1999pp}, namely the Teo-Wan solution \cite{Teo:2023wfd} and its generalization \cite{Tomizawa:2025tvb} have been constructed and shown to be regular. Since the single-center almost-BPS black hole with appropriate parameters lies in the same duality orbit as the Rasheed-Larsen solution \cite{Bena:2009ev}, it is natural to expect that the solution corresponding to (\ref{mabps-coset}) also has a parameter region in which it is regular. We leave a detailed investigation of this issue for future work.

\subsubsection{Almost-BPS black ring}

For the almost-BPS black ring, we can again show that the corresponding coset matrix is written in an exponential representation, but its underlying algebraic structure is more complicated.
By using the scalar functions (\ref{scalar-ringabps}) and (\ref{scalar-ringabps-mu}), we find that the coset matrix takes the form
\begin{align}\label{coset-aring}
    M_{\rm aBPS}(r,\theta)=Y_{\rm aBPS}\exp\left(\frac{1}{r} \sfA_1^{(1)}+f(r,\theta)\,\sfA_1^{(2)}+\frac{1}{r_2}\sfA_2^{(1)}+\left(2\frac{R-2r\cos\theta+R^{-1}r^2}{r_2}+\cos\theta-\frac{2r}{R}\right)\frac{1}{r_2^2}\sfA_2^{(2)}\right)\,,
\end{align}
where the scalar function $f(r,\theta)$ is given by
\begin{align}
    f(r,\theta)&=\frac{1}{2r_2^2}\biggl[15 \alpha (R-r\cos\theta) \left(1+\frac{q_0^2 R^2}{Q_6^2}\right)\frac{1}{r_2}+r\cos\theta \left(\frac{15 q_0^2 R^2-17 Q_6^2}{r_2 Q_6^2}-\frac{1}{r_1}\right)+\frac{2 +\frac{R^3}{r_2}}{r_1^2}+\frac{6 R}{r_1}\biggr]\no\\
    &\quad + \frac{R}{2r_2^3}\left(1-\frac{15 q_0^2 R^2}{Q_6^2}-\frac{16r_2 r_1}{R^2}+\frac{16 r_1^2}{R^2}\right)\,.
\end{align}
The $\mathfrak{so}(4,4)$ valued matrices $\sfA_i^{(1)}$ and $\sfA_i^{(2)}$\,$(i=1,2)$ are 
\begin{align}
    \sfA_1^{(1)}&=-Q_6 \mathbb{F}_0+\sum_{I=1}^{3}\frac{C_{IJK}d_Jd_K}{6 R^2}\mathbb{F}_I-\beta_1 \mathbb{F}_{p^0}\,,\\
    \sfA_1^{(2)}&=-\frac{2d_1d_2d_3 Q_6^2}{15 R^3}\,\mathbb{F}_{p^{0}}\,,\\
    \sfA_2^{(1)}&=\sum_{I=1}^{3}\left(Q_I+\frac{2Q_6}{3R^2}C_{IJK}d_Jd_K\right)\mathbb{F}_I-q_0\sum_{I=1}^{3}d_I \mathbb{F}_{p^I}+\sum_{I=1}^{3}d_Il_0^I\mathbb{E}_{q_0}-2 \sum_{I=1}^{3}d_I\mathbb{E}_{q_I}-\beta_2 \mathbb{F}_{p^0}\,,\\
    \sfA_2^{(2)}&=\sum_{I=1}^{3}\left(-\frac{C_{IJK}d_Jd_KQ_6}{3R^2}\right)\mathbb{F}_I+\frac{Q_6}{6 R^3}\left(R^2\sum_{I=1}^{3}d_IQ_I+\frac{22}{5} d_1d_2d_3 Q_6\right)\mathbb{F}_{p^0}\,,
\end{align}
with
\begin{align}
    \beta_1&=m_2+\frac{Q_6}{15 R^4}\left(5 R^2\sum_{I=1}^{3}d_I Q_I+22 d_1 d_2 d_3 Q_6\right)\,,\\
    \beta_2&=-m_1-\frac{1}{2}q_0\sum_{I=1}^{3}d_Il_0^I+\frac{Q_6}{15 R^4}\left(5 R^2 \left(\sum_{I=1}^{3}d_I Q_I\right)+6 d_1 d_2 d_3 Q_6\right)\,.
\end{align}
The matrices $\sfA_1^{(1)}$, $\sfA_2^{(1)}$, and $\sfA_2^{(2)}$ are nilpotent of degree three, while $\sfA_1^{(2)}$ is nilpotent of degree two, i.e.
\begin{align}
    (\sfA_1^{(1)})^3=0\,,\qquad (\sfA_1^{(2)})^2=0\,,\qquad   (\sfA_2^{(1)})^3=0\,,\qquad   (\sfA_2^{(2)})^3=0\,.
\end{align}
The charge matrices $\cQ_{(i)}$ at each center are represented as
\begin{align}
    \cQ_{(1)}&=Q_6\mathbb{F}_0-\frac{C_{IJK}d_Jd_K Q_6}{2R^2}\mathbb{F}_I-\frac{d_IQ_6}{R}\,\mathbb{F}_{p^I}+ \frac{d_1d_2d_3Q_6}{R^3}\mathbb{E}_{q_0}\,,\\
    \cQ_{(2)}&=-Q_I\mathbb{F}_I+d_I\left(q_0+\frac{Q_6}{R}\right)\,\mathbb{F}_{p^I}- \left(l_0^Id_I+\frac{d_1d_2d_3Q_6}{R^3}\right)\mathbb{E}_{q_0}+2\,d_I\,\mathbb{E}_{q_I}\,.
\end{align}
Some components of these charge matrices can be seen to encode the charges observed in \cite{Bena:2009ev}. Indeed, $\mathbb{F}_0$ in the charge matrix $\cQ_{(1)}$ at $r=0$ continues to represent the KK-monopole charge. By contrast, in the charge matrix $\cQ_{(2)}$ at the center $r_2=0$, the $\mathbb{F}_{p^I}$ and $\mathbb{E}_{q_I}$ components represent the ``effective'' dipole charges $d_I\left(q_0+\frac{Q_6}{R}\right)$  \cite{Bena:2009ev,Bena:2009en,Bena:2013gma} \footnote{In the BPS case, the integral of the dipole field strength $\Theta^I$ over a small $S^2$ surrounding a given center is proportional to $d_I$. 
In contrast, in the almost-BPS black ring, it takes the form $d_I\left(q_0+\frac{Q_6}{R}\right)$, which receives a correction proportional to the Gibbons--Hawking charge $Q_6$. 
Since they differ from the usual local dipole charges by contributions from the magnetic fluxes, the latter quantity is referred to as the effective dipole charge by following the terminology of Refs. \cite{Bena:2009ev,Bena:2009en,Bena:2013gma}.}and the M5-brane charges $d_I$, respectively. Furthermore, the sum of the $\mathbb{F}_I$ components of the two charge matrices give the M2-brane charges $Q_I+\frac{1}{2R^2}C_{IJK}d_Jd_K Q_6$.
Note that the $\mathbb{F}_{p^0}$ components of each charge matrix vanishes when we impose the condition (\ref{nonbps-ring-DM}) for the absence of the Dirac--Misner strings.

\medskip

Explicit examples of multi-center almost-BPS solutions, including black rings and multi-black rings, were constructed in~\cite{Bena:2009en}. 
However, unlike in the almost-BPS black hole case, the construction of the coset (or monodromy) matrices  corresponding to  multi-black ring solutions starting from a single black ring is not straightforward. 
This difficulty arises from the noncommutativity of the nilpotent matrices $\mathsf{A}_1^{(1)}, \mathsf{A}_2^{(1)}$, and $\mathsf{A}_2^{(2)}$:
\begin{align}
    [\mathsf{A}_1^{(1)},\mathsf{A}_2^{(1)}]&=2Q_6\left(\sum_{I=1}^{3}d_I\mathbb{F}_{p_I}+\frac{d_1d_2d_3}{R^2}\mathbb{E}_{q^0}-\frac{1}{2}\left(\sum_{I=1}^{3}l_0^Id_I-\frac{q_0d_1d_2d_3}{R^2}\right)\mathbb{F}_{p_0}\right)\,,\\
    [\mathsf{A}_2^{(1)},\mathsf{A}_2^{(2)}]&=-\frac{2d_1d_2d_3Q_6}{R}(q_0 \mathbb{F}_{p_0}+2\mathbb{E}_{q^0})\,.
\end{align}
We leave the development of a systematic procedure for constructing almost-BPS multi-black ring solutions from the perspective of the coset (or monodromy) matrix as a problem for future work.

\section{Monodromy matrices for multi-center extremal solutions}\label{sec:mono}

So far, we have studied the sigma model description of black hole solutions whose centers are located at arbitrary positions in the 3D Euclidean base space. In this section, in order to further investigate the mathematical structures underlying these black hole solutions, we restrict ourselves to the biaxisymmetric case in which the centers are aligned along the $x^3$-axis, i.e. the positions of the centers are given by $\vec{x}_j=(0,0,w_j)$. For such configurations, one can perform a dimensional reduction along the $\phi$-direction, and it is known that the 3D action (\ref{3d-sigma}) reduces to an integrable 2D symmetric coset sigma model coupled to 2D gravity.
Classical integrability of this 2D model has played an important role in constructing higher-dimensional biaxisymmetric non-extremal black hole solutions, for example through the inverse scattering method. 

\medskip

In this section, we show that the solution-generating technique based on the Breitenlohner-Maison (BM) linear system can be extended to the extremal multi-center solutions discussed in the previous section. In particular, we explicitly construct the monodromy matrices associated with these extremal multi-center solutions and show that, through their factorization, one can systematically derive the corresponding coset matrices, i.e., the gravitational solutions.
In the literature, this approach has been developed mainly for non-extremal black holes in \cite{Breitenlohner:1986um,Chakrabarty:2014ora,Katsimpouri:2012ky,Katsimpouri:2013wka,Katsimpouri:2014ara}. This is because the associated monodromy matrices in such cases contain only simple poles, for which the existing factorization techniques are well suited. In contrast, for general extremal black holes, the monodromy matrix develops higher-order poles, such as double poles, and therefore the procedure of \cite{Breitenlohner:1986um,Chakrabarty:2014ora,Katsimpouri:2012ky,Katsimpouri:2013wka,Katsimpouri:2014ara} cannot be applied directly.
However, in the extremal case, the residue matrices at each pole exhibit a relatively simple algebraic structure. Using this property, we show that the factorization of the monodromy matrix can be carried out without employing sophisticated mathematical techniques.

\subsection{Monodromy matrix}

A central object for constructing gravitational solutions based on the BM linear system is the monodromy matrix $\cM(w)$ \cite{Breitenlohner:1986um,Chakrabarty:2014ora,Katsimpouri:2012ky,Katsimpouri:2013wka,Katsimpouri:2014ara}. 
This matrix is constructed from the wave function of the BM linear system and is a matrix-valued meromorphic function of the complex spectral parameter $w\in\mathbb{C}$, satisfying
\begin{align}\label{m-con}
    \cM^{-1}=\eta \cM^{T}\eta\,,\qquad \cM^{\natural}=\cM\,, \qquad {\rm det}\,\cM=1\,.
\end{align}
Moreover, the spectral parameter must satisfy an algebraic relation 
\begin{align}\label{r-alg}
    \frac{1}{\la}-\la=\frac{2}{\rho}(w-z)
\end{align}
involving another spectral parameter $\lambda \in \mathbb{C}$ and the Weyl-Papapetrou coordinates $(z,\rho)$, as required by classical integrability.
For the precise definition and further details of the monodromy matrix, we refer the reader to the existing literature e.g. \cite{Breitenlohner:1986um,Chakrabarty:2014ora,Katsimpouri:2012ky,Katsimpouri:2013wka,Sakamoto:2025jtn}.
Once a monodromy matrix is given, it is known that the corresponding classical solution of the sigma model can be constructed by performing a factorization as follows,
\begin{align}\label{mm-fac}
    \cM(w(\la,z,\rho))=X_-(\la,z,\rho)M(z,\rho)X_+(\la,z,\rho)\,,
\end{align}
where the matrix-valued functions $X_+(\la,z,\rho)$ and $X_-(\la,z,\rho)=X_+^{\natural}(-1/\la,z,\rho)$ are required to satisfy the boundary condition
\begin{align}\label{Xpm-bc}
    X_+(0,z,\rho)=1_{8\times 8}\,.
\end{align}
The symbol $w(\la,z,\rho)$ in the left-hand side is to remind us that whenever $\cM(w)$ is rewritten as shown on the right-hand side, $w$ must always be substituted using its relation 
(\ref{r-alg}) with a branch 
\begin{align}\label{la-w}
    \la=\la(w;z,\rho)=\frac{1}{\rho}\left[(z-w)+ \sqrt{(z-w)^2+\rho^2}\right]\,.
\end{align}
Thus, factorizing a given monodromy matrix allows us to construct the coset matrix $M(z,\rho)$ that solves the equations of motion of the sigma model or equivalently those of supergravity.

\medskip

However, at present there is no general framework for systematically determining monodromy matrices corresponding to physically meaningful gravitational solutions. On the other hand, for known solutions, it is possible to construct the associated monodromy matrix by taking an appropriate limit of the corresponding coset matrix $M(z,\rho)$ in the limit $\rho \to 0^+$ in a region where $z$ is sufficiently negative:
\begin{align}\label{sub-rule}
    \cM(w)=\lim_{\rho \to 0^+}M(z=w,\rho)\qquad \text{for}\quad z<-R\,.
\end{align}
Here, $R$ is chosen as the radius of a semicircle in the upper half-plane $(z,\rho)$ that encloses all finite rods\footnote{In practice, the limit (\ref{sub-rule}) is evaluated for $z<-\infty$.}.
In this way, constructing monodromy matrices from known solutions provides insight into how the physical data of black holes are encoded within them, which is crucial for developing a solution-generating technique based on the BM linear system.

\medskip

It has been observed from some examples \cite{Chakrabarty:2014ora,Katsimpouri:2012ky,Katsimpouri:2013wka,Katsimpouri:2014ara,Sakamoto:2025jtn,Sakamoto:2025xbq,Sakamoto:2025sjq} that the monodromy matrices corresponding to asymptotically flat, 5D non-extremal black hole solutions take the universal form
\begin{align}
    \cM_{\text{non-extremal}}(w)=Y+\sum_{i=1}^{N}\frac{A_i}{w-w_i}\,.
\end{align}
Here, the constant matrix $Y=Y^{\natural
}$ characterizes the asymptotic structure of the gravitational solution and the all residue matrices $A_i$ have rank 2.
The number $N$ of simple poles expresses the number of the corner points of the rod structure (for the detail, see Refs.~\cite{Harmark:2004rm,Hollands:2007aj}), and the positions $w_i$ of simple poles are precisely identical with the locations of the corner points.
When a given monodromy matrix $\cM(w)$ has only simple poles with respect to $w$, we can systematically factorize the monodromy matrix by empolying the procedure developed in \cite{Breitenlohner:1986um, Chakrabarty:2014ora,Katsimpouri:2012ky,Katsimpouri:2013wka,Katsimpouri:2014ara} to construct the corresponding gravitational solutions. 

\medskip

For extremal black holes, the analytic structure of the monodromy matrix is modified, and higher-order poles in general appear. This can be understood from the fact that the extremal limit corresponds to the degeneration of the horizon rod to a point, which implies the collision of two simple poles in the monodromy matrix.
When the monodromy matrix contains higher-order poles, the standard procedure developed in \cite{Breitenlohner:1986um,Chakrabarty:2014ora,Katsimpouri:2012ky,Katsimpouri:2013wka,Katsimpouri:2014ara} cannot be applied directly. A prescription for the factorization in the presence of higher-order poles was discussed in \cite{Camara:2017hez}. Here, we show that the factorization can instead be carried out by a purely straightforward algebraic procedure, making use of the fact that the monodromy matrix corresponding to extremal black holes can be expressed in exponential representation, in analogy with the coset matrix.

\subsection{Bena--Warner's multi-center supersymmetric solution}

We first present the corresponding monodromy matrix for Bena--Warner's multi-center supersymmetric solutions and consider its factorization.
To this end, we take the limit (\ref{sub-rule}) for the coset matrix (\ref{bw-gauge-mat2}).
The evaluation of this limit is easily carried out in Weyl-Papapetrou coordinates, where the distance from the field point to the $j$-th center is expressed as
\begin{align}
    r_j=|\vec{x}-\vec{x}_j|=\sqrt{\rho^2+(z-w_j)^2}\,.
\end{align}
Hence, in the limit $\rho\to 0$ with $z<-R$, one then finds
\begin{align}
    \frac{1}{r_j}\to -\frac{1}{w-w_j}\,.
\end{align}
Accordingly, the monodromy matrix is obtained by implementing these replacements in the coset matrix (\ref{bw-gauge-mat2}), and its explicit expression is
\begin{align}\label{bw-monodromy}
    \cM_{\rm BW}(w)=Y_{\rm BW}\exp\left(-\sum_{j=1}^{N}\frac{\sfA_j}{w-w_j}\right)\,.
\end{align}
To clarify the pole structure, it is convenient to expand the monodromy matrix. 
Since the matrices $\mathsf{A}_j$ are nilpotent~(\ref{nil-con}), the expansion terminates after a finite number of terms, yielding
\begin{align}\label{mm-bw}
    \mathcal{M}_{\rm BW}(w)=Y_{\rm BW}+\sum_{j=1}^{N}\frac{A_j^{(1)}}{w-w_j}+\sum_{j=1}^{N}\frac{A_j^{(2)}}{(w-w_j)^2}\,.
\end{align}
In Ref.~\cite{Roy:2018ptt}, only simple poles appear in the corresponding monodromy matrix, as all centers satisfy the regularity conditions.
The residue matrices $A_j^{(1)}$ and $A_j^{(2)}$ are rank 4 and rank 2, respectively, and these matrices are given by
\begin{align}
    A_j^{(1)}&=Y_{\rm BW}\left(-\mathsf{A}_j+\frac{1}{2}\sum_{\substack{k=1\\ k\neq j}}^{N}\frac{1}{w_j-w_k}\{\sfA_{j},\sfA_{k}\}\right) \,,\\
    A_{j}^{(2)}&=Y_{\rm BW}\left(\frac{1}{2}\sum_{j=1}^{N}\sfA_{j}^2\right)\,.
\end{align}
When the regular condition (\ref{bubbing}) is imposed, the double poles in the monodromy matrix disappears and leave only simple poles. In addition, the matrix rank of the residue matrix $A_j^{(1)}$ becomes $\text{Rank}\,A_j^{(1)}=2$ \cite{Roy:2018ptt}.
These observations imply that the solution-generating procedure based on the monodromy matrix approach developed in \cite{Breitenlohner:1986um,Chakrabarty:2014ora,Katsimpouri:2012ky,Katsimpouri:2013wka,Katsimpouri:2014ara} can be applied which all centers satisfy the regular condition (\ref{bubbing}). Accordingly, Ref.~\cite{Roy:2018ptt} restricted attention to the monodromy matrix corresponding to the case of the bubbling solutions where the bubbling condition is imposed at all centers. However, even in the general situation in which the regular condition is necessary not imposed, the simple algebraic structure of (\ref{bw-gauge-mat}) allows one to factorize the corresponding monodromy matrix in a straightforward way, without employing the procedure of Ref.~\cite{Breitenlohner:1986um,Chakrabarty:2014ora,Katsimpouri:2012ky,Katsimpouri:2013wka,Katsimpouri:2014ara}.

\medskip

To see this, it is necessary to express the simple poles in the $w$-plane in terms of the $\la$-plane as
\begin{align}\label{w-la-map}
    \frac{1}{w-w_j}=\nu_j\left( \frac{\la_j}{\la-\la_j}+\frac{1}{1+\la \la_j}\right)\,,\qquad 
     \nu_j=-\frac{2}{\rho\left(\la_j+\la_j^{-1}\right)}\,.
\end{align}
The poles $\la=\la_j$ and $\la=\bar{\la}_j=-1/\la_j$ describe the positions of solition and anti-solition, respectively, and these are given by
\begin{align}\label{lambda-ex}
    \la_j=\frac{1}{\rho}\left((z-w_j)+\sqrt{(z-w_j)^2+\rho^2}\right)=-\frac{1}{\bar{\la}_{j}}\,.
\end{align}
Moreover, the matrices $\sfA_j$ in the exponent in (\ref{bw-monodromy}) satisfy nice algebraic relations
\begin{align}\label{bps-comm}
    [\sfA_i,[\sfA_j,\sfA_k]]=0\,.
\end{align}
Thanks to the relations, the factorization can be performed by simple algebraic manipulations, and then we can explicitly show 
\begin{align}\label{mm-bw-fac}
    \cM_{\rm BW}(w(\la,z,\rho))=X_-(\la,z,\rho)M_{\rm BW}(z,\rho)X_+(\la,z,\rho)\,,
\end{align}
where the matrix $X_{\pm}$ is given by
\begin{align}\label{bm-fac-x}
    X_+&=\exp\left( \sum_{j=1}^{N}\frac{\nu_j\la \la_j}{1+\la \la_j}\left(\sfA_j+\frac{1}{2}\sum_{\substack{k=1\\ k\neq j}}^{N}\left(\frac{\nu_k\la_k}{\la_{j,k}}-\frac{1}{w_j-w_k}\right)[\sfA_j,\sfA_k] \right)\right)\,,
\end{align}
and the matrix $X_{-}$ is evaluated by the relation $X_-(\la,z,\rho)=X_+^{\natural}(-1/\la,z,\rho)$.
The details of the computation are described in appendix \ref{bwfac-proof}.

\subsection{Almost-BPS black hole solution}

Next, we present the monodromy matrix corresponding to the non-supersymmetric extremal rotating black holes.
By taking the limit (\ref{sub-rule}) with the coset matrix (\ref{abps-coset}), we find
\begin{align}\label{abps-mono-single}
    \cM_{\rm aBPS}(w)=Y_{\rm aBPS}\exp\left(-\frac{1}{w} \sfA^{(1)}+\frac{1}{w^2}\tilde{\sfA}^{(2)}\right)\,,
\end{align}
where $\cos\theta$ is replaced with $-1$ in this limit.
This matrix also has a double pole structure 
\begin{align}\label{mm-abps}
    \cM_{\rm aBPS}(w)=Y_{\rm aBPS}+\frac{A^{(1)}}{w}+\frac{A^{(2)}}{w^2}\,,
\end{align}
where the residue matrices are
\begin{align}
    A^{(1)}&=-Y_{\rm aBPS}\,\sfA^{(1)}\,,\qquad A^{(2)}=Y_{\rm aBPS}\left(\tilde{\sfA}^{(2)}+\frac{1}{2}(\sfA^{(1)})^2\right)\,.
\end{align}
The residue matrices are nilpotent satisfying
\begin{align}
   (A^{(1)})^3=0\,,\qquad  (A^{(2)})^2=0\,,
\end{align}
and the their matrix ranks are
\begin{align}
    \text{Rank}\,A^{(1)}=4\,,\qquad \text{Rank}\,A^{(2)}=2\,.
\end{align}

\medskip

Unlike in the BPS case, the exponent in the monodromy matrix (\ref{abps-mono-single}) contains an additional $1/w^2$ term. However, because we find that the matrices $\sfA^{(1)}$ and $\tilde{\sfA}^{(2)}$ commute, the monodromy matrix (\ref{abps-mono-single}) can be easily factorized.
The $1/w^{2}$ term can be decomposed into the following three parts by using relation (\ref{w-la-map}) with $w_j=0$:
\begin{align}
    \frac{1}{w^2}&=-\frac{z}{(z^2+\rho^2)^{3/2}}+\biggl[\frac{\nu_0 \la}{1+\la \la_0}\frac{\rho}{z^2+\rho^2}+\left(\frac{\nu_0\la \la_0}{1+\la \la_0}\right)^2\biggr]\no\\
    &\quad +\biggl[-\frac{\nu_0 }{\la-\la_0}\frac{\rho}{z^2+\rho^2}+\left(\frac{\nu_0\la_0}{\la- \la_0}\right)^2\biggr]\,,
\end{align}
where $\la_0$ is defined as
\begin{align}
    \la_0=\frac{1}{\rho}\left(z+\sqrt{z^2+\rho^2}\right)=-\frac{1}{\bar{\la}_{0}}\,.
\end{align}
Here, the second parentheses collects the terms that have a pole at $\la=\la_0$. The first parentheses, on the other hand, is obtained by replacing $\la$ with $-1/\la$ in terms in the second parentheses, and thus collects the terms that have a pole at $\la=-1/\la_0$. As a result, the remaining term, namely $-z(z^2+\rho^2)^{-3/2}$, is left as the part that has no pole with respect to $\la$.
Finally, by exploiting the relations 
\begin{align}
    \left(\tilde{\sfA}^{(1)}\right)^{\natural}\,Y_{\rm aBPS}=Y_{\rm aBPS}\sfA^{(1)}\,,\qquad \left(\tilde{\sfA}^{(2)}\right)^{\natural}\,Y_{\rm aBPS}=Y_{\rm aBPS}\,\tilde{\sfA}^{(2)}\,,
\end{align} the $\la$-dependent terms can be extracted from the exponent, and we finally obtain the factorized form,
\begin{align}
    \cM_{\rm aBPS}(w)=X_-(\la,z,\rho)M_{\rm aBPS}(z,\rho)X_+(\la,z,\rho)\,,
\end{align}
where the matrix $X_{+}(\la;z,\rho)$ is given by
\begin{align}
\begin{split}
    X_+&=\exp\left(-\frac{\nu_0 \la \la_0}{1+\la \la_0}\sfA^{(1)}+\biggl[\frac{\nu_0 \la}{1+\la \la_0}\frac{\rho}{z^2+\rho^2}+\left(\frac{\nu_0\la \la_0}{1+\la \la_0}\right)^2\biggr]\tilde{\sfA}^{(2)}\right)\,.
\end{split}
\end{align}
Under this factorization, the $\la$-independent part $M_{\rm aBPS}(z,\rho)$ precisely reproduces the coset matrix (\ref{abps-coset}) for the almost-BPS black hole with a single center. This shows that the factorization method works successfully in this case.

\subsection{Almost-BPS black ring solution}

Finally, we derive the monodromy matrix for the almost-BPS black ring solution. By taking the limit (\ref{sub-rule}) with the coset matrix (\ref{coset-aring}), we obtain the monodromy matrix
\begin{align}\label{abps-ring-mono}
    \cM_{\rm aBPS}(w)=Y_{\rm aBPS}\exp\left(-\frac{1}{w} \sfA_1^{(1)}+\frac{1}{w^2}\sfA_1^{(2)}-\frac{1}{w-R}\sfA_2^{(1)}+\frac{1}{(w-R)^2}\sfA_2^{(2)}+\frac{1}{(w-R)^3}\sfA_2^{(3)}\right)\,.
\end{align}
In this limit, the function $f_2$ generates a double pole at $w=0$ and a third-order pole at $w=R$. To separate these contributions, we introduced a new matrix $A^{(3)}_2$ as
\begin{align}
    \sfA_2^{(3)}&=-R\,\sfA_1^{(2)}\,.
\end{align}
At first glance, the expression for this monodromy matrix appears to have a double pole at $w=0$ and a third-order pole at $w=R$. Indeed, for a generic value of $\alpha$, expanding the exponential with respect to $w$ gives
\begin{align}\label{abps-ring-g}
    \cM_{\rm aBPS}(w)=Y_{\rm aBPS}+\frac{A_1^{(1)}}{w}+\frac{A_2^{(1)}}{w-R}+\frac{A_2^{(2)}}{(w-R)^2}-\frac{2d_1^2d_2d_3}{R^3(w-R)^3}(q_0^2R^2+Q_6^2)\left(\alpha-\frac{q_0^2R^2}{q_0^2R^2+Q_6^2}\right)Y_{\rm aBPS}E_{5,1}\,.
\end{align}
Here, $E_{i,j}$ is a $8\times 8$ matrix whose only nonzero entry is 1 in the $(i,j)$-component.
Interestingly, the condition of the absence of the third-order pole is precisely same as the regularity condition (\ref{regular-alpha}) of the horizon area. 
Hence, the monodromy matrix for the regular solution is described by
\begin{align}\label{abps-ring}
    \cM_{\rm aBPS}(w)=Y_{\rm aBPS}+\frac{A_1^{(1)}}{w}+\frac{A_2^{(1)}}{w-R}+\frac{A_2^{(2)}}{(w-R)^2}\,.
\end{align}
The matrix ranks of the residue matrices are
\begin{align}
    \text{Rank}\,A_1^{(1)}=4\,,\qquad \text{Rank}\,A_2^{(2)}=2\,,
\end{align}
and nilpotent with the degree
\begin{align}
   (A_1^{(1)})^3=0\,,\qquad  (A_2^{(1)})^5=0\,,\qquad  (A_2^{(2)})^2=0\,.
\end{align}
In particular, $A_2^{(1)}$ is a nilpotent matrix of higher order than $\sfA_{2}^{(i)}$, in contrast to the single-center case.
It is noted that the pole structure of (\ref{abps-ring}) is same with the one of the BPS black ring, but the degree of nilpotency of $A_2^{(1)}$ associated with the horizon center is different.

\medskip

As in the previous examples, the monodromy matrix~(\ref{abps-ring-mono}) admits an exponential representation, and its factorization can therefore, in principle, be carried out using the Baker--Campbell--Hausdorff formula. 
However, whereas in the BPS case the nilpotent matrices $\mathsf{A}_i$ satisfy the simple commutation relations~(\ref{bps-comm}), in the almost-BPS black ring case, the algebra becomes significantly more intricate:
\begin{align}
     [\mathsf{A}_{i_1}^{(j_1)}, [\mathsf{A}_{i_2}^{(j_2)}, [\mathsf{A}_{i_3}^{(j_3)}, [\mathsf{A}_{i_4}^{(j_4)}, [\mathsf{A}_{i_5}^{(j_5)}, \mathsf{A}_{i_6}^{(j_6)}]]]]] = 0, \quad 
     \left([\mathsf{A}_{i_1}^{(j_1)}, [\mathsf{A}_{i_2}^{(j_2)}, [\mathsf{A}_{i_3}^{(j_3)}, [\mathsf{A}_{i_4}^{(j_4)}, \mathsf{A}_{i_5}^{(j_5)}]]]] \neq 0 \right).
\end{align}
Therefore, although factorization is possible in principle, the explicit computation is expected to be highly cumbersome. 
In the present work, we do not carry out the explicit factorization, and defer the details to future work.

\section{List of supersymmetric black hole solutions}\label{sec:example}

In the previous section, we constructed the monodromy matrices associated with solutions in the Bena--Warner family. In this section, we study, through explicit examples of supersymmetric black hole solutions in 5D minimal supergravity, how the rod structure and the regularity conditions at each center are reflected in the matrices $\sfA_i$ appearing in the coset matrix and the corresponding monodromy matrix. 
A detailed analysis of the moduli space of supersymmetric solutions in 5D minimal supergravity can be found in Ref.~\cite{Breunholder:2017ubu}, and we will utilize these results as needed.
The examples we consider are the BMPV black hole \cite{Breckenridge:1996is,Gauntlett:1998fz}, the supersymmetric black ring \cite{Elvang:2004rt,Gauntlett:2004wh}, the supersymmetric black lens \cite{Kunduri:2014kja,Tomizawa:2016kjh}, and the Kunduri-Lucietti black hole with a non-trivial domain of outer communication \cite{Kunduri:2014iga}.

\subsection{5D asymptotically flat supersymmetric solutions}

We begin by briefly summering the notation for the parameters that characterize asymptotically flat supersymmetric solutions in 5D minimal supergravity.
In this restricted supergravity theory, we consider configurations in which the three types of M2/M5-brane charges are equal to each other, or equivalently we restrict the parameters of the harmonic functions to $k_i=k_i^1=k_i^2=k_i^3$ and $l_i=l_i^1=l_i^2=l_i^3$. The condition for asymptotically 5D Minkowski spacetime becomes
\begin{align}\label{flatness-mini}
        q_0=0\,,\quad k_0=0\,,\quad l_0=1\,,\quad \sum_{j=1}^{N}q_j=\pm 1\,,\quad m_0=-\frac{3}{2}\sum_{j=1}^{N}k_j\,.
\end{align}
The assumption of equal charges reduces the eight associated
harmonic functions to four:
\begin{align}
\begin{split}\label{hf-source-flat}
    V&=\sum_{j=1}^{N}\frac{q_j}{r_j}\,,\qquad
    K:=K^I=\sum_{j=1}^{N}\frac{k_j}{r_j}\,,\qquad
    L:=L^I=1+\sum_{j=1}^{N}\frac{l_j}{r_j}\,,\qquad
    M=m_0+\sum_{j=1}^{N}\frac{m_j}{r_j}\,.
\end{split}
\end{align}
By solving the Hodge duality relations (\ref{5d-BW-V}), (\ref{5d-BW-xi}) and (\ref{omegabw}), the 1-form fields can be written as
\begin{align}
\begin{split}
    \varpi&=\sum_{i=1}^{N}q_i\cos \theta_i\,d\phi\,,\qquad \xi=-\sum_{i=1}^{N}k_i \cos\theta_i\,d\phi\,,\\
    \omega_{\rm BW}&=\sum_{i=1}^{N}s_i\cos\theta_i d\phi+\sum_{i=1}^{N}\sum_{j>i}\frac{C_{ij}}{w_i-w_j}(1+\cos\theta_i)\left(1-\frac{r_i+w_i-w_j}{r_j}\right)d\phi\,.
\end{split}
\end{align}
where
\begin{align}
    s_i&=\beta_i-\sum_{j\neq i}\frac{C_{ij}}{|w_i-w_j|}\,,\qquad
    \beta_i=-m_0 q_i-\frac{3}{2}k_i\,,\qquad
    C_{ij}=q_im_j-m_i q_j+\frac{3}{2}(k_il_j-l_i k_j)\,.
\end{align}
From the expression of $\omega_{\rm BW}$, the condition for absence of the Dirac--Misner strings between the centers
$z=w_i$ on the z-axis $(i=1,\dots,N)$ is given by
\begin{align}\label{flat-noCTC}
    -s_i=m_0 q_i+\frac{3}{2}k_i+\sum_{j\neq i}\frac{C_{ij}}{|w_i-w_j|}=0\,.
\end{align}

\subsubsection*{Rod structure}

In order to completely specify a gravitational solution, one must fix the rod structure that characterizes the topology of the spacetime. For supersymmetric solutions, this data is determined by a choice of the parameters $\{q_j\}_{j=1,\dots,N}$ of the harmonic function $V$ \cite{Breunholder:2017ubu}. 
To make this relation explicit, let us remind the notion of the rod structure and the corresponding rod diagram. These are defined in the Weyl-Papapetrou coordinate system.
Consider the 5D metric and extract the $3\times 3$ matrix corresponding to the Killing directions, excluding the components associated with the Weyl-Papapetrou coordinates $(z,\rho)$. We refer to this $3\times 3$ matrix as the Killing metric. The rods are then characterized by the intervals along the $z$-axis at $\rho=0$ where the determinant of the Killing metric vanishes.
At such points, a particular linear combination of the Killing directions degenerates. The degenerating direction is identified by the zero-eigenvalue eigenvector of the Killing metric.

\medskip

For a supersymmetric solution with $N$ centers, the $z$-axis is divided into $N+1$ segments. 
We denote the semi-infinite intervals by $I_- = (-\infty, w_1]\,, I_+ = [w_N, \infty)$, and the finite intervals by $I_1 = [w_1, w_2], I_2 = [w_2, w_3], \dots, I_{N-1} = [w_{N-1}, w_N]$.
The rod vectors associated with each segment are given by linear combinations of $\partial_{\phi}$ and $\partial_{\psi}$ with integer coefficients. This was first shown for
black lenses in \cite{Tomizawa:2016kjh}, and later generalized to more general cases in \cite{Breunholder:2017ubu}:
\begin{align}\label{rod-vec}
  v_{\pm}=\partial_{\phi}-\varpi_{\pm}\partial_{\psi}\,,\qquad  v_i=\partial_{\phi}-\varpi_i\partial_{\psi}\,,\qquad \varpi_i\,, \varpi_{\pm}\in \mathbb{Z}\,.
\end{align}
The integer coefficients $\varpi_{\pm}$ and $\varpi_i$ are obtained by evaluating the 1-form field $\varpi$ on the corresponding rod, and are given explicitly as
\begin{align}
    \varpi_{\pm}=\varpi\lvert_{I_{\pm}}=\pm 1\,,\qquad \varpi_i=\varpi\lvert_{I_i}=\sum_{j=1}^{i}q_j-\sum_{j=i+1}^{N}q_j=2\sum_{j=1}^{i}q_j-1\,,
\end{align}
where we used the asymptotic flatness condition (\ref{flatness-mini}) for $\{q_i\}_{i=1,\dots,N}$. By introducing the basis of the two component vectors by $(1,0)=\partial_{\phi}+\partial_{\psi}$ and $(0,1)=\partial_{\phi}-\partial_{\psi}$, the rod vectors (\ref{rod-vec}) are expressed as
\begin{align}
    v_-=(1,0)\,,\qquad v_i=\left(1-\sum_{j=1}^{i}q_j,\sum_{j=1}^{i}q_j\right)\,,\qquad v_+=(0,1)\,,\qquad i=1,2,\dots,N\,.
\end{align}

\medskip

For stationary, biaxisymmetric vacuum solutions, the regularity conditions that must be satisfied between adjacent rods are obtained in \cite{Hollands:2007aj}.
First, for two adjacent rods with $2\pi$-normalized rod vectors $v_i$ and $v_j$, one must have
\begin{align}
    \det(v_i^Tv_j^T)=\pm 1\,.
\end{align}
Moreover, if the two rods are separated only by a single center corresponding to a horizon, then the corresponding rod vectors satisfy
\begin{align}
    \det(v_i^Tv_j^T)=p\in \mathbb{Z}\,.
\end{align}
In this case, the horizon topology is a three-sphere $S^3$ for $p=\pm1$; for $p=0$, it is a ring with topology $S^2\times S^1$;  and for all other values of $p$, the horizon topology is the lens space $L(p,1)$.
In the supersymmetric case, the determinants of adjacent rod vectors are given by \cite{Breunholder:2017ubu}
\begin{align}
    \det(v_-^Tv^T_1)=q_1\,,\qquad \det(v_i^Tv_{i+1}^T)=q_{i+1}\,,\qquad \det(v_N^Tv_+^T)=q_{N}\,.
\end{align}
Thus, the choice of the parameters $\{q_i\}_{i=1,\dots,N}$ and the positions $\{w_i\}_{i=1,\dots,N}$ determines the rod structure for the 5D asymptotically flat supersymmetric solutions.

\medskip

Furthermore, the additional constraints, which depend on whether each center describes a corner or a horizon, must be imposed.
If the center $(\rho,z)=(0,w_i)$ is a corner i.e. $q_i=\pm 1$, the parameters satisfy
\begin{align}\label{corner-mini-con}
    &l_i=-q_i^{-1}k_i^2\,,\qquad m_i=\frac{1}{2}k_i^3\,,\qquad
    q_i+\sum^{N}_{\substack{j=1\\j\neq i}}\frac{2k_ik_j-q_i(q_jk_i^2-l_j)}{|w_i-w_j|}>0\,.
\end{align}
Unlike in $U(1)^3$ supergravity, the condition $q_i=\pm 1$ is imposed. As we will see later, it is also required for the nilpotency of $\sfA_j$.
If the center $(\rho,z)=(0,w_i)$ is a horizon, the parameters satisfy
\begin{align}
    &q_i\in\mathbb{Z}\,,\qquad q_il_i+k_i^2>0\,,\qquad -q_i^2m_i^2-3q_ik_il_im_i+j_il_i^3-2k_i^3m_i+\frac{3}{4}k_i^2l_i^2>0\,.
\end{align}
The third condition guarantees that the area of cross-sections $A_{h,i}$ of the horizon is real
\begin{align}
    A_{h,i}\propto \sqrt{-q_i^2m_i^2-3q_ik_il_im_i+j_il_i^3-2k_i^3m_i+\frac{3}{4}k_i^2l_i^2}\,.
\end{align}

\subsubsection*{From geometric and charge data to monodromy matrix}

Let us now examine how the regularity conditions given above are encoded in the properties of the matrices $\sfA_i$ that characterize the monodromy matrix. 
For the configuration under consideration and (\ref{sfA-mat}), the matrix $\sfA_i$ associated with the $i$-th center takes
\begin{align}
    \mathsf{A}_j&=-q_j \mathbb{F}_0-3l_{j}\mathbb{F}_I-\beta_j \mathbb{F}_{p^0}+3k_{j}\mathbb{F}_{p^I}-2m_j\mathbb{E}_{q_0}\,.
\end{align}
Its square is given by
\begin{align}
    \mathsf{A}_j^2= \begin{pmatrix}
        0&0&0&0&0&0&0&0\\
        0&0&0&0&0&0&0&0\\
        0&0&0&0&0&0&0&0\\
        0&0&0&0&0&0&0&0\\
        2(2m_jk_j-l_j^2)&0&2m_jq_j+k_jl_j&0&0&0&0&0\\
        0&0&0&0&0&0&0&0\\
        2m_jq_j+k_jl_j&0&-2(q_jl_j+k_j^2)&0&0&0&0&0\\
        0&0&0&0&0&0&0&0
    \end{pmatrix}
    \,.
\end{align}
The expression implies that, if the $i$-th center describes a corner i.e. the parameters satisfy the condition (\ref{corner-mini-con}), then $\sfA_i$ is a nilpotent matrix with degree two i.e. $\sfA_i^2=0$.
It should be noted that, unlike in the case of 5D $U(1)^3$ supergravity, the matrix $\sfA_j$ becomes nilpotent of degree two only when both the regularity conditions $l_i=-q_i^{-1}k_i^2$ and $m_i=\frac{1}{2}k_i^3$, together with the condition of being free from orbifold singularities,
\begin{align}
    q_j=\pm 1\,,
\end{align}
are imposed.
On the other hand, if the $i$-th center describes a horizon, then $\sfA_i$ is nilpotent with the degree three i.e. $\sfA_i^3=0$, and the $(7,3)$ component is always negative. In particular, $q_j \in \mathbb{Z}$, which characterizes the topology of the horizon cross-section, is included in the coefficient of $\mathbb{F}_0$. 
In this sense, the monodromy matrix encodes the topology of both the horizon-cross section or the DOC in an algebraic manner.

\subsection{Examples of 5D asymptotically flat supersymmetric black holes}

We illustrate, through explicit examples of supersymmetric black holes in five-dimenisonal minimal supergravity, the form of the nilpotent matrices $\mathsf{A}_i$ associated with each center and clarify their relation to the above discussion.
The list of the solutions we consider here is as follows:
\begin{itemize}
    \item BMPV black hole \cite{Breckenridge:1996is,Gauntlett:1998fz}
    \item (Multi-)supersymmetric black ring \cite{Elvang:2004rt,Gauntlett:2004wh} 
    \item Supersymmetric black lens \cite{Kunduri:2014kja,Tomizawa:2016kjh}
    \item Kunduri-Lucietti black hole with non-trivial DOC \cite{Kunduri:2014iga}
\end{itemize}

\subsubsection{BMPV black hole}
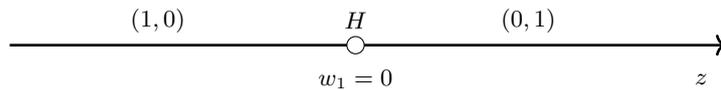
\begin{figure}
\begin{center}
\begin{tikzpicture}[scale=0.65]
\node[font=\small ] at (-6,0.5) {$(1,0)$};
\node[font=\small ] at (1.5,0.5) {$(0,1)$};
\node[font=\small ] at (-2.0,0.5) {$H$};
\node[font=\small ] at (-2,-0.7) {$w_1=0$};
\node[font=\small ] at (5,-0.7) {$z$};
\draw[->,black,line width = 1] (-9,0) -- (5.5,0);
\draw[fill=white] (-2,0) circle (5pt);
\end{tikzpicture}
\caption{Rod diagram for the BMPV black hole. The white circle represents the center on the $z$-axis corresponding to a horizon. Since the horizon has a spherical topology, we take $q_1=1$.}\label{single-rod}
\end{center}
\end{figure}
The BMPV black hole (Breckenridge-Myers-Peet-Vafa) is a solution of a 5D BPS rotating black hole that carries a charge with angular momentum \cite{Breckenridge:1996is}. 
The corresponding harmonic functions are 
\begin{align}\label{bmpv-harmonic}
    V=\frac{1}{r}\,,\qquad K=\frac{k_1}{r}\,,\qquad 
    L=1+\frac{l_1}{r}\,,\qquad M=m_0\,.
\end{align}
In this solution, the asymptotic flatness condition (\ref{flat-con}) is identified with the condition (\ref{flat-noCTC}) for the absence of the Dirac--Misner string,
\begin{align}\label{bmpv-no-dm}
    m_0=-\frac{3}{2}k_1\,.
\end{align}
Furthermore, the condition (\ref{noCTC2}) ensuring the absence of the CTCs is \cite{Gauntlett:2004wh}\footnote{Using the notation of \cite{Gauntlett:2004wh}, namely $k_1=-\frac{q}{2}$ and $l_1 = \frac{Q^2-q^2}{4}$, the condition (\ref{BMPV-CTC}) becomes
\begin{align}\label{BMPV-CTC}
    4Q^3>q^2(3Q-q^2)^2\,.
\end{align}}
\begin{align}\label{BMPV-CTC0}
    \left(\frac{3}{4}k_1^2+l_1\right)l_1^2>0\,.
\end{align}
The first condition (\ref{noCTC1}) is automatically satisfied if we impose (\ref{BMPV-CTC0}). Note that setting $k_1 = 0$ reduces the solution to the Strominger-Vafa black hole \cite{Strominger:1996sh}, which describes a 5D static supersymmetric black hole.

\medskip

The corresponding coset matrix is characterized by a nilpotent matrix $\sfA_1$ with degree three, which is given by
\begin{align}
      \mathsf{A}_1&=-\mathbb{F}_0-\sum_{I=1}^{3}l_{1}\mathbb{F}_I+\sum_{I=1}^{3}k_{1}\mathbb{F}_{p^I}\,.
\end{align}
Since the solution has a single center, the condition (\ref{bmpv-no-dm}) for the absence of the Dirac--Misner string is equivalent to the vanishing of the coefficient of $\mathbb{F}_{p^0}$.

\subsubsection{(Multi-)supersymmetric black ring}

The supersymmetric black ring has been constructed in Ref.~\cite{Elvang:2004rt} as a BPS solution of 5D minimal supergravity, and the explicit expressions of the corresponding harmonic functions was subsequently presented in Ref.~\cite{Gauntlett:2004wh}.
These harmonic functions involve two centers and are given by
\begin{align}
\begin{split}
    V&=\frac{1}{r_1}\,,\qquad K=-\frac{q}{2}\frac{1}{r_2}\,,\qquad
    L=1+\frac{Q-q^2}{4}\frac{1}{r_2}\,,\qquad M=\frac{3q}{4}-\frac{3qR^2}{16}\frac{1}{r_2}\,,
\end{split}
\end{align}
where the parameters are chosen to satisfy the asymptotic 5D flatness condition (\ref{flat-con}), and the two centers are placed in
\begin{align}
  w_1=0\,,\qquad w_2=-\frac{R^2}{4}\,.
\end{align}
Here, the real positive parameters $Q$ and $q$ are proportional to the total electric charge and to the dipole charge of the black ring, respectively.
The second center $r_2=0$ describes the horizon, and it has topology $S^1\times S^2$ which follows from $q_2=0$. The radii of the $S^1$ and $S^2$ factors are given by
\begin{align}
\sqrt{3\left[\frac{(Q-q^2)^2}{4q^2}-R^2\right]}
\end{align}
and $q/2$, respectively.
Imposing the conditions (\ref{noCTC1}) and (\ref{noCTC2}) for the absence of CTCs then leads to the inequalities
\begin{align}
Q \ge q^2\,,\qquad (Q-q^{2})^{2} > 4q^{2} R^{2}\,.
\end{align}
In particular, the second inequality guarantees the positivity of the radius of the $S^1$ part of the ring.

\medskip

The residue matrices of the associated monodromy matrix are characterized, using (\ref{sfA-mat}), by
\begin{align}
      \mathsf{A}_1&=-\mathbb{F}_0+\frac{3}{4}q\,\mathbb{F}_{p^0}\,,\qquad 
      \mathsf{A}_2=-\frac{Q-q^2}{4}\sum_{I=1}^{3}\mathbb{F}_I-\frac{3}{4}q\, \mathbb{F}_{p^0}-\frac{q}{2}\sum_{I=1}^{3}\mathbb{F}_{p^I}+\frac{3}{8}qR^2\,\mathbb{E}_{q_0}\,.
\end{align}
Note that the total charge $Q$ is carried by only the second center whereas the local dipole charge $q$ of the ring appears in both centers.
These matrices satisfy the nilpotency conditions
\begin{align}
    (\mathsf{A}_1)^2=0\,,\qquad (\mathsf{A}_2)^3=0\,.
\end{align}
In particular, only in the limit where the dipole charge vanishes, $q=0$, does one have $(\mathsf{A}_2)^2=0$.
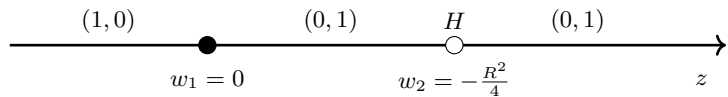
\begin{figure}
\begin{center}
\begin{tikzpicture}[scale=0.65]
\node[font=\small ] at (-7,0.5) {$(1,0)$};
\node[font=\small ] at (-2.5,0.5) {$(0,1)$};
\node[font=\small ] at (2.5,0.5) {$(0,1)$};
\node[font=\small ] at (0.0,0.5) {$H$};
\node[font=\small ] at (-5.0,-0.7) {$w_1=0$};
\node[font=\small ] at (0,-0.7) {$w_2=-\frac{R^2}{4}$};
\node[font=\small ] at (5,-0.7) {$z$};
\draw[->,black,line width = 1] (-9,0) -- (5.5,0);
\draw[fill=black] (-5.0,0) circle (5pt);
\draw[fill=white] (0,0) circle (5pt);
\end{tikzpicture}
\caption{Rod diagram for 5D supersymmetric black ring. The white circles and black circles denote the centers on the $z$-axis corresponding to a corner and a horizon, respectively. As the first center represents the corner and the second the horizon with a ring topology, $(q_1,q_2)$ are taken as $(q_1,q_2)=(1,0)$.}\label{ring-rod}
\end{center}
\end{figure}

\medskip

Furthermore, this single black ring solution can be extended to a multi-center configuration in which multiple rings are arranged concentrically, and the corresponding harmonic functions are given in \cite{Gauntlett:2004wh}
\begin{align}
    V&=\frac{1}{r_1}\,,\quad \quad  K=-\frac{1}{2}\sum_{i=2}^{N+1}\frac{q_i}{r_i}\,,\quad
    L=1+\frac{1}{4}\sum_{i=2}^{N+1}\frac{(Q_i-q_i^2)}{r_i}\,,\quad M=\frac{3}{4}\sum_{i=1}^{N+1}q_i+\frac{3}{4}\sum_{i=2}^{N+1}\frac{q_iw_i}{r_i}\,,
\end{align}
where we again impose the conditions
\begin{align}
    Q_i\ge q_i^2\,,\qquad (Q_i-q_i^2)^2\geq 4q_i^2R_i^2\,,
\end{align}
and the locations of the centers are
\begin{align}
    w_1=0\,,\qquad w_i=-\frac{R_i^2}{4}\,,\qquad (i=2,\dots,N+1)\,.
\end{align}
The horizon topology for each ring is $S^1\times S^2$, where $S^1$ and $S^2$ have radii 
\begin{align}
\sqrt{3\biggl[\frac{(Q_i-q_i^2)^2}{4q_i^2}-R_i^2\biggr]}    
\end{align}
and $q_i/2$, respectively.
The residue matrices $\mathsf{A}_i$ associated with the corresponding monodromy matrices can then be written explicitly as
\begin{align}
      \mathsf{A}_1&=-\mathbb{F}_0+\frac{3}{4}q\,\mathbb{F}_{p^0}\,,\qquad 
      \mathsf{A}_i=-\frac{Q_i-q_i^2}{4}\sum_{I=1}^{3}\mathbb{F}_I-\frac{3}{4}\sum_{j=1}^{N+1}q_j\, \mathbb{F}_{p^0}-\frac{q_i}{2}\sum_{I=1}^{3}\mathbb{F}_{p^I}+\frac{3}{8}q_iR_i^2\,\mathbb{E}_{q_0}\,,
\end{align}
and find the nilpotency properties
\begin{align}
    (\mathsf{A}_1)^2=0\,,\qquad (\mathsf{A}_i)^3=0\,.
\end{align}

\subsubsection{Supersymmetric black lens}

Next, we consider a supersymmetric black lens solution with the horizon topology $L(N,1)=S^3/\mathbb{Z}_{N}$.
The associated eight harmonic functions are given by \cite{Kunduri:2014kja,Tomizawa:2016kjh}
\begin{align}
\begin{split}
    V&=\sum_{i=1}^{N}\frac{q_i}{r_i}=\frac{N}{r_1}-\sum_{i=2}^{N}\frac{1}{r_i}\,,\qquad K=\sum_{i=1}^{N}\frac{k_i}{r_i}\,,\qquad
    L=1+\sum_{i=1}^{N}\frac{l_i}{r_i}\,,\qquad M=m_0+\sum_{i=1}^{N}\frac{m_i}{r_i}\,.
\end{split}
\end{align}
When $N=1$, the set of the harmonic functions reduces to that of the BMPV black hole \cite{Breckenridge:1996is}.
By using the shift symmetry \cite{Bena:2005va}, we can set $m_1=0$ without loss of generality. 
The asymptotic 5D flatness condition (\ref{flat-con}) becomes
\begin{align}\label{lens-reg}
    m_0=-\frac{3}{2}\sum_{i=1}^{N}k_i\,.
\end{align}
The condition (\ref{noCTC2}) for the absence of CTCs around the horizon $r_1=0$ then implies the inequality
\begin{align}\label{noCTC-lens-horizon}
l_1 > -\frac{3k_1^2}{4N}\,.
\end{align}
Note that the first condition (\ref{noCTC1}) is automatically satisfied once (\ref{noCTC-lens-horizon}) is imposed.
In order to remove singularities at $r_i=0\,(i=2,\dots,N)$ , we impose
\begin{align}
    l_i=k_i^2\,,\qquad m_i=\frac{1}{2}k_i^3\,,
\end{align}
which follow from the condition (\ref{bubbing}).
The absence of the Dirac--Misner string at each corner is \cite{Tomizawa:2016kjh}
\begin{align}
    m_0-\frac{3}{2}k_i+\sum_{\substack{j=1\\j\neq i}}^{N}\frac{1}{|w_i-w_j|}\biggl[3k_i^2k_j+2k_i^3q_j-\frac{3}{2}(k_il_j+l_ik_j+k_il_iq_j)+m_j\biggr]=0\,.
\end{align}
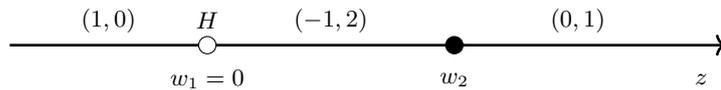
\begin{figure}
\begin{center}
\begin{tikzpicture}[scale=0.65]
\node[font=\small ] at (-7,0.5) {$(1,0)$};
\node[font=\small ] at (-2.5,0.5) {$(-1,2)$};
\node[font=\small ] at (2.5,0.5) {$(0,1)$};
\node[font=\small ] at (-5.0,0.5) {$H$};
\node[font=\small ] at (-5.0,-0.7) {$w_1=0$};
\node[font=\small ] at (0,-0.7) {$w_2$};
\node[font=\small ] at (5,-0.7) {$z$};
\draw[->,black,line width = 1] (-9,0) -- (5.5,0);
\draw[fill=white] (-5.0,0) circle (5pt);
\draw[fill=black] (0,0) circle (5pt);
\end{tikzpicture}
\caption{Rod diagram for 5D supersymmetric black lens with $L(2,1)$. The white circles and black circles denote the centers on the $z$-axis corresponding to a corner and a horizon, respectively. As the first center represents the corner and the second the horizon with a ring topology, $(q_1,q_2)$ are taken as $(q_1,q_2)=(2,0)$.}\label{lens-rod}
\end{center}
\end{figure}

From the general expression (\ref{sfA-mat}) of $\sfA_1$ and $\sfA_i (i=2,3,\dots,N)$, the residue matrices of the monodromy matrix $\cM_{\rm BW}(w)$ are described by
\begin{align}
      \mathsf{A}_1&=-N\,\mathbb{F}_0-l_1\sum_{I=1}^{3}\mathbb{F}_I+\left(Nm_0+\frac{3}{2}k_1\right)\, \mathbb{F}_{p^0}+k_1\sum_{I=1}^{3}\mathbb{F}_{p^I}\,,\\
      \mathsf{A}_i&=\,\mathbb{F}_0-k_i^2\sum_{I=1}^{3}\mathbb{F}_I-\left(m_0-\frac{3}{2}k_i\right)\, \mathbb{F}_{p^0}+k_i\sum_{I=1}^{3}\mathbb{F}_{p^I}-k_i^3\,\mathbb{E}_{q_0}\,,
\end{align}
subject to the flatness condition (\ref{lens-reg}) and the inequality (\ref{noCTC-lens-horizon}).
In this configuration, we find
\begin{align}
    \sfA_1^3=0\,,\qquad \sfA_i^2=0\,,\qquad (i=2,3,\dots,N)\,.
\end{align}
This nilpotent property of the matrices $\sfA_j$ is compatible with the rod structure of the black lens solutions.

\subsubsection{Kunduri-Lucietti black hole with non-trivial DOC}

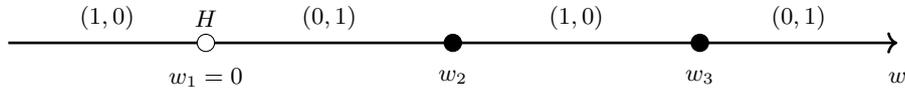
\begin{figure}
\begin{center}
\begin{tikzpicture}[scale=0.65]
\node[font=\small ] at (-7,0.5) {$(1,0)$};
\node[font=\small ] at (-2.5,0.5) {$(0,1)$};
\node[font=\small ] at (2.5,0.5) {$(1,0)$};
\node[font=\small ] at (7,0.5) {$(0,1)$};
\node[font=\small ] at (-5.0,-0.7) {$w_1=0$};
\node[font=\small ] at (0,-0.7) {$w_2$};
\node[font=\small ] at (5.0,-0.7) {$w_3$};
\node[font=\small ] at (9.0,-0.7) {$w$};
\node[font=\small ] at (-5.0,0.5) {$H$};
\draw[->,black,line width = 1] (-9,0) -- (9,0);
\draw[fill=white] (-5.0,0) circle (5pt);
\draw[fill=black] (0,0) circle (5pt);
\draw[fill=black] (5.0,0) circle (5pt);
\end{tikzpicture}
\caption{Rod diagram for 5D Kunduri-Lucietti black hole with non-trivial DOC. }\label{KL-rod}
\end{center}
\end{figure}

The exterior region of the BMPV black hole horizon is topologically trivial in the sense that a spatial slice $\Sigma$ is diffeomorphic to $\mathbb{R}^3 \setminus \mathbb{B}^3$, where $\mathbb{B}^3$ denotes the black hole interior. Nevertheless, the topological censorship theorem of Friedman~\cite{Friedman:1993ty} allows for a broader class of black hole spacetimes. In particular, it states that, assuming the averaged null energy condition, the domain of outer communication (DOC) of an asymptotically flat spacetime is simply connected. In four dimensions, this severely restricts the topology of the exterior region, implying that $\mathrm{DOC} \cap \Sigma$ must be diffeomorphic to $\mathbb{R}^3 \setminus \mathbb{B}^3$. By contrast, in higher dimensions, simple connectedness does not preclude the presence of non-trivial topology, and the spatial section $\mathrm{DOC} \cap \Sigma$ may admit non-trivial homology. Indeed, it was shown in Ref.~\cite{Hollands:2010qy} that in five dimensions the exterior region can have the topology
$\bigl[\mathbb{R}^4 \# n(S^2 \times S^2) \# m(\pm \mathbb{C}P^2)\bigr] \setminus \mathbb{B}^4$.
Explicit examples realizing such non-trivial structures have since been constructed. In 5D minimal supergravity, Kunduri and Lucietti~\cite{Kunduri:2014iga} obtained a four-parameter family of supersymmetric black hole solutions with spherical horizon topology and a non-trivial 2-cycle in the exterior, corresponding to $\mathrm{DOC} \cap \Sigma \simeq \bigl[\mathbb{R}^4 \# (S^2 \times S^2)\bigr] \setminus \mathbb{B}^4$.
Non-supersymmetric solutions describing spherical black holes with
$\mathrm{DOC} \cap \Sigma \simeq \bigl[\mathbb{R}^4 \# \mathbb{C}P^2\bigr] \setminus \mathbb{B}^4$
have also been constructed in Refs.~\cite{Suzuki:2023nqf,Suzuki:2024phv,Suzuki:2024abu}.

\medskip

Here, we focus on the supersymmetric solutions of \cite{Kunduri:2014iga}, which are described by harmonic functions with three centers, one corresponding to the horizon and the remaining two corresponding to merely smooth points.
The harmonic functions take the form
\begin{align}
\begin{split}\label{KL-harmonic}
    V&=\frac{1}{r_1}-\frac{1}{r_2}+\frac{1}{r_3}\,,\qquad K^I=\sum_{i=1}^{3}\frac{k_i}{r_i}\,,\qquad
    L^I=1+\sum_{i=1}^{3}\frac{l_i}{r_i}\,,\qquad M=m_0+\sum_{i=1}^{3}\frac{m_i}{r_i}\,,
\end{split}
\end{align}
where we assume $ w_1=0<w_2<w_3$.
By using the shift symmetry \cite{Bena:2005va}, we can set $m_1=0$ without loss of generality. 
Since the centers at $r_2=0$ and $r_3=0$ correspond to smooth points without brane sources, the regular conditions (\ref{bubbing}) i.e.
\begin{align}
    l_2=(k_2)^2\,,\qquad l_3=-(k_3)^2\,,\qquad  m_2=\frac{k_2^3}{2}\,,\qquad m_3=\frac{k_3^3}{2}
\end{align}
are imposed.
On the other hand, for the center at $r_1=0$ to describe a horizon, the following inequalities are required \cite{Kunduri:2014iga}:
\begin{align}
    j:=k_1^2+l_1>0\,,\qquad j-\frac{k_1^2(k_1^2+\frac{3}{2}l_1)^2}{j^2}>0\,.
\end{align}
Finally, the conditions (\ref{local_noCTC-0}) for the absence of Dirac--Misner string singularities become
\begin{align}
    w_2(k_2+k_3)^3+(w_3-w_2)\left(3k_1k_2^2+k_2^3-3(k_1+k_3+2k_2)w_2-3k_2l_1\right)&=0\,,\\
    w_3(k_2+k_3)^3+(w_3-w_2)(3k_1k_3^2-k_3^3-3(k_1+k_2)w_3+3k_3l_1)&=0\,.
\end{align}
It has been shown that there are black holes in this family with identical conserved changes to the BMPV black hole, which means a violation of black hole uniqueness. In particular, there exist the parameter region where the horizon area is larger than that of the BMPV black hole with the same charge.

\medskip

For these harmonic functions, the residue matrices of the corresponding monodromy matrix $\cM(w)$ are characterized by
\begin{align}
      \mathsf{A}_1&=-\mathbb{F}_0-l_1\sum_{I=1}^{3}\mathbb{F}_I+\left(2m_0+\frac{3}{2}k_1\right)\, \mathbb{F}_{p^0}+k_1\sum_{I=1}^{3}\mathbb{F}_{p^I}\,,\\
      \mathsf{A}_2&=+\mathbb{F}_0-l_2\sum_{I=1}^{3}\mathbb{F}_I+\left(m_0-\frac{3}{2}k_2\right)\, \mathbb{F}_{p^0}+k_2\sum_{I=1}^{3}\mathbb{F}_{p^I}-2m_j\,\mathbb{E}_{q_0}\,,\\
    \mathsf{A}_3&=-\mathbb{F}_0-l_3\sum_{I=1}^{3}\mathbb{F}_I+\left(m_0+\frac{3}{2}k_3\right)\, \mathbb{F}_{p^0}+k_3\sum_{I=1}^{3}\mathbb{F}_{p^I}-2m_j\,\mathbb{E}_{q_0}\,.
\end{align}
These matrices satisfy the nilpotency conditions
\begin{align}
\sfA_1^3=0\,,\qquad \sfA_2^2=0\,,\qquad \sfA_3^2=0\,,    
\end{align}
which are consistent with the fact that the center at $r_1=0$ describes a horizon, while the other two centers correspond to smooth points without brane sources.

\newpage

\section{Monodromy matrices for extremal limits of Rasheed-Larsen solution}\label{sec:RL}

All monodromy matrices associated with the BPS and almost-BPS extremal black holes studied so far exhibit a nilpotent algebraic structure. In fact, the relationship between extremal black holes and nilpotent algebras has been discussed in many previous works \cite{Gunaydin:2005mx,Gaiotto:2007ag,Bossard:2009my,Bossard:2009at,Bossard:2009mz,Bossard:2009we,LindmanHornlund:2010gen,Kim:2010bf,Bossard:2011kz}.
This observation may suggest that nilpotent algebras universally underlie extremality. The example examined in this section, however, shows that this expectation does not hold in general.

\medskip

As an explicit example, we consider the Rasheed-Larsen solution describing the 4D dyonic Kerr black hole \cite{Rasheed:1995zv,Larsen:1999pp}. This solution contains two inequivalent extremal limits within a single solution, namely (i) slowly rotating  extremal limit, and (ii) fast rotating extremal limit, thereby making it clear that the algebraic structure of the monodromy associated with extremality is not uniquely determined. In fact, while the former lies in the same duality orbit as the almost-BPS solution, the latter exhibits an idempotent, rather than nilpotent, structure.

\subsection{Rasheed-Larsen solution}

We begin by presenting the explicit expression of the 4D dyonic Kerr black hole, which is a solution of 4D Einstein-Maxwell-dilaton gravity.
This solution arises from the Kaluza-Klein reduction of a vacuum solution of 5D pure Einstein gravity.
To construct the corresponding monodromy matrix, it is convenient to start from the 5D vacuum configuration.

\medskip

The corresponding 5D metric of the Rasheed-Larsen solution is given by \cite{Rasheed:1995zv,Larsen:1999pp}
\begin{align}\label{5dRL_metric}
    ds_5^2=\frac{H_2}{H_1}(dx^5+{\bm A})^2-\frac{H_3}{H_2}(dt+{\bm B})^2+H_1\left(\frac{dr^2}{\Delta^2}+d\theta^2+\frac{\Delta}{H_3}\sin^2\theta d\phi^2\right)\,,
\end{align}
where the angular coordinates $\theta$ and $\phi$ take values in the ranges $0\leq \theta \leq \pi$ and $0\leq \phi \leq 2\pi$, and the scalar functions $H_{1,2,3}$ and $\Delta$ are
\begin{align}
\begin{split}
H_1 &= r^2 + a^2 \cos^2\theta 
+ r(p - 2m) 
+ \frac{p(p - 2m)(q - 2m)}{2(p + q)} 
-  \frac{p\sqrt{(q^2 - 4m^2)(p^2 - 4m^2)}}{2m(p + q)} a \cos\theta\,,\\
H_2 &= r^2 + a^2 \cos^2\theta 
+ r(q - 2m) 
+ \frac{q(p - 2m)(q - 2m)}{2(p + q)} 
+  \frac{q\sqrt{(q^2 - 4m^2)(p^2 - 4m^2)}}{2m(p + q)} a \cos\theta\,,\\
H_3 &= r^2 + a^2 \cos^2\theta - 2mr\,,\\
\Delta &= r^2 + a^2 - 2mr\,.
\end{split}
\end{align}
Here, $m,a,p,q$ are real parameters and $m,p,q$ are constrained by
\begin{align}
    p,q \ge 2m > 0\,.
\end{align}
This solution is characterized by four parameters $(m,a,p,q)$, which determine the physical mass $M$,
angular momentum $J$, electric charge $Q$ and magnetic charge $P$ according to
\begin{align}\label{rl-phys}
    M=\frac{p+q}{4}\,,\quad J=\frac{\sqrt{pq}(pq+4m^2)}{4m(p+q)}a\,,\quad Q^2=\frac{q(q^2-4m^2)}{4(p+q)}\,,\quad P^2=\frac{p(p^2-4m^2)}{4(p+q)}\,.
\end{align}
For the solution to have a regular horizon, the following bound must be satisfied:
\begin{align}
    \left(\frac{P}{2M}\right)^{\frac{2}{3}}+\left(\frac{Q}{2M}\right)^{\frac{2}{3}}\leq 1\,.
\end{align}

\medskip

The 5D metric is expressed in terms of the 4D metric $g_{\mu\nu}  \,(\mu,\nu=0,1,2,3)$ as
\begin{align}
    ds_5^2=e^{-\frac{2\phi}{\sqrt{3}}}(dx^5+{\bm A})^2+e^{\frac{\phi}{\sqrt{3}}}g_{\mu\nu}dx^{\mu}dx^{\nu}\,.
\end{align}
The dimensional reduction along the $x^5$-direction leads to the 4D geometry described by \footnote{Here, we follow the notation used in Larsen’s paper; While we cannot directly verify that the expression indeed solves the Einstein equation, we can confirm that, by flipping the sign of the $A_t$ component of the Kaluza-Klein gauge field $A$, the resulting 4D geometry or equivalently, the corresponding 5D geometry becomes Ricci flat.}
\begin{align}
\begin{split}
    ds_4^2&=g_{\mu\nu}dx^{\mu}dx^{\nu}=-\frac{H_3}{\sqrt{H_1H_2}}(dt+{\bm B})^2+\sqrt{H_1H_2}\left(\frac{dr^2}{\Delta^2}+d\theta^2+\frac{\Delta}{H_3}\sin^2\theta d\phi^2\right)\,,\label{RL_4dmetric}\\
{\bm B}&=\sqrt{pq}\frac{(pq+4m^2)r-m(p-2m)(q-2m)}{2m(p+q)H_3}a \sin^2\theta d\phi\,,\\
{\bm A}&=\biggl[2Q\left(r+\frac{p-2m}{2}\right)+\sqrt{\frac{q^3(p^2-4m^2)}{4m^2(p+q)}}a\,\cos\theta\biggr]\,H_2^{-1}dt\\
&\quad-\biggl[2P(H_2+a^2\sin^2\theta)\cos\theta+\sqrt{\frac{p(q^2-4m^2)}{4m^2(p+q)^3}}\\
&\quad\times((p+q)(pr-m(p-2m))+q(p^2-4m^2) )a\sin^2\theta
\biggr]H_2^{-1}d\phi\,,\\
e^{-\frac{2}{\sqrt{3}}\phi}&=\frac{H_2}{H_1}\,.
\end{split}
\end{align}
Under taking the limit $p\to 2m, q \to 2m$, the 4D metric (\ref{RL_4dmetric}) reduces to the Kerr black hole solution.

\subsubsection{Extremal limits of Rasheed-Larsen solution}

This 4D black hole solution has two different extremal limits, depending on the ratio of the angular momentum $J$ and the product $PQ$ of the electric and magnetic charges:
\begin{itemize}
    \item Slowly rotating extremal limit $|J|<|PQ|$\,: This extremal limit is realized by taking the limit $m\to 0, a\to 0$ with $j=a/m=$fixed. This limit gives an ergo-free extremal rotating black hole, which can be mapped to an almost-BPS solution via a $U$-duality transformation. The entropy is 
    \begin{align}
        S=2\pi\sqrt{P^2Q^2-J^2}\,.
    \end{align}
    \item Fast rotating extremal limit $|J|>|PQ|$\,: This limit is performed by taking $a\to m$. The extremal limit leads to an extremal rotating black hole with an ergoregion. In contrast to the slowly rotating extremal limit, the angular velocity of the horizon does not vanish, and hence an ergoregion exists around the horizon. The entropy is 
    \begin{align}
        S=2\pi\sqrt{J^2-P^2Q^2}\,.
    \end{align}
\end{itemize}

\subsubsection*{Slowly rotating extremal limit}

We consider a slowly rotating extremal limit
\begin{align}\label{sre}
    m\to0\,,\qquad a\to 0\,,\qquad j=a/m=\text{fixed}\,.
\end{align}
The 5D metric is simplified by
\begin{align}\label{sr-exbh}
ds^2_{5} &= \frac{H_2}{H_1}\,\bigg[ dx_5 + \left(2\Bigl(r + \frac{p}{2}\Bigr)+pj\cos\theta\right)\frac{Q}{H_2}\,dt -\biggl\{ 2H_2\cos\theta+q\Bigl(r + \frac{pq}{p+q}\Bigr)\,j\sin^2\theta\biggr\}\frac{P}{H_2}d\phi\bigg]^2 \no\\
&\quad - \frac{r^2}{H_2}\Bigl(dt + \tfrac{2j PQ\sin^2\theta\,d\phi}{r}\Bigr)^2 + H_1\Bigl(\frac{dr^2}{r^2} + d\theta^2 + \sin^2\theta\,d\phi^2\Bigr)\,,
\end{align}
where the two functions $H_1$ and $H_2$ are
\begin{align}
H_1&=r^2+p r+\frac{p^2q(1-j\cos\theta)}{2(p+q)}\,,\quad
H_2=r^2+q r+\frac{q^2p(1+j\cos\theta)}{2(p+q)}\,.
\end{align}
This solution asymptotically approaches the 5D Kaluza-Klein space at $r\to \infty$:
\begin{align}
    ds^2_{5} &\simeq -dt^2+r^2(dr^2+d\theta^2+\sin^2\theta d\phi^2)+(dx^5-2P \cos\theta d\phi)^2\,.
\end{align}
The corresponding 4D geometry is
\begin{align}
\begin{split}
    ds_4^2&=- \frac{r^2}{\sqrt{H_1H_2}}\Bigl(dt + \tfrac{2j PQ\sin^2\theta\,d\phi}{r}\Bigr)^2 + \frac{\sqrt{H_1H_2}}{r^2}\Bigl(dr^2 + r^2(d\theta^2 + \sin^2\theta\,d\phi^2)\Bigr)\,,\\
    {\bm A}&=\left(2\Bigl(r + \frac{p}{2}\Bigr)+pj\cos\theta\right)\frac{Q}{H_2}\,dt
    -\biggl\{ 2H_2\cos\theta+ q\Bigl(r + \frac{pq}{p+q}\Bigr)\,j\sin^2\theta\biggr\}\frac{P}{H_2}d\phi\,,\\
    e^{\frac{2\phi}{\sqrt{3}}}&=\frac{H_2}{H_1}\,.
\end{split}
\end{align}
The physical quantities become
\begin{align}\label{slow-phy}
    M=\frac{p+q}{4}\,,\qquad J=j PQ\,,\qquad P^2=\frac{p^3}{4(p+q)}\,,\qquad Q^2=\frac{q^3}{4(p+q)}\,.
\end{align}

\subsection{Coset space description}

We embed the 5D metric (\ref{5dRL_metric}) of the Rasheed-Larsen black hole solution as the ansatz (\ref{5d-d1d5p-e-gen}) of the 5D $U(1)^3$ supergravity with $A^I=0$. As in the extremal black hole case, we construct the coset matrix taking values in the symmetric coset (\ref{sym-coset}) corresponding to this black hole solution. The dimensional reduction to three dimensions is performed in the same manner as in section \ref{sec:sigma}, and the scalar fields parametrizing the symmetric coset (\ref{sym-coset}) are obtained using the same procedure. The resulting expressions for the sixteen scalar fields are given below,
\begin{align}
\begin{split}\label{4dybh-scalar}
    e^{2U}&=\sqrt{\frac{H_2}{H_1}}\frac{H_3}{\sqrt{H_1H_3-A_t^2}}\,,\qquad 
      x^I=0\,,\qquad
   y^I=f h^I=\sqrt{\frac{H_1H_3-A_t^2}{H_1H_2}}\\
    \zeta^0&=\frac{H_2A_t }{H_1H_3-A_t^2}\,,\qquad
    \zeta^I=0\,,\qquad
    \tilde{\zeta}_{0}=\frac{\tilde{\zeta}'_0}{H_1H_2}\,,\qquad     \tilde{\zeta}_{I}=0\,,\\
   \sigma&=-\frac{\sigma'}{4m^3(p+q)^2H_1(H_1H_3-A_t^2)}\,,
\end{split}
\end{align}
where $A_t$ is the $t$-component of the Kaluza-Klein gauge field ${\bm A}$ presented in (\ref{RL_4dmetric}), and the numerator $\tilde{\zeta}'_0$ of $\tilde{\zeta}_0$ is
\begin{align}
    \tilde{\zeta}'_0&=PQ\left[\left(\frac{q \left(4 m^2+p q\right)}{p+q}+2 (r (q+r)-m (q+2 r))\right)+\frac{a^2\left(2 m^2 (p-q)-p q^2\right)}{m^2 (p+q)} \cos^2\theta\right]\no\\
    &\quad-2J\cos\theta \left(r (q+r)-m (q+2 r)+a^2 \cos^2\theta\right)-a m q\sqrt{pq}\cos\theta\,,
\end{align}
while the numerator $\sigma'$ of $\sigma$ is given in (\ref{sigma-n}).
The corresponding coset matrix $M_{\rm RL}(r,\theta)$ can be computed by substituting these scalar fields into (\ref{ginv-M}).

\subsection{Monodromy matrix}

We now present the monodromy matrix corresponding to the Rasheed-Larsen black hole solution (\ref{5dRL_metric}). Since this solution is generically non-extremal, in contrast to the extremal solutions considered in the previous section, we must adopt a definition of the Weyl coordinates different from that used in (\ref{WP-def}). We hence introduce the Weyl coordinates by the following definition:
\begin{align}
    \rho=\sqrt{\Delta}\sin\theta\,,\qquad z=(r-m)\cos\theta\,.
\end{align}
 By substituting the coset matrix $M_{\rm RL}(z,\rho)$ into the formula (\ref{sub-rule}), we have
\begin{align}\label{monodromy-RL}
    \cM_{\rm RL}(w)=1+\frac{A^-}{w-w^-}+\frac{A^+}{w-w^+}\,,
\end{align}
where the locations of the simple poles are
\begin{align}
    w^-=-\alpha\,,\qquad w^+=\alpha\,.
\end{align}
It is convenient to multiply $\eta'$ by $A_{\pm}$ since the products $\eta'A^{\pm}$ become symmetric matrices.
Their explicit forms are given by
\begin{align}
    \eta'A^{\pm}&=
    \begin{pmatrix}
        A_{11}^{\pm}&0&0&F_{14}^{\pm}A_{11}^{\pm}&0&0&F_{17}^{\pm}A_{11}^{\pm}&0\\
        0&0&0&0&0&0&0&0\\
        0&0&A_{33}^{\pm}&0&F_{35}^{\pm}A_{33}^{\pm}&0&0&F_{38}^{\pm}A_{33}^{\pm}\\
        F_{14}^{\pm}A_{11}^{\pm}&0&0&(F_{14}^{\pm})^2A_{11}^{\pm}&0&0&F_{14}^{\pm}F_{17}^{\pm}A_{11}^{\pm}&0\\
        0&0&F_{35}^{\pm}A_{33}^{\pm}&0&(F_{35}^{\pm})^2A_{33}^{\pm}&0&0&F_{35}^{\pm}F_{38}^{\pm}A_{33}^{\pm}\\
        0&0&0&0&0&0&0&0\\
        F_{17}^{\pm}A_{11}^{\pm}&0&0&F_{17}^{\pm}F_{14}^{\pm}A_{11}^{\pm}&0&0&(F_{17}^{\pm})^2A_{11}^{\pm}&0\\
        0&0&F_{38}^{\pm}A_{33}^{\pm}&0&F_{38}^{\pm}F_{35}^{\pm}A_{33}^{\pm}&0&0&(F_{38}^{\pm})^2A_{33}^{\pm}
    \end{pmatrix}\,,
\end{align}
where we have introduced the basic ingredients
\begin{align}
     A_{11}^{\pm}&=\frac{q}{2}\mp \frac{\sqrt{q} }{\sqrt{p}\alpha}\left(\frac{m}{a}J-\frac{a}{m}PQ\right)\,,\qquad
     A_{33}^{\pm}=\frac{p}{2}\mp\frac{\sqrt{p} }{\sqrt{q} \alpha }\left(\frac{m}{a}J+\frac{a}{m}PQ\right)\,,
\end{align}
and the explicit expressions of the factors $F_{14}^{\pm},F_{17}^{\pm},F_{35}^{\pm},F_{38}^{\pm}$ are written down in appendix \ref{sec:RL-mono}.
From the form of these symmetric matrices, it is clear that the residue matrices $A^{\pm}$ are rank-2 matrices. This is the same algebraic structure that appears in vacuum solutions of other 5D non-extremal black holes \cite{Katsimpouri:2012ky,Sakamoto:2025jtn,Sakamoto:2025xbq,Sakamoto:2025sjq}.
Since the monodromy matrix (\ref{monodromy-RL}) has only simple poles with rank-2 residue matrices, its factorization can be systematically performed by employing the procedure developed in \cite{Chakrabarty:2014ora,Katsimpouri:2012ky,Katsimpouri:2013wka,Katsimpouri:2014ara}.
However, as our main interest is in the extremal limit, we will not perform the explicit factorization here.

\subsection{Extremal limits of monodromy matrix}

We apply two different extremal limits to the monodromy matrix corresponding to the Rasheed-Larsen black hole solution obtained in the previous subsection, and investigate how the associated algebraic structures differ in each case.

\subsubsection{Slowly rotating extremal limit}

We first consider the slowly rotating extremal limit (\ref{sre}).
In this limit, the two poles of the monodromy matrix (\ref{monodromy-RL}) collide, giving rise to a reduced monodromy matrix with a second-order pole
\begin{align}
    \cM_{\rm Slow}(w)=1+\frac{A^{(1)}}{w}+\frac{A^{(2)}}{w^2}\,.
\end{align}
The residue matrices $A^{(1)}$ and $\widetilde{A}^{(2)}=A^{(2)}-\frac{1}{2}(A^{(1)})^2$ are elements of $\mathfrak{so}(4,4)$ and can be expanded as
\begin{align}
    A^{(1)}&=\left(\frac{q}{2}-p\right)\mathbb{H}_0-\frac{1}{2}  q\sum_{j=1}^{3}\mathbb{H}_j-2P(\mathbb{E}_0-\mathbb{F}_0)-2Q(\mathbb{E}_{q_0}-\mathbb{F}_{q_0})\,,\\
    \widetilde{A}^{(2)}&=\frac{j p q (2 p+q)}{4(p+q)}\mathbb{H}_0-\frac{j p q^2}{4(p+q)}\sum_{j=1}^{3}\mathbb{H}_j-\frac{qJ}{Q}(\mathbb{E}_0-\mathbb{F}_0)-\frac{p J}{P}(\mathbb{E}_{q_0}-\mathbb{F}_{q_0})-2 J (\mathbb{E}_{p_0}+\mathbb{F}_{p_0})\,.
\end{align}
Here, the physical quantities $P,Q,J$ are given in (\ref{slow-phy}).
We find that these matrices are nilpotent
\begin{align}
    ( A^{(1)})^3=0\,,\qquad ( \widetilde{A}^{(2)})^2=0
\end{align}
with the matrix rank $\text{Rank}\,A^{(1)}=4$ and $\text{Rank}\,\widetilde{A}^{(2)}=2$.
As in the extremal black hole case discussed previously, we observe that the matrices $A^{(1)}$ and $\widetilde{A}^{(2)}$ commute, the monodromy matrix can be rewritten as
\begin{align}
    \cM_{\rm Slow}(w)=\exp\left(\frac{1}{w}A^{(1)} +\frac{1}{w^2}\widetilde{A}^{(2)}\right)\,.
\end{align}
In particular, the factorization can also be performed straightforwardly.
The corresponding coset matrix is
\begin{align}
    M_{\rm Slow}(z,\rho)=\exp\left(-\frac{1}{r}A^{(1)}-\frac{\cos\theta}{r^2}\tilde{A}^{(2)}\right)\,.
\end{align}
From this expression, we can read off the following sixteen scalar fields, which parametrize the symmetric coset (\ref{sym-coset}):
\begin{align}
\begin{split}\label{slow-scalar}
    e^{2U}&=\frac{\sqrt{p}\,r^2}{\sqrt{H_1 \left(H_1 (p+q)-q (p+r)^2\right)}}\,,\qquad   x^I=0\,,\qquad
   y^I=f h^I=\sqrt{\frac{H_1(p+q)-q (p+r)^2}{H_1 p}}\\
    \zeta^0&=-\frac{p Q (p+2 r+p j \cos\theta)}{H_1 (p+q)-q (p+r)^2}\,,\qquad
    \zeta^I=0\,,\qquad
    \tilde{\zeta}_{0}=\frac{2 PQ(1-j \cos \theta)}{H_1}\,,\qquad     \tilde{\zeta}_{I}=0\,,\\
   \sigma&=-\frac{2P^3}{ p^2H_1 \left(H_1(p+q)-q (p+r)^2\right)}
   \biggl[j q \cos\theta \left(2 r \left(4 p^2+2 p (q+r)-q (q-2 r)\right)-j p q \cos\theta (2 p+q)\right)\\
   &\qquad+2 q^2 r (2 p+q)+p q^2 (2 p+q)-8 r^3 (p+q)-4 r^2 (2 p-q) (p+q)\biggr]\,.
\end{split}
\end{align}
One can explicitly verify that these expressions are consistent with the slowly rotating extremal black hole solution (\ref{sr-exbh}).

\subsubsection{Fast rotating extremal black hole}

Next, we consider the fast rotating extremal limit.
The monodromy matrix for the fast rotating extremal black hole is given by
\begin{align}\label{fast-mono}
    \cM_{\rm Fast}(w)=1+\frac{A^{(1)}}{w}+\frac{A^{(2)}}{w^2}\,,
\end{align}
where the residue matrices $A^{(1)}$ and $\widetilde{A}^{(2)}=A^{(2)}-\frac{1}{2}(A^{(1)})^2$ are 
\begin{align}
    A^{(1)}&=\left(\frac{q}{2}-p\right)\mathbb{H}_0-\frac{1}{2}  q\sum_{j=1}^{3}\mathbb{H}_j+2P(\mathbb{E}_0-\mathbb{F}_0)-2Q(\mathbb{E}_{q_0}-\mathbb{F}_{q_0})\,,\\
    \widetilde{A}^{(2)}&=\frac{2PQ}{\sqrt{pq}}\left(\left(p+\frac{q}{2}\right)\mathbb{H}_0-\frac{1}{2}q\sum_{j=1}^{3}\mathbb{H}_j\right)-\frac{p^2Q}{\sqrt{pq}}(\mathbb{E}_0-\mathbb{F}_0) -\frac{q^2P}{\sqrt{pq}}(\mathbb{E}_{q_0}-\mathbb{F}_{q_0}) -2J(\mathbb{E}_{p^0}+\mathbb{F}_{p^0})\,.
\end{align}
Here, the physical quantities are given in (\ref{rl-phys}) with $a\to m$.
Interestingly, these matrices are not nilpotent but instead are idempotent-type matrices satisfying
\begin{align}
    (A^{(1)})^3=4m^2A^{(1)}\,,\qquad ( \widetilde{A}^{(2)})^3=4m^4 \widetilde{A}^{(2)}\,.
\end{align}
Moreover, their ranks are given by
\begin{align}
    \text{Rank}\,A^{(1)}=4\,,\qquad \text{Rank}\,\widetilde{A}^{(2)}=4\,,\qquad \text{Rank}\,A^{(2)}=2\,.
\end{align}
To the best of our knowledge, there is no existing literature explicitly identifying extremal black holes with idempotent elements of $\mathfrak{so}(4,4)$.  At present, it is unclear whether the monodromy matrix (\ref{fast-mono}) can be expressed in exponential representation, as in other extremal solutions. Even if such a representation exists, it is expected to be more complicated than in the nilpotent case, making its factorization
more challenging. It may be possible to apply the procedure developed in \cite{Camara:2017hez} to perform the factorization; however,
we leave this issue for future work.

\subsection{$U$-duality transformation from almost-BPS solution}

The slowly rotating extremal black hole (\ref{sr-exbh}) can be mapped to the almost-BPS black hole solution (\ref{hf-nonbps}) with a single center in Taub-NUT space under a certain $U$-duality transformation \cite{Bena:2009ev}. The duality transformation acts as an element of $SO(4,4)$ on the coset matrix or equivalently the monodromy matrix.
For completeness, we provide below the explicit form of the appropriate $SO(4,4)$ transformation at the level of the coset matrix\footnote{In previous works, this relationship has been shown in the 4D Lorentzian STU model through duality transformations. In contrast, in our formulation, the coset matrix the 4D model is the Euclidean STU model, and therefore a different $U$-duality transformation from those used in the existing literature must be employed. }.

\medskip

Whether two coset matrices are related to each other by a $U$-duality transformation strongly depends on the choice of gauge fixing for the scalar fields.
For our purpose, we perform a (formal) coordinate transformation
\begin{align}
    \psi\to \psi- \sqrt{\frac{p+q}{q}}\,t\,,
\end{align}
which corresponds to a constant shift for the scalar fields $\zeta^0$ given in (\ref{slow-scalar}). 
The constant shifts of $\tilde{\zeta}_0$ and $\sigma$ represent ambiguity from integration constants in the solution of the Hodge duality relations.
After this coordinate transformation, the dimensional reduction leads to the following sixteen scalar fields in a much simpler expression:
\begin{align}
\begin{split}
    e^{2U}&=\sqrt{\frac{-q\,r^2}{pH_1}}\,,\qquad   x^I=0\,,\qquad
   y^I=f h^I=\sqrt{\frac{-p\,r^2}{qH_1}}\,,\\
    \zeta^0&=-\sqrt{\frac{q}{p+q}}\frac{ (p (q+r)+q r)}{p r}\,,\qquad
    \zeta^I=0\,,\qquad
    \tilde{\zeta}_{0}=-\frac{\sqrt{p}r^2}{\sqrt{q}H_1}\,,\qquad     \tilde{\zeta}_{I}=0\,,\\
   \sigma&=\frac{2rP((p+q)r+pq)}{p^2H_1}\,.
\end{split}
\end{align}
Here, in order to simplify the corresponding coset matrix, we fixed integration constants of $\tilde{\zeta}_0$ and $\sigma$ that appear when the Hodge duality relations are integrated.

\medskip

Then, the coset matrix becomes
\begin{align}
    M_{\rm Slow}'(r)=Y_{\rm Slow}'\exp\left(\frac{1}{r}\sfA^{(1)}-\frac{\cos\theta}{r^2}\tilde{\sfA}^{(2)}\right)\,,
\end{align}
where the asymptotic matrix $Y'_{s}$ is
\begin{align}
    Y'_{\rm Slow}=\left(
\begin{array}{cccccccc}
 1 & 0 & 0 & \frac{\sqrt{p+q}}{\sqrt{q}} & 0 & 0 & -\frac{\sqrt{q}}{\sqrt{p}} & 0 \\
 0 & 1 & 0 & 0 & 0 & 0 & 0 & 0 \\
 0 & 0 & 1 & 0 & -\frac{\sqrt{p}}{\sqrt{q}} & 0 & 0 & \frac{\sqrt{p+q}}{\sqrt{p}} \\
 -\frac{\sqrt{p+q}}{\sqrt{q}} & 0 & 0 & -\frac{p}{q} & 0 & 0 & 0 & 0 \\
 0 & 0 & -\frac{\sqrt{p}}{\sqrt{q}} & 0 & 0 & 0 & 0 & 0 \\
 0 & 0 & 0 & 0 & 0 & 1 & 0 & 0 \\
 -\frac{\sqrt{q}}{\sqrt{p}} & 0 & 0 & 0 & 0 & 0 & 0 & 0 \\
 0 & 0 & -\frac{\sqrt{p+q}}{\sqrt{p}} & 0 & 0 & 0 & 0 & -\frac{q}{p} \\
\end{array}
\right)\,,
\end{align}
and the residue matrices are simplified to
\begin{align}
    \sfA^{(1)}&=-2P\,\mathbb{F}_{0}+2Q\,\mathbb{E}_{q_0}\,,\qquad
    \sfA^{(2)}=2PQj\,\mathbb{F}_{p^0}\,.
\end{align}
From this explicit expression of the coset matrix $M_{\rm Slow}'(r)$, the duality transformation relating it to $M_{\rm aBPS}(r)$ is given below,
\begin{align}
    M_{\rm Slow}'(r)=g^{\natural}M_{\rm aBPS}(r)g\,,\qquad g=g_{H}g_{E_{q}}g_{F_{q}}\in SO(4,4)\,,
\end{align}
where 
\begin{align}
    g_{H}&=\exp\left(-\frac{1}{2}\sum_{I=1}^{3}\log\left(\frac{p }{q}\frac{Q_I}{P} \right)\mathbb{H}_{I}\right)\,,\qquad 
    g_{E_{q}}=\exp\left(-\frac{1}{2}\sqrt{\frac{q}{p}}\sum_{I=1}^{3}\mathbb{E}_{q_I}\right)\,,\qquad
    g_{F_{q}}=\exp\left(-\sqrt{\frac{p}{q}}\sum_{I=1}^{3}\mathbb{F}_{q_I}\right)\,,
\end{align}
and the parameters $(l_0^I,Q_I,q_0,Q_6,m_0,\alpha)$ in the almost-BPS black hole solution (\ref{hf-nonbps}) are taken as
\begin{align}
\begin{split}
    &l_0^I=\frac{p+q}{pq}\bar{Q}_I\,,\qquad Q_I=\bar{Q}_I\,,\qquad q_0=4P\frac{p+q}{pq}\,,\qquad Q_6=4P\,,\\
    &m_0=-\frac{p(p-q)}{2q^2P^{3/2}}\sqrt{\bar{Q}_1\bar{Q}_2\bar{Q}_3}\,,\qquad 
    \alpha=-2 \sqrt{P\bar{Q}_1\bar{Q}_2\bar{Q_3}}\, j\,.
\end{split}
\end{align}
In this way, the slowly rotating extremal black hole (\ref{sr-exbh}) lies in the same duality orbit as the almost-BPS black hole solution (\ref{hf-nonbps}).

\newpage

\section{Conclusion and discussion}\label{sec:conclusion}

In this work, we have developed a monodromy-matrix formulation for extremal, stationary biaxisymmetric solutions in 5D $U(1)^3$ supergravity, based on the Breitenlohner--Maison (BM) linear system. 
Our primary focus has been on solutions constructed over a Gibbons--Hawking base, including both supersymmetric Bena--Warner multi-center solutions and non-supersymmetric almost-BPS solutions.
In the BPS case, all fields are determined by eight harmonic functions, while in the almost-BPS case the equations are more involved and only admit analytic solutions in special configurations. 
Regularity conditions, such as the absence of curvature singularities, orbifold singularities, Dirac--Misner string singularities, and closed timelike curves, impose nontrivial constraints on the parameters appearing in the harmonic functions and play a crucial role in determining physically admissible solutions. 
In particular, these constraints tightly restrict the allowed parameter space and are essential for ensuring the global consistency of the resulting geometries.

\medskip
Under dimensional reduction to three dimensions, the system can be reformulated as a sigma model with target space $SO(4,4)/(SO(2,2)\times SO(2,2))$. 
In this framework, gravitational solutions are encoded in coset matrices constructed from scalar fields obtained via dualization. 
For the Bena--Warner solutions, we explicitly construct the coset matrices and show that they admit an exponential representation in terms of nilpotent matrices associated with each center. 
In particular, when regularity conditions are imposed, these matrices become nilpotent of degree two, leading to significant simplifications in both the algebraic structure and the resulting expressions for the fields, as originally pointed out by Virmani et al. \cite{Roy:2018ptt}.
The corresponding monodromy matrices are then obtained from the BM linear system. 
For generic extremal solutions, the monodromy matrices exhibit higher-order poles in the spectral parameter, in contrast to the simple-pole structure typically encountered in non-extremal cases. 
In particular, Bena--Warner multi-center solutions generically give rise to double poles, reflecting the extremal nature of the configurations. 
Despite this complication, we have demonstrated that the underlying algebraic structure of the relevant $\mathfrak{so}(4,4)$ elements allows for an explicit factorization of the monodromy matrices through elementary manipulations. 
This provides a systematic and constructive method for reconstructing the original gravitational solutions from their monodromy data, thereby extending the applicability of integrable techniques to a broader class of extremal solutions.

\medskip
We have extended this analysis to almost-BPS solutions, including a single-center rotating extremal black hole and a two-center black ring in Taub--NUT space. 
In the single-center black hole solution, the monodromy matrix retains a relatively simple structure due to commuting residue matrices, which allows for a straightforward factorization. 
In contrast, the two-center black ring solution exhibits a more intricate algebraic structure, and the monodromy matrix develops a third-order pole in the spectral parameter. 
Remarkably, this higher-order pole disappears precisely when the parameters are tuned to ensure regularity of the horizon, illustrating that physical regularity conditions are directly reflected in the analytic structure of the monodromy matrix. 
This observation indicates that not only the residue matrices of the monodromy matrix but also its analytic structure encode detailed information about the regularity and physical properties of spacetime fields. 
In particular, the analytic behavior of the monodromy matrix provides a powerful diagnostic tool for identifying physically acceptable solutions.

\medskip
Finally, we have analyzed the extremal limits of the Rasheed--Larsen solution, focusing on both the slowly rotating and fast-rotating branches. 
While the former is associated with nilpotent algebra similar to almost-BPS solutions, the latter exhibits a distinct structure involving idempotent elements of $\mathfrak{so}(4,4)$, indicating a qualitatively different algebraic characterization. 
We have also constructed an explicit $SO(4,4)$ duality transformation relating the slowly rotating extremal solution to an almost-BPS solution with a single center at the level of the coset matrix, thereby clarifying the relation between different extremal configurations within a unified framework. 
These results demonstrate that the monodromy-matrix approach provides a powerful and unified framework for analyzing extremal black holes, revealing deep connections between algebraic structures, analytic properties, and physical regularity conditions.

\medskip
From a more general perspective, Ref.~\cite{Bossard:2011kz} shows that the almost-BPS system does not exhaust the entire class of non-BPS extremal solutions for which the Einstein equations admit a factorization into a system of first-order differential equations. 
In particular, there exist more general non-BPS solutions beyond the almost-BPS ones that still exhibit such a factorized structure associated with nilpotent orbits of higher degree. 
Understanding the monodromy-matrix description of such solutions remains an important open problem. 
Another natural direction is to investigate whether almost-BPS black hole solutions, as well as more general non-BPS solutions, can be generalized to configurations with different asymptotics, such as asymptotically flat almost-BPS multi-black rings and black saturn configurations, in analogy with almost-BPS multi-center solutions in Taub--NUT space~\cite{Bena:2009en,Bena:2009ev}.
It would therefore be interesting to clarify the monodromy description of these solutions and to examine whether there exist almost-BPS or more general non-BPS solutions that preserve regularity under such generalizations. 
These directions are expected to provide further insights into the role of integrable structures in higher-dimensional gravity and supergravity theories.

\newpage

\appendix 

\section*{Appendix}

\section{Solving (anti-)self-duality relation}\label{sec:dual-sol}

In this appendix, we solve the self-duality relation (\ref{bps-eq}) and the anti-self-duality relation (\ref{abps-eq}) for explicit examples.

\medskip

Before solving the duality relations, we summarize our conventions for the Hodge star operator on the 4D Gibbons--Hawking space (\ref{4dGH}).  
To this end, we decompose the metric of the 4D Gibbons--Hawking space as
\begin{align}
    ds_4^2=(e^\psi)^2+ds_{3}^2\,,\qquad ds^2_{3}=\sum_{i=1}^{3}(e^i)^2\,,
\end{align}
where the 1-form $e^\psi$ expresses the $U(1)$ fiber direction and are given by
\begin{align}
    e^\psi=\frac{1}{\sqrt{V}}(d\psi+\varpi)\,,\qquad e^i=\sqrt{V}dx^i\,.
\end{align}
The volume form is taken as
\begin{align}
    \text{Vol}_4=+ e^{\psi}\wedge e^1\wedge e^2 \wedge e^3= +e^{\psi}\wedge \text{Vol}_3\,.
\end{align}
The action of the Hodge star operator is defined by
\begin{align}
    \star_{4}(e^{a_1}\wedge e^{a_2} \dots \wedge e^{a_p})=\frac{1}{(4-p)!}\epsilon^{a_1a_2\dots a_{p}}{}_{a_{p+1}\dots a_{4}}e^{a_{p+1}}\wedge \dots \wedge e^{a_4}\,,
\end{align}
where the totally antisymmetric tensor is normalized as $\epsilon^{\psi123}=+1$.

\subsection{Self-duality case}

We first consider the general solution to the self-duality equation (\ref{bps-eq}).
In terms of the orthonormal frame $e^a$, the 2-forms  $\Theta^I=dB^I$ on the 4D Gibbons--Hawking space can be expanded as
\begin{align}\label{theta-ansatz}
    \Theta^I=\alpha_2^I+e^{\psi}\wedge \beta_1^I\,,
\end{align}
where $\alpha_2$ and $\beta_1$ are arbitrary  2-form and 1-form on the 3D Euclidean space.
The action of the Hodge star operator $\star_4$ on each term is
\begin{align}
    \star_4\alpha_2=-\star_3\alpha_2^I\wedge e^\psi\,,\qquad \star_4(e^{\psi}\wedge \beta_1^I)=\star_3 \beta_1^I\,.
\end{align}
Imposing the self-duality condition $\star_4\Theta^I=\Theta^I$, we obtain
\begin{align}
    \star_4\Theta^I=-\star_3\alpha_2^I\wedge e^\psi+\star_3\beta_1^I=\alpha_2^I+ e^{\psi}\wedge \beta_1^I=\Theta^I\,,
\end{align}
which leads to
\begin{align}
    \beta_1^I=\star_3\alpha_2^I\,.
\end{align}
For later convenience, we redefine $\beta^I_1$ as
\begin{align}
    \bar{\beta}^I_1:=V^{-1}\beta_1^I\,,
\end{align}
and then $\Theta^I$ can be written as
\begin{align}
    \Theta^I=(d\psi+\varpi)\wedge \bar{\beta}^I_1+V\star_3\bar{\beta}^I_1\,.
\end{align}
Since $\Theta^I$ must be closed $d\Theta^I=0$, we have constraints
\begin{align}
    d\Theta^I=-(d\psi+\varpi)\wedge d\bar{\beta}^I+\star_3dV\wedge \bar{\beta}_1^I+d(V\star_3\bar{\beta}^I_1)=0\,,
\end{align}
where we used $\star_3d\varpi=dV$.
The first term means $d\bar{\beta}^I=0$, and we introduce the scalar functions $\gamma^I$ such that $\bar{\beta}^I=d\gamma^I$.
The remaining term can be rewritten as
\begin{align}
    d\star_3d(V \gamma^I)=0\,,
\end{align}
where we used $d\star_3dV=0$.
This equations have the solution
\begin{align}
    \gamma^I=-\frac{1}{V}K^I\,,
\end{align}
where $K^I$ is a harmonic function on $\mathbb{R}^3$.
Hence, we obtain
\begin{align}\label{Theta-sol-bps}
    \Theta^I=d(VK^I)\wedge(d\psi+\varpi)-V\star_3d(V^{-1}K^I)\,.
\end{align}
This expression implies that the 1-form potential $B^I$ takes the form
\begin{align}
    B^I=V^{-1}K^I(d\psi+\varpi)+\xi^I\,.
\end{align}
The condition the 1-form $\xi^I$ must satisfy is obtained by considering the exterior derivative on $B^I$, which gives
\begin{align}
    \Theta^I&=d(V^{-1}K^I)\wedge (d\psi+\varpi)+VK^I d\omega+d\xi^I\,,
\end{align}
Acting the Hodge star operator for $\mathbb{R}^3$ on the 3D base part and comparing it with (\ref{Theta-sol-bps}) lead to
\begin{align}
    V^{-1}K^I \star_3d\varpi+\star_3d\xi^I=V^{-1}K^IdV+\star_3d\xi^I= -Vd(V^{-1}K^I)\,.
\end{align}
We hence have
\begin{align}
    \star_3d\xi^I=-dK^I\,.
\end{align}

\medskip

Next, we solve the second equation of (\ref{bps-eq}). The left-hand side can be expanded as
\begin{align}\label{nabla4-z}
    d\star_4dZ_I&=-d(\star_3dZ_I)\wedge (d\psi+\varpi)+\dots\no\\
    &=-\frac{1}{2}C_{IJK}d\star_3d\left(\frac{K^JK^K}{V}\right) \wedge (d\psi+\varpi)+\dots\,.
\end{align}
On the other hand, the wedge product of $\Theta^J$ and $\Theta^K$ is expressed as
\begin{align}
    \Theta^J\wedge\Theta^K=-d\star_3d\left(\frac{K^JK^K}{V}\right) \wedge (d\psi+\varpi)+\dots\,.
\end{align}
Hence, we find
\begin{align}
    Z_I=L^I+\frac{1}{2}C_{IJK}V^{-1}K^JK^K\,,
\end{align}
where $L^I$ are harmonic functions on $\mathbb{R}^3$.

\medskip

Finally, we will solve the third equation of (\ref{bps-eq}).
By substituting expression (\ref{omega-ex}) for $\omega$ and extracting only the terms that include the $U(1)$ fiber direction, we obtain the following equation:
\begin{align}
   d\mu\wedge (d\psi+\varpi)+\star_4 [\mu\,d\varpi+d\omega_{\rm BW}]=Z_I dK^I\wedge (d\psi+\varpi)\,.
\end{align}
The second term in the left-hand side can be rewritten as 
\begin{align}
  \star_4 [\mu\,d\varpi+d\omega_{\rm BW}]
   &=-\frac{\mu}{V} \star_3 d\omega_{3}\wedge(d\psi+\varpi)-\frac{1}{V} \star_3 d\omega_{\rm BW}\wedge(d\psi+\varpi)\no\\
   &=\frac{1}{V} \left(-\mu\,dV- \star_3 d\omega_{\rm BW}\right)\wedge(d\psi+\varpi)\,.
\end{align}
In this way, we find
\begin{align}
   \star_3d\omega_{\rm BW}= Vd\mu-\mu dV-VZ_{I}dK^I\,.
\end{align}

\subsection{Anti-self-duality case}

Next, we will solve the anti-self-duality equation (\ref{abps-eq}). We can again employ the ansatz (\ref{theta-ansatz}) of $\Theta^I$.
Imposing the anti-self-duality condition $\star_4\Theta^I=-\Theta^I$, we obtain
\begin{align}
    \star_4\Theta^I=-\star_3\alpha_2^I\wedge e^\psi+\star_3\beta_1^I=-\alpha_2^I- e^{\psi}\wedge \beta_1^I=-\Theta^I\,,
\end{align}
which leads to
\begin{align}
    \beta_1^I=-\star_3\alpha_2^I\,.
\end{align}
As in the BPS case, we introduce $\bar{\beta}^I_1$ as
\begin{align}
    \bar{\beta}^I_1:=V^{-1}\beta_1^I\,,
\end{align}
and hence $\Theta^I$ can be written as
\begin{align}\label{theta-an}
    \Theta^I=(d\psi+\varpi)\wedge \bar{\beta}^I_1-V\star_3\bar{\beta}^I_1\,.
\end{align}
Since $\Theta^I$ must be closed $d\Theta^I=0$, we have constraints
\begin{align}
    d\Theta^I=-(d\psi+\varpi)\wedge d\bar{\beta}^I_1+\star_3dV\wedge \bar{\beta}_1^I-d(V\star_3\bar{\beta}^I_1)=0\,,
\end{align}
where we used $\star_3d\varpi=dV$ i.e. $d\varpi=\star_3 dV$. From the first term, we obtain $d\bar{\beta}^I=0$ and then denote $\bar{\beta}^I$ by $\bar{\beta}^I_1=-dK^{I}$, where $K^I$ are new scalar functions on $\mathbb{R}^3$. The remaining term leads to
\begin{align}
    -\star_3dV\wedge dK^I+dV\wedge \star_3dK^I+Vd(\star_3dK^I)=0\,.
\end{align}
Since the first two terms are identically canceled out, we obtain $d(\star_3dK^I)=0$, and hence the scalar functions $K^I$ are harmonic functions on $\mathbb{R}^3$.
This result implies that the 1-form potentials $B^I$ take the form
\begin{align}
    B^I=K^I(d\psi+\varpi)+\xi^I\,.
\end{align}
and the dipole field strength $\Theta^I$ becomes
\begin{align}
    \Theta^I&=dK^I\wedge (d\psi+\varpi)+K^I\star_3dV +d\xi^I\,.
\end{align}
On the other hand, using the expression (\ref{theta-an}) for $\Theta^I$ with $\bar{\beta}^I_1=-dK^{I}$, we obtain
\begin{align}
    \Theta^I
    &=dK^I\wedge (d\psi+\varpi)+V\star_3dK^I\,.
\end{align}
Comparing these expressions, we find that the condition the 1-form $\xi^I$ must satisfy 
\begin{align}\label{dxi-abps}
    \star_3d\xi^I=-K^IdV+VdK^I\,.
\end{align}
Next, we consider the second equation of (\ref{abps-eq}).
The wedge product of $\Theta^I$ and $\Theta^J$ is expressed as
\begin{align}
    \Theta^J\wedge \Theta^K=Vd\star_3d(K^JK^K)\wedge (d\psi+\varpi)+\dots\,.
\end{align}
Here, we used the fact that $K^I$ are harmonic functions, and the dots represents the collection of the terms without the $U(1)$ fiber direction. By using $d\star_4dZ_I=-d(\star_3dZ_I)\wedge (d\psi+\varpi)+\dots$ in (\ref{nabla4-z}), we obtain
\begin{align}
    d\star_3dZ_I=\frac{1}{2}C_{IJK}Vd\star_3d(K^JK^K)\,.
\end{align}
Finally, we will solve the third equation of (\ref{abps-eq}).
By substituting expression (\ref{omega-ex}) for $\omega$ and extracting only the terms that include the $U(1)$ fiber direction, we obtain the following equation:
\begin{align}
   d\mu\wedge (d\psi+\varpi)-\star_4 [\mu\,d\varpi+d\omega_{\rm aBPS}]=-Z_I dK^I\wedge (d\psi+\varpi)\,.
\end{align}
The second term in the left-hand side can be rewritten as 
\begin{align}
  \star_4 [\mu\,d\varpi+d\omega_{\rm aBPS}]
   &=-\frac{\mu}{V} \star_3 d\omega_{3}\wedge(d\psi+\varpi)-\frac{1}{V} \star_3 d\omega_{\rm aBPS}\wedge(d\psi+\varpi)\no\\
   &=\frac{1}{V} \left(-\mu\,dV- \star_3 d\omega_{\rm aBPS}\right)\wedge(d\psi+\varpi)\,.
\end{align}
In this way, we find
\begin{align}
    d(V\mu)+\star_3d\omega_{\rm aBPS}=-VZ_{I}dK^I\,.
\end{align}

\section{Proof of factorization of monodromy matrix for BW type solutions}\label{bwfac-proof}

In this appendix, we will show that the monodromy matrix $\cM_{\rm BW}(w)$ for BW-type solutions admits a factorization of the form (\ref{mm-fac}).

\medskip

We verify the factorized expression (\ref{mm-fac}) of the monodromy matrix $\cM_{\rm BW}(w)$ by explicitly evaluating the matrix product on the right-hand side of (\ref{mm-fac}).
To this end, we first rewrite the coset matrix (\ref{bw-gauge-mat2}).
Using the expression (\ref{lambda-ex}) of $\la_j$, we find the relation 
\begin{align}
    \frac{1}{r_i}=\frac{2}{\rho}\frac{\la_i}{1+\la_i^2}=-\nu_i\,.
\end{align}
Then, the coset matrix (\ref{bw-gauge-mat2}) can be rewritten as
\begin{align}\label{bw-gauge-mat3}
    M_{\rm BW}(z,\rho)=Y_{\rm BW}\exp\left(-\sum_{j=1}^{N}\nu_j\sfA_j\right)\,.
\end{align}
Since the matrices $\sfA_i$ satisfy
\begin{align}
    [\sfA_i,[\sfA_i,\sfA_j]]=0\,,
\end{align}
it is enough to use the truncated Baker--Campbell--Hausdorff formula 
\begin{align}
    e^{A}e^{B}=e^{A+B+\frac{1}{2}[A,B]}\,,
\end{align}
which is valid under the assumption $[A,[A,B]]=[B,[B,A]]=0$.
Using the identities 
\begin{align}
    Y_{\rm BW}^{-1}\sfA_j^{\natural}Y_{\rm BW}=\sfA_j\,,\qquad Y_{\rm BW}^{-1}[\sfA_i,\sfA_j]^{\natural}Y_{\rm BW}=-[\sfA_i,\sfA_j]\,,
\end{align}
we can rewrite the right-hand side in (\ref{mm-fac}) as 
\begin{align}
   &X_-M_{\rm BW}(z,\rho)X_+\no\\
   &=Y_{\rm BW}\exp\left(-\sum_{j=1}^{N}\frac{\nu_j\la_j}{\la- \la_j}\left(\sfA_j-\frac{1}{2}\sum_{\substack{k=1\\ k\neq j}}^{N}\left(\frac{\nu_k\la_k}{\la_{j,k}}-\frac{1}{w_j-w_k}\right)[\sfA_j,\sfA_k] \right)\right)\no\\
   &\quad \times \exp\left(-\sum_{j=1}^{N}\nu_j\sfA_j\right)
   \exp\left( \sum_{j=1}^{N}\frac{\nu_j\la \la_j}{1+\la \la_j}\left(\sfA_j+\frac{1}{2}\sum_{\substack{k=1\\ k\neq j}}^{N}\left(\frac{\nu_k\la_k}{\la_{j,k}}-\frac{1}{w_j-w_k}\right)[\sfA_j,\sfA_k] \right)\right)\no\\
   &=Y_{\rm BW}\exp\left(-\sum_{j=1}^{N}\frac{\nu_j\la_j}{\la- \la_j}\left(\sfA_j-\frac{1}{2}\sum_{\substack{k=1\\ k\neq j}}^{N}\left(\frac{\nu_k\la_k}{\la_{j,k}}-\frac{1}{w_j-w_k}\right)[\sfA_j,\sfA_k] \right)\right)\no\\
   &\quad\times \exp\Biggl( -\sum_{j=1}^{N}\frac{\nu_j}{1+\la \la_j}\sfA_j+\frac{1}{2}\sum_{\substack{j,k=1\\ k\neq j}}^{N}\frac{\nu_j\la \la_j}{1+\la \la_j}\left(\frac{\nu_k\la_j}{\la_{j,k}}-\frac{1}{w_j-w_k}\right)[\sfA_j,\sfA_k]\Biggr)\no\\
   &=Y_{\rm BW}\exp\Biggl( -\sum_{j=1}^{N}\nu_j\left(\frac{\la_j}{\la- \la_j}+\frac{1}{1+\la \la_j}\right)\sfA_j
   +\frac{1}{2}\sum_{\substack{j,k=1\\ k\neq j}}^{N}\cK_{jk}[\sfA_j,\sfA_k]\Biggr)\,,\label{fac-ex}
\end{align}
where $\cK_{jk}$ is defined by
\begin{align}
   \cK_{jk}= \frac{\nu_j\la_j}{\la- \la_j}\frac{\nu_k}{1+\la \la_k}
   +\frac{\nu_j\la_j}{\la- \la_j}\left(\frac{\nu_k\la_k}{\la_{j,k}}-\frac{1}{w_j-w_k}\right)
   +\frac{\nu_j\la \la_j}{1+\la \la_j}\left(\frac{\nu_k\la_j}{\la_{j,k}}-\frac{1}{w_j-w_k}\right)\,.
\end{align}
The first term in the exponent of Eq.~(\ref{fac-ex}) becomes
\begin{align}
    -\sum_{j=1}^{N}\nu_j\left(\frac{\la_j}{\la- \la_j}+\frac{1}{1+\la \la_j}\right)\sfA_j=-\sum_{j=1}^{N}\frac{\sfA_j}{w-w_j}\,.
\end{align}
Hence, the remaining task is to show that the second term in the exponent of Eq.~(\ref{fac-ex}) vanishes.
This can be easily shown as follows: 
\begin{align}
    &\sum_{\substack{j,k=1\\ k\neq j}}^{N}\cK_{jk}[\sfA_j,\sfA_k]\no\\
    &=\sum_{\substack{j,k=1\\ k\neq j}}^{N}\biggl(\frac{\nu_j\la_j\nu_k}{1+\la_j\la_k}\left(\frac{1}{\la-\la_j}-\frac{\la_k}{1+\la \la_k}\right)
   \no\\
    &\qquad +\frac{\nu_j\la_j}{\la- \la_j}\left(\frac{\nu_k\la_k}{\la_{j,k}}-\frac{1}{w_j-w_k}\right)
    +\frac{\nu_j\la \la_j}{1+\la \la_j}\left(\frac{\nu_k\la_j}{\la_{j,k}}-\frac{1}{w_j-w_k}\right)
    \biggr)[\sfA_j,\sfA_k]\no\\
    &=\sum_{\substack{j,k=1\\ k\neq j}}^{N}\biggl(\frac{\nu_j\la_j\nu_k}{1+\la_j\la_k}\left(\frac{1}{\la-\la_j}-\frac{\la_k}{1+\la \la_k}\right)
   \no\\
    &\qquad +\frac{\nu_j\la_j}{\la- \la_j}\left(\frac{\nu_k\la_k}{\la_{j,k}}-\frac{1}{w_j-w_k}\right)
    +\left(1-\frac{1}{1+\la \la_j}\right)\nu_j\left(\frac{\nu_k\la_j}{\la_{j,k}}-\frac{1}{w_j-w_k}\right)
    \biggr)[\sfA_j,\sfA_k]\no\\
    &=\sum_{\substack{j,k=1\\ k\neq j}}^{N}\biggl( \frac{\nu_j\la_j}{\la- \la_j}\left(\frac{\nu_k}{1+\la_j\la_k}+\frac{\nu_k\la_k}{\la_{j}-\la_{k}}-\frac{1}{w_j-w_k}\right)
   \no\\
    &\qquad
     -\frac{\nu_j}{1+\la \la_j}\left(\frac{\nu_k\la_j}{\la_{j}-\la_{k}}-\frac{1}{w_j-w_k}\right)+\nu_j\left(\frac{\nu_k\la_j}{\la_{j,k}}-\frac{1}{w_j-w_k}\right)\no\\
     &\qquad  -\frac{\nu_k}{1+\la \la_k}\frac{\nu_j\la_j\la_k}{1+\la_j\la_k}
    \biggr)[\sfA_j,\sfA_k]\no\\
    &=\sum_{\substack{j,k=1\\ k\neq j}}^{N}\biggl( \frac{\nu_j\la_j}{\la- \la_j}\left(\nu_k\left(\frac{\la_k}{\la_{j}-\la_{k}}+\frac{1}{1+\la_j\la_k}\right)-\frac{1}{w_j-w_k}\right)
   \no\\
    &\qquad
     -\frac{\nu_j}{1+\la \la_j}\left(\nu_k\left(-\frac{\la_j\la_k}{1+\la_j\la_k}+\frac{\la_j}{\la_{j}-\la_{k}}\right)-\frac{1}{w_j-w_k}\right)\no\\
     &\qquad +\nu_j\left(\frac{\nu_k\la_j}{\la_{j}-\la_{k}}-\frac{1}{w_j-w_k}\right)
    \biggr)[\sfA_j,\sfA_k]\no\\
    &=\sum_{\substack{j,k=1\\ k\neq j}}^{N}\biggl( \nu_j\left[\nu_k\left(-\frac{\la_j\la_k}{1+\la_j\la_k}+\frac{\la_j}{\la_{j}-\la_{k}}\right)-\frac{1}{w_j-w_k}\right]+\frac{\nu_j\nu_k\la_j\la_k}{1+\la_j\la_k}
    \biggr)[\sfA_j,\sfA_k]\no\\
    &=0\,.
\end{align}
In the last step we used the identity
\begin{align}
    \nu_k\left(-\frac{\la_j\la_k}{1+\la_j\la_k}+\frac{\la_j}{\la_{j}-\la_{k}}\right)-\frac{1}{w_j-w_k}=0\,,
\end{align}
which follows directly from (\ref{lambda-ex}).

\section{Details of a Scalar Field and Monodromy Matrix in Rasheed-Larsen Black Hole}

In this appendix, we collect the explicit expression of the scalar field $\sigma$ in (\ref{4dybh-scalar}), and some components of the monodromy matrix (\ref{monodromy-RL}) for the Rasheed-Larsen rotating black hole solution.

\subsection{Explicit expression of scalar field $\sigma$}\label{sec:sigma-ex}

Here, we write down the explicit expressions of the scalar field $\sigma$ for the Rasheed-Larsen rotating black hole solution.
The scalar field $\sigma$ is given by
\begin{align}
    \sigma&=-\frac{\sigma'}{4m^3(p+q)^2H_1(H_1H_3-(\omega_t^a)^2)}\,.
\end{align}
The numerator of $\sigma$ is given by
\begin{align}\label{sigma-n}
    \sigma'&=\frac{\sqrt{p}}{\sqrt{q}}a Q \cos\theta\biggl[
    8 a^4 \cos^4\theta m^2 p (p+q)^2\no\\
    &\quad+a^2 \cos^2\theta \Bigl(16 m^5 (2 p-q) (p+q)+8 m^4 \left(2 p^3+p^2 (q-4 r)-2 p q r-q^2 (q-2 r)\right)\no\\
    &\quad-4 m^3 p (p+q) \left(4 p^2+8 r (p+q)+2 p q+q^2\right)\no\\
    &\quad+4 m^2 p \left(r (p+q) \left(4 p^2+2 p q+q^2\right)+q \left(2 p^3+2 p^2 q+3 p q^2+q^3\right)+4 r^2 (p+q)^2\right)-p^3 q^3 (2 p+q)\Bigr)\no\\
    &\quad+m^2 \Bigl(4 r^3 (p+q) \left(4 m^2 (q-2 p)-8 m p (p+q)+p \left(4 p^2+2 p q+q^2\right)\right)\no\\
    &\quad-8 r (2 m-p) (p+q) \left(q^2 \left(-5 m^2-2 m p+p^2\right)-2 m q \left(m^2-m p+p^2\right)+4 m^2 p (m+p)+m q^3\right)\no\\
    &\quad+4 r^2 (p+q) \left(2 q \left(-6 m^3+7 m^2 p-3 m p^2+p^3\right)-q^2 (2 m-p) (5 m+4 p)+12 m p (2 m-p) (m+p)-p q^3\right)\no\\
    &\quad+q (p-2 m)^2 (2 m-q) \left(4 m p^2-q^2 (4 m+p)-2 p q (m+p)\right)+8 p r^4 (p+q)^2\Bigr)
    \biggr]\no\\
    &\quad+m P\biggl[2 a^4 \cos^4\theta (p+q) \left(8 m^3 (p+q)+m^2 (4 p (2 p-q)-8 r (p+q))+p q^2 (q-2 p)\right)\no\\
    &\quad+a^2 \cos^2\theta \Bigl(-32 m^2 r^3 (p+q)^2\no\\
    &\quad+(2 m-p) (2 m-q) 
    \Bigl(4 m^3 \left(2 p^2+p q+q^2\right)+2 m^2 q \left(-2 p^2-3 p q+q^2\right)\no\\
    &\quad+2 m q^2 \left(-4 p^2-2 p q+q^2\right)-p q^3 (2 p+q)\Bigr)\no\\
    &\quad+2 r^2 (p+q) \left(48 m^3 (p+q)-4 m^2 q (4 p+q)+p q^2 (q-2 p)\right)\no\\
    &\quad-2 r \Bigl(8 m^4 \left(6 p^2+9 p q+5 q^2\right)-8 m^3 q (p+q) (4 p+q)+2 m^2 q \left(4 p^3-2 p^2 q-p q^2+q^3\right)\no\\
    &\quad-2 m p q^2 (2 p-q) (p+q)+p q^3 \left(-2 p^2+p q+2 q^2\right)\Bigr)\Bigr)\no\\
    &\quad+m^2 \Bigl(2 r \left(-4 m^2 (4 p+3 q)+4 m (2 p-q) (p+q)+q^2 (2 p+q)\right)\no\\
    &\quad+(2 m-p) \left(4 m^2-q^2\right) (2 p+q)+4 r^2 (p+q) (6 m-2 p+q)-8 r^3 (p+q)\Bigr) \no\\
    &\quad\times\left(4 m^2 q-2 m (p+q) (q+2 r)+p q^2+2 r^2 (p+q)+2 q r (p+q)\right)\biggr]
\end{align}

\subsection{Explicit expressions of monodromy matrix}\label{sec:RL-mono}

Here, we present explicit expressions for the factors $F^{\pm}_{ij}$ appearing in the residue matrices of the monodromy matrix (\ref{monodromy-RL}). Their expressions are given as follows:
\begin{align}
    F_{14}^{\pm}&=\frac{2 (p+q) }{ \sqrt{q}\cF_3^{\pm}}\left(\frac{a q}{\sqrt{p}} P-\frac{m  }{\sqrt{q}}Q(p\pm 2 \alpha)\right)\,,\\
     F_{17}^{\pm}&=-\frac{2 \sqrt{p}}{\cF_1^{\pm}(\cF_2^{\pm})^2 \cF_3^{\pm} \sqrt{q}}\left((F_{17}^{(1)}\pm \alpha F_{17}^{(2)})-\frac{2(p+q)}{\sqrt{p} \sqrt{q}}P Q(F_{17}^{(3)}\pm\alpha F_{17}^{(4)})\right)\,,\\
     F_{35}^{\pm}&=\frac{4 m \sqrt{q} (p+q)^2}{\cF_1^{\pm}\cF_3^{\pm} \sqrt{p}}\left(m\pm\frac{2 \alpha J}{a\sqrt{pq}}\right) \left(a m+\frac{2\alpha}{\sqrt{pq}}PQ\right)\,,\\
     F_{38}^{\pm}&=\frac{2 (p+q)}{\cF_1^{\pm}(\cF_2^{\pm})^2 \sqrt{p}}\left(\frac{P}{\sqrt{p}}(F_{38}^{(1)}\pm \alpha F_{38}^{(2)})+\frac{a Q }{\sqrt{q}}(F_{38}^{(3)}\pm \alpha F_{38}^{(4)})\right)\,,
\end{align}
where
\begin{align}
    \cF_1^{\pm}&=a \sqrt{p^2-4m^2}\sqrt{q^2-4m^2}+4 m^3+m p q\mp 2 m \alpha (p+q)\,,\\
    \cF_2^{\pm}&=m^2 \left(4 m^2+p^2\right) \left(4 m^2-q^2\right)-a^2 \left(16 m^4-8 m^2 q^2+p^2 q^2\right)+2 a m p q \sqrt{p^2-4m^2}\sqrt{q^2-4m^2}\no\\
    &\quad \mp4 m \alpha \left(a q \sqrt{p^2-4m^2}\sqrt{q^2-4m^2}+m p \left(4 m^2-q^2\right)\right)\,,\\
    \cF_3^{\pm}&=-a \sqrt{p^2-4m^2}\sqrt{q^2-4m^2}+4 m^3+m p q\mp2 m \alpha (p+q)\,,
\end{align}
and
\begin{align}
\begin{split}
    F_{17}^{(1)}&=2 a m \Bigl(a^4 \left(4 m^2+p q\right)^2 \left(16 m^6 \left(5 p^2-10 p q-3 q^2\right)+8 m^4 q \left(5 p^3-6 p^2 q+5 p q^2+2 q^3\right)+5 m^2 p^2 q^2 (p-q)^2-p^4 q^4\right)\no\\
    &\quad-2 a^2 m^4 \left(4 m^2-q^2\right) \Bigl(64 m^6 \left(5 p^2-10 p q-3 q^2\right)+16 m^4 \left(5 p^4+20 p^3 q-21 p^2 q^2+2 p q^3+2 q^4\right)\no\\
    &\quad+4 m^2 p^2 q (p-q) \left(10 p^2+29 p q+q^2\right)+p^4 q^2 (p-q) (5 p+11 q)\Bigr)\no\\
    &\quad+m^4 \left(q^2-4 m^2\right)^2 \left(16 m^6 \left(5 p^2-10 p q-3 q^2\right)+40 m^4 p^2 \left(p^2+2 p q-q^2\right)+m^2 p^4 \left(p^2+22 p q+25 q^2\right)+p^6 q^2\right)\Bigr)\,,\\
    F_{17}^{(2)}&=-a^5 m \left(4 m^2+p q\right)^3 \left(16 m^4 (p-3 q)+8 m^2 q \left(p^2-p q+2 q^2\right)+p^2 q^2 (p-3 q)\right)\no\\
    &\quad+2 a^3 m^3 \left(4 m^2-q^2\right) \left(4 m^2+p q\right) \Bigl(64 m^6 (p-3 q)+16 m^4 \left(5 p^3-p^2 q-8 p q^2+2 q^3\right)\no\\
    &\quad+4 m^2 p^2 q \left(10 p^2-3 p q-9 q^2\right)+p^4 q^2 (5 p+q)\Bigr)\no\\
    &\quad-a m^5 \left(q^2-4 m^2\right)^2 \left(64 m^6 (p-3 q)+16 m^4 p \left(10 p^2-5 p q-11 q^2\right)+20 m^2 p^3 \left(p^2+7 p q+2 q^2\right)+p^5 q (5 p+17 q)\right)\,,\\
    F_{17}^{(3)}&=-a^6 \left(4 m^2+p q\right)^3 \left(16 m^4+8 m^2 q (p-2 q)+p^2 q^2\right)\no\\
    &\quad+a^4 m^2 \left(4 m^2+p q\right) \left(768 m^8+128 m^6 \left(5 p^2+6 p q-5 q^2\right)+32 m^4 q \Bigl(10 p^3+3 p^2 q-10 p q^2+2 q^3\right)\no\\
    &\quad+8 m^2 p^2 q^2 \left(5 p^2-6 p q-9 q^2\right)-5 p^4 q^4\Bigr)\no\\
    &\quad-a^2 m^4 \left(4 m^2-q^2\right) \Bigl(768 m^8+320 m^6 (4 p-q) (p+q)+16 m^4 p \left(5 p^3+40 p^2 q+8 p q^2-13 q^3\right)\no\\
    &\quad+4 m^2 p^3 q \left(5 p^2+33 p q+12 q^2\right)+5 p^5 q^3\Bigr)\no\\
    &\quad+m^6 \left(q^2-4 m^2\right)^2 \left(64 m^6+80 m^4 p (2 p+q)+20 m^2 p^3 (p+2 q)+p^5 q\right)\,,\\
    F_{17}^{(4)}&=-2 m^2 \Bigl(a^4 \left(4 m^2+p q\right)^2 \left(16 m^4 (5 p+q)-8 m^2 q \left(-5 p^2+5 p q+2 q^2\right)+p^2 q^2 (5 p-7 q)\right)\no\\
    &\quad-2 a^2 m^2 \left(4 m^2-q^2\right) \Bigl(64 m^6 (5 p+q)+16 m^4 \left(5 p^3+15 p^2 q-4 p q^2-2 q^3\right)\no\\
    &\quad+4 m^2 p^2 q \left(10 p^2+21 p q-q^2\right)+p^4 q^2 (5 p+9 q)\Bigr)\no\\
    &\quad+m^4 \left(q^2-4 m^2\right)^2 \left(16 m^4 (5 p+q)+40 m^2 p^2 (p+q)+p^4 (p+5 q)\right)\Bigr)\,,
\end{split}
\end{align}

\begin{align}
\begin{split}
    F_{38}^{(1)}&=a^4 \bigl(256 m^9 (4 p+q)+256 m^7 q (3 p+q) (p-2 q)+32 m^5 q^2 \left(6 p^3-11 p^2 q+8 p q^2+4 q^3\right)\no\\
    &\quad+16 m^3 p^2 q^3 (p-q) (p-2 q)-3 m p^4 q^5\bigr)\no\\
    &\quad-2 a^2 m^3 \left(4 m^2-q^2\right) \left(64 m^6 (4 p+q)+16 m^4 \left(2 p^3+9 p^2 q-10 p q^2-4 q^3\right)+4 m^2 p^2 q \left(2 p^2+4 p q-9 q^2\right)+p^4 q^3\right)\no\\
    &\quad+m^5 \left(q^2-4 m^2\right)^2 \left(16 m^4 (4 p+q)+8 m^2 p^2 (2 p+3 q)+p^4 q\right)\,,\\
    F_{38}^{(2)}&=-2 m \bigl(a^4 \left(4 m^2+p q\right) \left(64 m^6+16 m^4 q (3 p-8 q)+4 m^2 q^2 \left(3 p^2-4 p q+8 q^2\right)+p^2 q^3 (p-4 q)\right)\no\\
    &\quad-2 a^2 m^2 \left(4 m^2-q^2\right) \left(64 m^6+16 m^4 (p+2 q) (3 p-2 q)+4 m^2 p q \left(6 p^2-3 p q-10 q^2\right)+3 p^4 q^2\right)\no\\
    &\quad+m^4 \left(4 m^2-q^2\right)^2 \left(16 m^4+8 m^2 p (3 p+2 q)+p^3 (p+4 q)\right)\bigr)\,,\\
    F_{38}^{(3)}&=a^4 \left(4 m^2+p q\right) \left(64 m^6 (p-4 q)+16 m^4 q \left(3 p^2-4 p q+8 q^2\right)+4 m^2 p^2 q^2 (3 p-8 q)+p^4 q^3\right)\no\\
    &\quad+2 a^2 m^2 \bigl(-256 m^8 (p-4 q)-64 m^6 (p-2 q) \left(3 p^2+4 p q-3 q^2\right)\no\\
    &\quad-32 m^4 p q \left(3 p^3-p q^2+6 q^3\right)-4 m^2 p^3 q^2 (3 p-7 q) (p+q)+p^5 q^4\bigr)\no\\
    &\quad+m^4 (4m^2-q^2) \left(64 m^6 (p-4 q)+16 m^4 p \left(6 p^2-8 p q-13 q^2\right)+4 m^2 p^3 \left(p^2+12 p q+2 q^2\right)+3 p^5 q^2\right)\,,\\
    F_{38}^{(4)}&=8 a^2 m^2 \left(64 m^6 \left(p^2-3 p q-q^2\right)+16 m^4 q \left(3 p^3-4 p^2 q+4 p q^2+2 q^3\right)+4 m^2 p^2 q^2 \left(3 p^2-5 p q+q^2\right)+p^4 q^3 (p-2 q)\right)\no\\
    &\quad-8 m^4 (4m^2-q^2)\left(16 m^4 \left(p^2-3 p q-q^2\right)+4 m^2 p^2 (p-q) (p+3 q)+p^4 q (p+2 q)\right)\,.
\end{split}
\end{align}

\end{document}